\documentclass[12pt]{emulateapj}
\usepackage{apjfonts}
\usepackage{natbib}
\usepackage{amsfonts}
\usepackage{amsmath}
\usepackage[usenames]{color}
\usepackage{epsfig}
\usepackage{graphicx}
\usepackage{amssymb}

\newcommand{\vecn}{{\bf n}}
\newcommand{\vecr}{{\bf r}}
\newcommand{\p}{\partial}
\newcommand{\ve}{\varepsilon}
\newcommand{\vk}{\varkappa}
\newcommand{\be}{\begin{equation}}
\newcommand{\ee}{\end{equation}}

\newcommand{\bea}{\begin{eqnarray}}
\newcommand{\eea}{\end{eqnarray}}

\shortauthors{Abdikamalov et al.}
\shorttitle{Monte Carlo Neutrino Transport}

\begin{document}

\slugcomment{Submitted to ApJ.}

\title{A New Monte Carlo Method for Time-Dependent Neutrino
  Radiation Transport}  

\author{
Ernazar Abdikamalov\altaffilmark{1},
Adam Burrows\altaffilmark{2}, 
Christian D. Ott\altaffilmark{1,3,4,5},
Frank L\"offler\altaffilmark{4},
Evan O'Connor\altaffilmark{1},
Joshua C. Dolence\altaffilmark{2}, and
Erik Schnetter\altaffilmark{6,7,4,8}}

\altaffiltext{1}{TAPIR, California Institute of Technology, MC 350-17,
  1200 E California Blvd., Pasadena, CA 91125, USA} 
\altaffiltext{2}{Department of Astrophysical Sciences, Princeton
  University, Peyton Hall, Ivy Lane, Princeton, NJ 08544, USA}  
\altaffiltext{3}{Kavli Institute for the Physics and Mathematics of
  the Universe, Todai Institutes for Advanced Study, the University of
  Tokyo, Kashiwa, Japan 277-8583 (Kavli IPMU, WPI)}
\altaffiltext{4}{Center for Computation \& Technology, Louisiana
State University, 216 Johnston Hall, Baton Rouge, LA 70803, USA}
\altaffiltext{5}{Alfred P. Sloan Research Fellow}
\altaffiltext{6}{Perimeter Institute for Theoretical Physics,
  Waterloo, ON, Canada} 
\altaffiltext{7}{Department of Physics, University of Guelph, Guelph,
  ON, Canada} 
\altaffiltext{8}{Department of Physics \& Astronomy, Louisiana State
  University, Baton Rouge, LA 70803, USA}

\email{abdik@tapir.caltech.edu}

\begin{abstract}
Monte Carlo approaches to radiation transport have several attractive
properties such as simplicity of implementation, high accuracy, and
good parallel scaling. Moreover, Monte Carlo methods can handle
complicated geometries and are relatively easy to extend to multiple
spatial dimensions, which makes them potentially interesting in
modeling complex multi-dimensional astrophysical phenomena such as
core-collapse supernovae. The aim of this paper is to explore Monte
Carlo methods for modeling neutrino transport in core-collapse
supernovae. We generalize the implicit Monte Carlo photon transport
scheme of Fleck \& Cummings and gray discrete-diffusion scheme of
Densmore et al. to energy-, time-, and velocity-dependent neutrino   
transport. Using our 1D spherically-symmetric implementation, we show
that, similar to the photon transport case, the implicit scheme enables
significantly larger timesteps compared with explicit time
discretization, without sacrificing accuracy, while the
discrete-diffusion method leads to significant speed-ups at high
optical depth. Our results suggest that a combination of spectral,
velocity-dependent, implicit Monte Carlo and discrete-diffusion Monte
Carlo methods represents a robust approach for use in neutrino
transport calculations in core-collapse supernovae. Our
velocity-dependent scheme can easily be adapted to photon transport.
\end{abstract} 

\keywords{Hydrodynamics, Neutrinos, Radiative Transfer, Stars:
  Evolution, Stars: Neutron, Stars: Supernovae: General} 

\section{Introduction}
\label{sec:intro}

Core-collapse supernovae (CCSNe) are among the most energetic
explosions in the Universe. They mark the end of massive
star evolution and are powered by the release of gravitational energy
in the collapse of the stellar core to a proto-neutron star
(PNS). 
Despite  
decades of effort, the details of the explosion mechanism remain obscure
and represent a formidable computational challenge. Simulations
in spherical symmetry with the latest nuclear and neutrino physics,
sophisticated neutrino transport, and up-to-date progenitor models
fail to explode, suggesting that multi-dimensional effects
are probably crucial for producing explosions~\citep{Herant:92,
  Herant:94,Burrows:95,Janka:96}. 
Indeed, modern 2D (axisymmetric) simulations, while still ambiguous
and problematic, exhibit fluid instabilities and turbulence that lead to more
favorable conditions for explosion~\citep{Marek:09,Ott:08,Yakunin:10}. 
Moreover, recent calculations by~\cite{Nordhaus:10,
Takiwaki:12, Hanke:11} show that the role of the third spatial
dimension cannot be neglected, and conclusive CCSN simulations will
have to be carried out in full 3D. 

One of the most important ingredients in modeling CCSNe is 
neutrino transfer. Neutrinos play a crucial role in
transporting energy from the PNS to the material behind the supernova
shock, influencing the hydrodynamic and thermodynamic conditions 
of the explosion. At the same time, accurate neutrino transport 
is one of the most complicated and computationally
expensive aspects of numerical CCSN modeling.   

The transport methods used in previous 1D and 2D simulations of CCSNe
exhibit drawbacks that are likely to become particularly
pronounced in 3D calculations. For example, the ray-by-ray
method~\cite[used, e.g., in][]{Marek:09, Bruenn:06, Takiwaki:12}
solves a series of coupled 1D transport calculations along a number of
radial rays. While computationally less expensive compared with a full
3D scheme, this method does not incorporate lateral transport,
exaggerates local heating  and cooling, and cannot easily follow
off-center motions. The $S_N$ scheme~\cite[used, e.g.,
  in][]{Liebendoerfer:04, Ott:08, Sumiyoshi:12}, while adequate for 1D
calculations, suffers from so-called ray-effects in the higher dimensional
case~\citep{Castor:04} and involves a complex solution and
parallelization scheme~\cite[][]{McClarren:08, Swesty:06b}.    
Although improving on these drawbacks is the topic of ongoing
  research~\citep[][]{Godoy:12}, it is worthwhile to
  explore alternative approaches to neutrino transport.
One such approach is the Monte Carlo method and the aim of this paper
is to explore its use in core-collapse supernova simulations.     

This paper is organized as follows. 
In Section~\ref{sec:ccsn_status}, 
we summarize the current status of the 
CCSN simulations, after which we present a more detailed introduction
to Monte Carlo transport methods methods (Section~\ref{sec:mcrt_intro}). Then, in
Section~\ref{sec:MCImplementation}, we describe a
simple Monte Carlo method for solving the equations of time-dependent
radiative transfer. For this, we restrict ourselves to the 1D
spherically-symmetric problem with a static matter background
that is assumed to emit, absorb, and scatter radiation
In Section~\ref{sec:IMC_photon}, we describe some key aspects of a
widely used method for the time discretization of
the nonlinear photon transport equations
  by~\cite{Fleck:71}. In Section~\ref{sec:IMC_neutrino}, we extend
  this method to \emph{neutrino} transport and provide a Monte Carlo
  interpretation of the resulting equations. 
In Section~\ref{sec:ddmc},
  we generalize the discrete-diffusion Monte Carlo scheme
  of~\cite{Densmore:07} to energy-dependent
  neutrino transport. In
  Section~\ref{sec:velocity}, we describe the extension of 
this scheme to the case when matter is moving. In
Section~\ref{sec:tests}, we present tests of the numerical
implementation of these schemes, while in Section~\ref{sec:conclusion}
we provide conclusions and thoughts about future work. 

\section{Summary of current core-collapse supernova simulations}
\label{sec:ccsn_status}

Some basic aspects of the CCSN mechanism are well established. The collapse
of the evolved stellar core to a PNS and its
evolution to a compact cold neutron star (NS) involve huge amounts of
gravitational energy ($\sim$3$\times 10^{53} $ ergs). The explosion mechanism must convert 
a $\sim$$10^{51}$-erg fraction of this energy into the kinetic and internal
energy of the exploding stellar envelope to match observations of 
core-collapse supernovae. However, after four decades of research, the details of
this process remain obscure. 

The hydrodynamical shock wave produced by core bounce stalls
soon after formation, and it must be reenergized to lead to a 
supernova explosion~\citep{Bethe:90}. The delayed neutrino mechanism
relies on an imbalance between neutrino heating and cooling behind the
shock to deposit sufficient energy to revive the shock and drive the
explosion on a timescale of hundreds of milliseconds. However, in
spherical symmetry, this mechanism has been shown to fail for regular
massive stars~\citep{Burrows:95, Rampp:00, Liebendoerfer:01, Liebendoerfer:05,
Thompson:03, Sumiyoshi:05}, while for the low-mass $8.8 M_\odot$ progenitor
of~\cite{nomoto:88}, \cite{Kitaura:06} obtain a
spherical explosion after a short post-bounce
delay ~\citep[see also][]{Burrows:07}. This progenitor can explode in
1D because its envelope is extremely rarefied, but the explosion
energy is too low ($\lesssim 10^{50}$ erg) to match observations of typical CCSNe.

Increases in computer power in the 1990s enabled detailed numerical
simulations in 2D~\citep{Burrows:92, Herant:92, Herant:94, Burrows:95,
  Janka:96}, which demonstrated the existence and potential importance
of multi-D hydrodynamical instabilities and neutrino-driven convection
in the core-collapse supernova phenomenon. More recent calculations in 2D 
have shown that these instabilities and convection increase the dwell time of
matter in the gain region~\citep{Murphy:08}, a region where neutrino
heating exceeds neutrino cooling. This results in greater neutrino
energy deposition efficiency behind the shock and, thus, creates more
favorable conditions for explosion~\citep{Murphy:08, Burrows:93,
Janka:01, Thompson:05, Pejcha:12}, with some of these calculations
leading to weak delayed neutrino-driven explosions~\citep{Buras:06a,
Buras:06b, Bruenn:06, Mezzacappa:07, Bruenn:09, Marek:09,
Yakunin:10}.   

However, despite obtaining explosions, these simulations pose new
questions. First of all, the explosion energies obtained in these 2D
simulations are typically one or two orders magnitude smaller than the
canonical value. Moreover, these exploding models employ a soft
version of the nuclear equation of state (EOS) by~\cite{Lattimer:91} with
an incompressibility at nuclear densities, $K$, of $180$ MeV. Such
a soft EOS is ruled out by the recent observation of a $\sim$$2 M_\odot$ 
neutron star~\citep{Demorest:10}. \cite{Marek:09} did not obtain 
an explosion for their model with $K=263$
MeV~\citep{Hillebrandt:85}, suggesting that it may be harder
to obtain explosions with stiffer EOSs. Furthermore, while the
Garching and Tokyo groups have found marginal
explosions~\citep{Marek:09, Suwa:10}, the Oak Ridge group
reports stronger and earlier explosions for a 
wider range of progenitors~\citep{Bruenn:09,Yakunin:10},
though ~\cite{Burrows:06, Burrows:07a,Ott:08} did not see neutrino-driven
explosions for progenitors greater in mass than $\sim$8.8 M$_\odot$.  

Although it is not yet entirely clear why 2D simulations by different
groups produce different results, the marginality of explosion in~\cite{Marek:09}
and~\cite{Suwa:10} hints at the possible importance of the third spatial
dimension in explosion dynamics. Three-dimensional (3D) fluid
dynamics has different flow patterns than in 2D.
This fact could have an impact on the existence and the growth rate
of nonradial hydrodynamic instabilities in the supernova core, which
could alter the dynamics of the neutrino-driven explosion. Indeed, recent
simulations by \cite{Nordhaus:10}, \cite{Takiwaki:12}, and \cite{Hanke:11} found
significant differences in the explosion dynamics between 2D and 3D
simulations. 

\section{Deterministic and Monte Carlo Transport}
\label{sec:mcrt_intro}

Two fundamentally different computational approaches exist to solve
the radiation transport equations, each with well-established schools of
thought, and with advantages and drawbacks. 
They are the \emph{deterministic} approach and
the \emph{Monte Carlo} approach\footnote{Recently,
  hybrid methods that combine Monte Carlo and deterministic methods
  have also found some success~\citep{Wollaber:09}.}.    

Deterministic methods involve the discretization of the full or
approximate transport equation on a phase space grid, generating a
coupled system of algebraic equations. The
optimal way to represent 
the transport equation on these grids for a given situation is
frequently far from obvious and is a research topic in itself. For
example, one may chose finite-difference, finite element, or finite
volume representations. In the
momentum-angle variables, discrete ordinates (as in
the $\mathrm{S_N}$ method) or spherical harmonic expansions (as in the
$\mathrm{P_N}$ method) are often employed~\citep{Castor:04}. Once the 
equations have been discretized, the solution of the resulting system
of equations is completely ``determined'' for given initial and
boundary conditions. A numerical solution of this system produces the
global (i.e., over all of phase space) solution of the transport
equation, and provides numerical estimates of the radiation field in the
entire problem domain. The global nature of such solutions is one of
the main advantages of deterministic methods. However, the
discretization process introduces (often significant) truncation
errors; for example, a simple $\mathrm{P_N}$ may suffer from negative
energies~\citep{McClarren:08}. Reducing such errors is an area of
active research~\citep[e.g.,][]{McClarren:10} 

There are various deterministic approaches to both approximate [e.g., the
diffusion approximation~\citep{Pomraning:73}] and full multi-angle and
multi-energy transport. For the latter, one of the most widely used
approaches is the discrete-ordinates $\mathrm{S_N}$
method which solves the transport equation along several particular
directions in each spatial zone~\citep{Castor:04}. However, such
methods have several drawbacks. First and foremost, they suffer from
ray effects~\citep{Morel:03}. Because of the discrete nature of the
angular representation, this method introduces large spatial
oscillations in, e.g., energy density. Also, $\mathrm{S_N}$ methods
employ a very complex solution and parallelization procedures. For 
large systems, direct inversion of the transport operator can be very
inefficient, forcing one to resort to complicated iterative
approaches~\citep{Adams:02}. A further limitation of such
  methods that has emerged more recently is their somewhat limited
  parallel scalability. \cite{Swesty:06b} shows that a variant of
  the $\mathrm{S_N}$ scheme based on Krylov iterative frameork can
  scale well up to only $\sim 256$ cores for a uniform grid (although
  the exact details somewhat depend on problem parameters).
This is significant because any 3D radiation transport
calculation is likely to require parallel calculations on many
thousands of processors. Improving the parallel scalability of such
methods is an area of active research~\citep{Swesty:06b, Godoy:12}.   

In contrast with deterministic approaches, in Monte Carlo methods
one does
not \emph{solve} the transport equation; instead, such methods employ
pseudo-random number sequences to \emph{directly simulate} the
transport of radiation particles through matter. Since the maximum
number of particles that one can simulate is constrained by
computer
memory and CPU power, typically many fewer Monte Carlo particles 
than actual physical particles participate in the numerical transport process, implying that each
Monte Carlo particle represents some packet of many physical particles. 
Based on local emissivity, Monte
Carlo particles are sampled in various zones with various frequencies,
directions, and spatial coordinates. Then, each particle is tracked
through the problem domain until it crosses a boundary or is
absorbed. If a sufficiently large number of Monte Carlo particles is
simulated, then one can obtain an accurate estimate of the average
behavior of the system. This is the basic idea behind Monte Carlo
radiation transport\footnote{\cite{Zink:08} suggested a Monte Carlo
  discretization of the general relativistic transport equations. In
  this scheme, contrary to traditional Monte Carlo radiation
  transport, one \emph{solves} the transport equations \emph{using}
  Monte Carlo methods, instead of directly simulating radiation
  transport using pseudo-random numbers.}. 

In this paper, the term ``particle'' refers
to a single radiation particle, such as a photon or a neutrino. We
use the term ``MC particle'' or ``MCP'' to refer to a Monte Carlo particle
that represents a packet of physical particles. The number 
of physical particles represented by a given MCP
will be referred to as the weight of the particle.

Monte Carlo methods have some interesting features that may be
particularly advantageous in multi-dimensional transport
simulations. \emph{First}, such methods are generally easily adapted to work
with complicated geometries, meshes, and multiple spatial dimensions. This is
because the most geometry-dependent aspect of Monte Carlo methods
consists of the algorithms that deal with tracking MC particles
through spatial zones in the problem domain. Most of the rest of
the algorithm represents geometry-independent operations such as
equation of state and opacity calculations, and calculations that
depend on the length of particle paths. Once this tracking is
implemented for a given mesh type, the rest of a Monte Carlo algorithm
is relatively straightforward~\citep{Gentile:09}.

\emph{Second}, Monte Carlo methods model physical processes in a more
direct and simple way than deterministic methods. For example,
anisotropic scattering is handled easily, just by changing the Monte
Carlo particle's direction when a scattering event
occurs~\citep{Castor:04}. Since scattering is modeled by deflecting
the paths of MC particles in a simulation, it can represent the
angular behavior of a scattering kernel with more
fidelity~\citep{Gentile:09}. The angle of the scattered radiation
particle can be chosen from a probability distribution function that
can easily be constructed for scattering kernels of any (physically
reasonable) functional form. The same is also true for the operation
of selecting the energy of scattered particles in the case of
inelastic scattering. Incorporation of anisotropic and inelastic
scattering in deterministic methods is far more involved and much less
straightforward [but is nevertheless doable~\citep{Mezzacappa:93a}].     

The MC method can also be modified to account for material motion in a
relatively straightforward manner using a mixed-frame
formalism~\citep{Mihalas:82,Hubeny:07}
Emission takes places in the fluid frame and radiation
particles are Lorentz-transformed into the Eulerian lab frame, where
transport is performed. This method produces the correct distribution
for a fluid moving with relativistic velocity. Velocity-dependence in
deterministic methods is again much more complicated to implement.      

Monte Carlo methods also have the advantage that if the entire
problem domain (i.e., meshes, hydrodynamic and thermodynamic
variables, etc.) can fit into the memory of one CPU node, then 
parallelization is trivial and strongly scalable. One just simulates
copies of the problem on a number of processors, where each processor
carries a fraction of the total number of particles. Quantities
accumulated over all particles (e.g., the total emitted or
deposited energy or lepton number, etc.) are then summed over all
processors. This approach is usually referred to as mesh
replication~\citep{Gentile:09}. Even if the problem domain does not
fit into the memory of one CPU node, it is frequently possible to
decompose the domain of the problem onto separate nodes and maintain
a high degree of parallel
scalability~\citep{Brunner:06,Brunner:09}.   

There are, however, also some negative aspects of Monte Carlo
methods. The most serious property is the noise intrinsic to random
processes. Monte Carlo methods exhibit statistical fluctuations in
quantities such as radiation energy density and temperature. According
to the central limit theorem, this statistical error scales as
$N^{-1/2}$, where $N$ is the number of MC particles used in the
calculation~\citep{Kalos:08}. Because the noise (more rigorously, the
standard deviation of calculated quantities) decreases so slowly with
the number of MC particles, it can take many particles to produce a
sufficiently smooth solution, and this can make large simulations
computationally very expensive. Therefore, Monte Carlo methods are
likely to be more expensive than deterministic methods in
lower-dimensional cases. On the other hand, Monte Carlo methods are
likely to become more efficient with increasing number of dimensions,
at least for calculations of definite integrals~\citep[see, e.g., the
  discussion in Section IIB of][]{Zink:08}. However, numerical
integration is not the same as direct simulation of radiation
transport. Hence, such an argument should be taken with a grain of
salt. Therefore, it is \emph{a priori} unclear how a Monte Carlo
method with good parallel scaling may fare against deterministic
methods in terms of total computational cost in multi-dimensional
calculations. 

When emission and absorption of radiation leads to non-negligible
cooling or heating of matter through which radiation is propagating,
then the transport problem becomes \emph{nonlinear}. Such a scenario is
described by a system of non-linear equations with a number of 
unknowns: the radiation intensity, the material temperature (here and
hereafter we assume that material is well-described by temperature,
i.e., that it is in thermal equilibrium), and the leptonic composition
(if we are dealing with the transfer of neutrinos with lepton number). These equations
are coupled due to absorption and emission terms --  the material
cools through emission and heats through absorption. Similarly,
inelastic scattering also leads to nonlinear coupling between the
material temperature and radiation.

The state-of-the-art in both deterministic and Monte Carlo methods for
solving non-linear radiation transport problems involves linearizing
equations over a timestep and solving the resulting linear system
during the timestep. 
Performing this linearization produces a
linearization error during the timestep, but it enables the use of a large
portion of the existing arsenal of linear transfer
methods. 
Moreover, the linearization errors can be mitigated by
  performing iterations within a timestep~\citep[e.g.,][]{Burrows:00}.
Non-linear (or, more precisely, semi-linear)
solution schemes have also been proposed~\citep{NKaoua:91}, but they
are not widely explored in practical problems. 
In this paper, we consider only methods that involve a
linearization procedure within each timestep.  

The classic and widely used method for nonlinear Monte Carlo photon
transport is the method of Fleck and Cummings (1971, hereafter
FC71). This method is known as Implicit Monte Carlo (IMC). This method
reformulates the nonlinear transport equation so that the emission
term is treated semi-implicitly. This process leads to the effective reduction of the emission and absorption
opacity, and the appearance of a scattering term that effectively replaces
a fraction of the absorption and re-emission of radiation within a
timestep. 
This
reduces the coupling between radiation and matter within a
timestep, enabling much larger timesteps and significantly improving
the stability of the system~\citep{Wollaber:08}. Since its first
publication, the IMC method has successfully been used in photon
transport, in part because of its simplicity, versatility, and
robustness~\citep{Gentile:09}. In this paper, we generalize the IMC
method to neutrino transport.  

One drawback of the IMC method is that it becomes computationally
inefficient at high optical depth. This is because in such regimes
the radiation mean-free-path due to effective scattering becomes very
small, i.e., most of the computation is spent in modeling these
scatterings. Several methods have been suggested to overcome this
inefficiency. One of the simplest and most efficient such methods
is the discrete-diffusion Monte Carlo (DDMC) scheme of Densmore et
al. (2007), developed for the case of gray transport for non-moving
matter. In this paper, we extend the gray DDMC
scheme of~\cite{Densmore:07} to the multi-group case, and 
generalize for moving matter. We demonstrate that the combination of the 
IMC scheme at low optical depths with the DDMC scheme at high optical depths 
is an attractive approach for neutrino transport in CCSN simulations. 

We stress that in the present work our focus is on Monte Carlo
neutrino transport, as well as on energy and lepton number coupling
between radiation and matter. The issue of momentum coupling between
radiation and matter is not discussed, but is straightforward. The full
radiation-hydrodynamics scheme and associated simulations will be presented 
in a subsequent publication.

We point out that neutrino transport in
previous time-dependent simulations of CCNSe has been
performed using only deterministic
methods. Monte Carlo methods were used for the study of neutrino
equilibration in static uniform matter~\citep{Tubbs:78} and for calculations of stationary
neutrino transfer in static spherically symmetric supernova
matter~\citep{Janka:89}. The latter code has also been applied to the 
study of neutrino spectrum formation~\citep{Janka:89,Keil:03},
neutrino-antineutrino annihilation~\citep{Janka:91}, and for assessing
the quality of deterministic transport
solvers~\citep{Janka:92,Yamada:99}. Unlike these codes, our Monte
Carlo code is fully time-dependent and can handle energy and lepton
number coupling between matter and radiation, matter motion, as well
as diffusion at high optical depth.

Unless otherwise noted, in the following we use spherical polar coordinates and CGS
units. 

\section{A Simple Monte Carlo Method for Radiation Transport}
\label{sec:MCImplementation}

In this section, for completeness we describe some salient aspects of
a simple time-explicit Monte Carlo method for 
non-linear time-dependent
radiative transfer. For simplicity of illustration, we consider
\emph{static} matter that emits, absorbs, and scatters radiation. Our
description closely follows the presentation in Chapter 3 
of~\cite{Wollaber:08}. We start by writing the multi-D transport
equation for such a system~\citep{Pomraning:73}:  
\begin{eqnarray}
\label{eq:te_mD}
\frac{1}{c}\frac{\p I}{\p t} (\vecr,\vecn,\ve,t) + \vecn \cdot
 \nabla I (\vecr,\mu,\ve,t)  = \nonumber\\\nonumber\\
 \kappa_a(\ve,T) \left[B(\ve,T) - I (\vecr,\vecn,\ve,t)\right] -  
  \kappa_s(\ve,T) I (\vecr,\vecn,\ve,t) \nonumber\\\nonumber\\ 
+ \int_{4\pi} \int_0^\infty \vk_s(\ve',\vecn'\to \ve,
  \vecn) I(\vecr,\vecn',\ve',t) d\Omega' d\ve' \, , 
\end{eqnarray} 
which is coupled to the material energy equation\footnote{In the case
  of neutrinos with lepton number, the transport
  equation~(\ref{eq:te_mD}) is also coupled to the equation for the
  electron fraction $Y_e$ of the material.}:  
\begin{eqnarray}
\label{eq:temattercoupling_md}
\rho \frac{\p U_m}{\p t}({\bf r},T) = \int_{4\pi} \int_0^\infty
\kappa_a(\ve, T) \left[I(\vecr, \vecn, \ve, t) - B (\ve, T) \right]
d\Omega d\ve \nonumber\\\nonumber\\ + \int_{4\pi} \int_{4\pi}
\int_0^\infty \int_0^\infty \bigg[\frac{\ve}{\ve'} \vk_s(\ve',\vecn'
  \to \ve,\vecn) I (\vecr,\vecn',\ve',t) \nonumber\\\nonumber\\ -
  \vk_s(\ve, \vecn \to \ve',  \vecn') I (\vecr,\vecn,\ve,t) \bigg]
d\Omega d\Omega' d\ve d\ve' \, , \nonumber\\\nonumber\\
\end{eqnarray} 
where $I$ is the radiation specific intensity, $t$ is the time, $T$ is
the temperature, $\rho$ is the matter density, $c$ is the speed of
light, $\kappa_a$ is the total absorption opacity, and $\kappa_s$ is the
total scattering opacity. Also, $\vecr$ is the spatial coordinate, 
$\vecn$ is a unit vector in the radiation particle propagation direction, $\Omega$ is the solid angle,
and $\vk_s(\ve', \vecn' \to \ve, \vecn)$ is the differential scattering
opacity for scattering from energy and propagation direction $\{\ve',
\vecn'\}$ to $\{\ve, \vecn\}$ (for brevity the
argument $T$ of function $\vk_s$ is suppressed). $B$ is the Planck
function if the radiation particles are photons (or the Fermi-Dirac
function if we are dealing with fermions), $\ve$ is the energy of a
single physical radiation particle, and $U_m$ is the specific internal
energy of matter. Here and hereafter, we define opacity as the 
inverse mean-free-path of radiation particles.

From here on, we assume that the matter is distributed spherically
symmetrically and use the spherical polar coordinate
system. Therefore, our system is described by only radius ($0\le 
r\le R$) and one angular variable $\mu=\cos\theta$, where $\theta$ is
the angle between $r$ and the particle propagation direction. The
radiative transfer equation in spherical coordinates is given by:   
\begin{eqnarray}
\label{eq:te}
\frac{1}{c}\frac{\p I (r,\mu,\ve,t)}{\p t} + \frac{\p 
  I(r,\mu,\ve,t)}{\p r} + \frac{1-\mu^2}{r} \frac{\p I
  (r,\mu,\ve,t)}{\p \mu} \nonumber\\\nonumber\\ =
  \kappa_a(\ve,T) \left[ B(\ve,T) - I
  (x,\mu,\ve,t)\right] - \kappa_s(\ve,T) I
  (r,\mu,\ve,t) \nonumber\\\nonumber\\ + 2 \pi
  \int_{-1}^{+1} \int_0^\infty \vk_s(\ve',\mu' \to
  \ve,\ \mu) I(x,\mu',\ve',t) d\mu' d\ve' \, , 
\end{eqnarray} 
while the material energy equation~(\ref{eq:temattercoupling_md}) has
the following form: 
\begin{eqnarray}
\label{eq:temattercoupling}
\rho \frac{\p U_m}{\p t}(x,T) = 2\pi \int_{-1}^1 \int_0^\infty
\kappa_a(\ve, T) \big[I(x, \mu, \ve, t)
  \nonumber\\\nonumber\\ - B (\ve, T) \big] d\mu d\ve
+ S \, ,
\end{eqnarray}
where function $S$ represents the amount of energy exchange between
radiation and matter due to inelastic scattering:
\begin{eqnarray}
\label{eq:s}
S = (2\pi)^2 \int_0^\infty \!\!\!\! \int_0^\infty \!\!\! \int_{-1}^1
  \int_{-1}^1 \!\! \bigg[\frac{\ve}{\ve'} \vk_s(\ve', \mu' \!\!\to
  \ve,\mu) I (x,\mu',\ve',t) \nonumber\\\nonumber\\ - \vk_s(\ve, \mu
  \to \ve', \mu') I (x,\mu,\ve,t) \bigg] d\ve d\ve' d\mu d\mu' \, .
\end{eqnarray} 
Here again, $\vk_s(\ve', \mu' \to \ve, \mu)$ is the differential
scattering opacity for scattering from energy and angle $\{\ve',
\mu'\}$ to $\{\ve, \mu\}$. The initial conditions are:   
\begin{eqnarray}
\label{eq:teicI}
I (r,\mu,\ve,0) &=& I_i (r,\mu,\ve) \, , \\
\label{eq:teicT}
T(r,0) & = & T_i(r) \, ,
\end{eqnarray}
and the boundary conditions are 
\begin{equation}
\label{eq:tebcR}
I (R,\mu,\ve,0) = I_R (\mu,\ve,t)\, , \quad
-1\le\mu\le 0 \, .
\end{equation}
We assume that our spatial domain $r\in(0, R]$ is split into many
non-overlapping spatial zones with coordinates $r\in[r_{j-1/2},
r_{j+1/2}]$, where $j = \{1, \dots, N_r\}$, $r_{1/2} = 0$, and
$r_{N_r+1/2} = R$. The quantities that represent the properties of the
matter (such as temperature, opacity, emissivity, etc.) are
represented on these cells within each timestep $t_n\le t\le t_{n+1}$
with their cell-averaged values at $t=t_n$. In the following, we
describe a simple Monte Carlo method for solving
equations~(\ref{eq:te}-\ref{eq:tebcR}).   

We start by considering the possible sources and sinks of radiation
particles that enter the transport equation~(\ref{eq:te}). For instance,
in the first timestep, MCPs may be present initially, or are born due
to the boundary conditions or emission by the matter. The energies
of the emitted particles are subtracted from the material internal
energy. By the end of a timestep, some MCPs may have been absorbed in
the material. The energies of these MCPs are added to the material
internal energy and these MCPs are removed from computer memory. A
fraction of MCPs may leave the system via the outer
boundary. 
Other
MCPs may continue to exist -- these MCPs are usually stored in a
\emph{census} in computer memory in preparation for the next
timestep. At the beginning of the next timestep, these MCPs emerge from
the census (similar to the situation in the first timestep), while
boundary conditions and emission may supply additional MCPs. Using
this synopsis as a guide, a natural algorithm emerges with which to
perform the Monte Carlo sampling procedure: We use random numbers
to choose the positions, the propagation directions, and the energies
of the newly-born MCPs. Once this is done, random numbers are 
used to simulate the propagation of these MCPs through matter within a
timestep. This procedure is described in more detail in the
following. 

We first choose the weight of MCPs, i.e., the total number of
radiation particles contained in each MCP. For simplicity, 
we assume that each MCP represents $N_0$ radiation
particles. Each radiation particle within a given
MCP has the same position, propagation angle, and energy.  

The total number of
radiation particles emitted by matter is: 
\begin{equation}
{\cal N}_T = 8\pi^2\int_{t_n}^{t_{n+1}}\int_0^R\int_0^\infty
\frac{\kappa_a(\ve, T) B(\ve, T)}{\ve} r^2 dt
dr d\ve  \, .
\end{equation}
Since each MCP contains $N_0$ radiation particles, the total number of
MCPs emitted in this process is
\begin{equation}
N_T = \mathrm{RInt} \left( {\cal N}_T / N_0 \right) \, .
\end{equation}
Here $\mathrm{RInt}(x) $ is an operator that returns the largest
integer that is not greater than $x$ plus the quantity $K$, which is
chosen randomly to be $1$  with probability $p = \left\{ {\cal   N}_T
/ N_0 \right\} $, where the latter is the fractional part of ${\cal
  N}_T / N_0$. Otherwise, $K$ is selected to be $0$.  In practice,
this is done by sampling a (pseudo) random number $\xi$ with uniform
distribution on the interval $[0,1]$. If $\xi<p$ then $K=1$, otherwise
$K=0$.     

The particle's energy in each MCP is chosen according to the
functional form of $\kappa_a(\ve, T) B(\ve, T)$. Due to the isotropy
of the emission, the MCP angle is chosen uniformly on a unit
sphere. In practice, this is usually done by choosing $\mu$ uniformly
on the interval $[-1, 1]$ from   
\begin{equation}
\mu = 2\xi-1 \, .
\end{equation}
Since we use time-centered values of the emissivities within any
timestep $t_n\le t\le t_{n+1}$, the particles are emitted with uniform
probability within $t\in[t_n,t_{n+1}]$. Hence, the emission time of
the MCP is chosen as  
\begin{equation}
t = t_n + (t_{n+1} - t_n) \xi \, .
\end{equation}
In order to choose the MCP spatial location, one first recalls
that for transport problems space is represented by many connected,
non-overlapping spatial zones. We first choose the zone in which an
MCP is born, after which we select the spatial location of the MCP 
within that cell. More specifically, if $N_{T,j}$ is the total number
of particles emitted in zone $j$, then an MCP is born in that cell
with probability $N_{T,j}/N_T$. Note that here we assume that the 
weight of MCPs is given in terms of the total number of physical
particles represented by a single MCP. If the weight were given in
terms of the total energy of MCPs, then we would need to use the ratio of
the total \emph{energy} of particles emitted in each zone to that of
particles emitted in all of the zones. For a 1D zone defined by
$[r_{j-1/2}, r_{j+1/2}]$, the particle location $r$ is chosen
according to   
\be
r=\left[r_{j-1/2}^3 + \left(r_{j+1/2} - r_{j-1/2} \right)^3\xi
  \right]^{1/3} \, , 
\ee
which guarantees a uniform sampling within the cell 
  volume\footnote{However, in highly-diffusive regimes the sampling of
  the location of an MC particle should reflect the gradient of the
  thermal emissivity within the zone, i.e., particles should be born
  with higher probability at points within the cell where the
  emissivity is higher~\citep{Fleck:84}. Uniform sampling
  may lead to unphysical results (Densmore 2011, private
  communication).}.  

The number of particles that appear during $t\in[t_n, t_{n+1}]$ due to
a boundary source at $r=R$ is obtained by integrating the boundary
condition (\ref{eq:tebcR}) over the timestep $t\in[t_n, t_{n+1}]$, the
boundary surface area $4\pi R^2$, and the angle $\mu\in [-1,0)$: 
\begin{equation}
  N_B = \mathrm{RInt} \left[ - \frac{8\pi^2 R^2}{N_0}
    \int_{t_n}^{t_{n+1}}\int_0^\infty \int_{-1}^0 \frac{\mu
      I_R(\mu,\ve,t)}{\ve} dt d\ve d\mu \right] \, ,  
\end{equation}
where the particle location on the boundary, the direction, the
energy, and its time of emission are selected according to the
functional forms of $I_R$.

During the first timestep, particles may also be present due to
initial conditions, i.e.:  
\begin{equation}
  N_{IC} = \mathrm{RInt} \left[  \frac{8\pi^2}{cN_0} \int_0^R
    \int_{-1}^1 \int_0^\infty I_i (r, \mu, \ve) r^2 dr d\mu d\ve
    \right] \, .  
\end{equation}
Here, the particle's cell, the spatial location, direction, and energy
are again selected randomly, this time using the functional form of $I_i$.

Thus, the total number of MCPs contained in the problem during the
first timestep is  
\begin{equation}
  N_{TOT} = N_T + N_B + N_{IC} \, ,
\end{equation}
while in subsequent timesteps this number is
\begin{equation}
  N_{TOT} = N_T + N_B + N_{C} \, ,
\end{equation}
where $N_{C}$ is the number of particles in the census from previous
timesteps.   

Once an MC particle is introduced into the problem, the next
task is to transport it through the system (and update all the
relevant quantities along the way). There are essentially  
three types of events that must be considered and which can affect the
transport of the MCP: 
\begin{itemize}
  \item The MC particle could collide with the matter (e.g., with an
  atom, a nucleus, or an electron), 
  \item the MC particle could leave one cell and enter an adjacent cell
  with different opacities, or
  \item the MC particle could travel without collisions inside the cell
    until the end of the timestep (i.e., while $t<t_{n+1}$). 
\end{itemize}
There are three different distances associated with these three
possibilities: the distance to collision $d_c$, the distance to the
cell boundary $d_b$, and the distance $d_t$ that the particle would
travel until $t=t_{n+1}$\footnote{In an alternative Monte Carlo
  approach, the so-called continuous absorption method can be used
  (e.g., FC71). This is explained in more detail in
  Section~\ref{sec:con_absorption}.}.
The distance to the cell boundary can be
calculated using elementary geometric considerations and is given by    
\begin{equation}
\label{eq:d_b}
d_b=\left\{
\begin{array}{ll}
\left|\left[r_{j+1/2}^2-r^2(1-\mu^2)\right]^{1/2} - r \mu\right|, & 
\mathrm{if} \ j=1 \ \mathrm{or} \\ & \mu>0 \, , \ \sin\theta \ge
\frac{R_{j-1/2}}{r} \, , \\ 
& \\ & \\
\left|\left[r_{j-1/2}^2-r^2(1-\mu^2)\right]^{1/2} + r \mu\right|, &
\mathrm{if} \ \mu < 0 \, , \ \sin\theta <
\frac{R_{j-1/2}}{r} \, . \\ & \\
\end{array}
\right.
\end{equation}
The distance traveled until the end of the timestep, $d_t$, is simply
given by
\begin{equation}
  \label{eq:d_t}
  d_t = c (t_{n+1} - t) \, .
\end{equation}
In the simplest case, the distance to a collision can be calculated
probabilistically and is given by 
\begin{equation}
  \label{eq:d_c}
  d_c = - \frac{\ln \xi}{\kappa_a + \kappa_s} \, ,
\end{equation}
where $\xi$ is a random number with a uniform distribution on the
interval $(0,1]$~\citep[See, e.g.,][for the derivation of this
  formula.]{Wollaber:08}. Once the three distances are calculated, the next step is to
determine which one is the smallest of the three. Depending 
on which is smallest, the MCP is then moved to either the collision location,
the cell spatial boundary, or the time boundary. Accordingly, the
MCP location and time are updated using the operation:
\begin{eqnarray}
  \label{eq:xupdate}
  r &\to& \sqrt{r^2 - 2 r d \mu + d^2} \, ,\\ \nonumber\\
  \label{eq:tupdate}
  t &\to& t + d / c \, ,
\end{eqnarray}
where $d=\mathrm{min}\{d_c, d_b, d_t\}$ is the minimum distance. If
$d=d_b$, then we check whether this boundary is the outer boundary of
the computational domain. If that is the case, then the MCP
leaves the system (and, thus, information about the MCP is erased
from computer memory). Otherwise, the transport sampling process
begins again in the new spatial zone (with a new opacity). If $d=d_t$,
the MCP is stored in computer memory for the next timestep.   

If $d=d_c$, the type of collision event must be determined. The MCP
is absorbed with probability of $p_a = \kappa_a / (\kappa_a +
\kappa_s)$ and scattered with probability of $p_s = 1 - p_a = \kappa_s
/ (\kappa_a + \kappa_s) $. If it is absorbed, then the MCP energy is 
deposited into the cell and information about the MCP is erased
from computer memory. If it is scattered, then, once the particle
location and time are updated according to
equations~(\ref{eq:xupdate}-\ref{eq:tupdate}), the new angle and (if
the scattering is inelastic) energy of the particle are selected
randomly from the functional form of the scattering
kernel.

Finally, at the end of the timestep, the material temperature in each
spatial zone is updated according to
equation~(\ref{eq:temattercoupling}) using the information about how 
many particles (of which energy) are emitted, absorbed, or scattered
in each zone. This process is then repeated for each new timestep, for
each MCP.

\subsection{The continuous absorption method}
\label{sec:con_absorption}

The continuous absorption method is a \emph{variance reduction}
mechanism that is typically used in practical implementations of
IMC~\citep{Wollaber:08}. In this method, one calculates \emph{four}
different distances (instead of the three distances in the method
described above): the distance to the boundary, $d_b$, the distance
traveled by the MCP until the end of the timestep, $d_t$, the distance
to scattering, $d_s$, and the distance to absorption, $d_a$. The
distances $d_b$ and $d_t$ are again calculated using
equations~(\ref{eq:d_b}) and (\ref{eq:d_t}), respectively. The
distance to scattering is calculated probabilistically (similarly to
equation~\ref{eq:d_c}):      
\begin{equation}
  \label{eq:d_s}
  d_s = - \frac{\ln \xi}{\kappa_s} \, ,
\end{equation}
where $\xi$ is a random number with a uniform distribution on the
interval $(0,1]$. On the other hand, the distance to absorption is
calculated \emph{deterministically} in the following way. When an MCP
propagates a distance $dx$ through a material with absorption opacity
$\kappa_a$, then the number of radiation particles $N(t)$ in this MCP
at time $t$ decreases according to the law
\begin{equation}
\label{eq:n_of_t}
N(t)  =  N(0) e^{-\kappa_a dx} \, ,
\end{equation}
where $N(0)$ is the initial number of radiation particles in the
MCP. An MCP is assumed to be absorbed when only
a small user-defined fraction $\varsigma$ of the initial radiation
particles remains in the MCP. The parameter
$\varsigma$ is usually chosen to be $0.01$ (FC71).  

The transport algorithm in the continuous absorption method is again
based on the calculation of the smallest of the distances. However, as
mentioned above, in this case, we are dealing with four different distances:
$d_a$, $d_b$, $d_s$, and $d_t$. If $d_a$ is the smallest of the four,
then we deposit all the particle energy (and lepton number, if we are
dealing with neutrinos with lepton number) into its spatial cell. If the  
minimum distance is $d_s$, $d_b$, or $d_t$, then we move the MCP to
its new location according to equations~(\ref{eq:xupdate}) and
(\ref{eq:tupdate}). We then calculate what fraction of the MCP is
absorbed according to equation~(\ref{eq:n_of_t}) during its
propagation to its new location, and deposit the energy (and the
lepton number, if the particles are neutrinos with lepton number) of
the absorbed fraction into the spatial cell. After that we perform
a scattering if $d_s$ is the smallest of the four, or move to a
new cell if $d_b$ is the smallest.

\section{The Fleck \& Cummings method for Implicit Monte Carlo photon
  transport}
\label{sec:IMC_photon}

The first step in almost all of the commonly-used methods for solving
non-linear transport equations is to linearize them over a timestep $t_n
\le t \le t_{n+1}$. As mentioned in Section~\ref{sec:intro}, this
linearization introduces discretization errors (that grow with the
size of the timestep), but it allows use of the large number of 
techniques developed for solving linear radiation transport. 
One of the most well-known and widely-used linearization techniques is
the Implicit Monte Carlo method suggested by FC71.

Consider a 1D spherically symmetric problem with static matter that
can emit, absorb, and scatter radiation (the generalization to multi-D
is conceptually trivial). The transport equation for such 
a system is given by equation~(\ref{eq:te}), while the material energy
equation is given by equation~(\ref{eq:temattercoupling}).  
The IMC method reformulates the transport equation~(\ref{eq:te}) using
the material energy equation~(\ref{eq:temattercoupling}), so that the
emissivity in the  former equation is treated implicitly. This leads
to the appearance of two new terms in the transport equation, which
look like sink and source terms due to some scattering process. This
scattering is called effective scattering by FC71, and it models
absorption and re-emission of a photon within a timestep. The
introduction of effective scattering reduces the stiffness of the
non-linear coupling between the matter temperature and the
radiation, significantly improving the stability of the system of the
equations relative to the case when there is no effective
scattering~\citep{Larsen:87}. 

The central point in this reformulation of the transport equation
is to approximate the radiation source term $\kappa_a B$ using the value of
the intensity at the current time $t \in [t_n, t_{n+1}]$ and 
using the values of the other quantities at the beginning of timestep
$t=t_n$. In that case, the coupling between the two equations would
simplify. Specifically, equation~(\ref{eq:te}) can be solved
independently of equation~(\ref{eq:temattercoupling}) within a
timestep, while the result of solving equation~(\ref{eq:te}) can
then be used to solve equation~(\ref{eq:temattercoupling}) within the
same timestep. This approximation allows for much larger timesteps
than the mean absorption and re-emission timescale (a very short
interval in highly-diffusive regions), without compromising
accuracy\footnote{However, too large timesteps may lead to unphysical
  solutions~\citep{Larsen:87, Martin:01, Densmore:04}}. 
Although the emission term is treated semi-implicitly in the IMC method, the term
``implicit'' is, strictly speaking, a misnomer since the rest of the
problem parameters must be (explicitly) evaluated prior to performing
the timestep. 
Since the original work by Fleck \& Cummings, the IMC method
has been widely and successfully used for solving many radiative
transfer problems~\citep{Gentile:01, Gentile:09, McClarren:09,
  Kasen:11}. As a prelude to the extension of the IMC method to
\emph{neutrinos} in the next section, here we describe some of the key 
aspects of the Fleck \& Cummings method for \emph{photons}.   

We start by introducing a new set of variables:
\begin{eqnarray}
  \label{eq:u_r} 
    {U_r} & = & \frac{4\pi}{c} \int_0^\infty B d \ve \, , \\ 
  \label{eq:b}
  b & = & \frac{B}{4\pi \int_0^\infty B d\ve} \, , \\ 
  \label{eq:sigma_p}
  \kappa_p & = & \frac{\int_0^\infty \kappa_a B d\ve}{\int_0^\infty B 
  d\ve} \, ,  
\end{eqnarray}
and
\begin{eqnarray}  
  \label{eq:beta}
  \beta &=& \frac{1}{\rho} \frac{\p U_r}{\p U_m} \, ,
\end{eqnarray}
where $U_r$ is the radiation energy density if in thermodynamic
equilibrium, and $\kappa_p$ is the Planck mean opacity. Using these new functions,
we rewrite the transport equation (\ref{eq:te}) in the following
way\footnote{Here and hereafter, we do not consider changes in
  $U_r$, $\rho$, and other thermodynamic variables due to the motion of
  matter. This issue will be addressed in a future
  publication on full radiation-hydrodynamics calculations.}:  
\begin{eqnarray}
\label{eq:te2}
\frac{1}{c} \frac{\p I(\ve, \mu)}{\p t} + \mu \frac{\p
I(\ve, \mu)}{\p r} + \frac{1-\mu^2}{r} \frac{\p I(\ve,
  \mu)}{\p \mu} \nonumber\\\nonumber\\ = \kappa_a b c U_r - \kappa_a 
I(\ve, \mu) - \kappa_s I(\ve, \mu) 
\nonumber\\\nonumber\\ + 2 \pi \int_{-1}^{+1} \int_0^\infty
\vk_s(\ve',\mu' \to \ve,\ \mu)
I(\mu',\ve') d\mu' d\ve' \, , 
\end{eqnarray}
while the material energy equation can be transformed into the
following:  
\begin{equation}
\label{eq:temattercoupling2b} 
\frac{1}{\beta}\frac{\p U_r}{\p t} + \kappa_p c U_r = 2\pi \int_{-1}^1
\int_0^\infty \kappa_a I \, d\mu d\ve + S \, . 
\end{equation}

In most applications, the information (such as temperature) is known
at the beginning of timestep $t=t_n$ from the previous timestep  
or from the initial conditions, and one needs to find the solution at
the end of timestep $t=t_{n+1}$. We approximate the functions
$\{\kappa_a, \kappa_p, \kappa_s, \vk_s, b, \beta\}$ with constants
$\{\tilde\kappa_a, \tilde\kappa_p, \tilde\kappa_s, \tilde\vk_s, \tilde
b, \tilde\beta\}$ that are time-centered values of $\{\kappa_a,
\kappa_p, b, \beta\}$ within $[t_n, t_{n+1}]$. Obviously, such an
approximation loses its validity if these functions change rapidly
within a timestep. In many practical applications, these functions
are usually given by their values at the beginning of the timestep,
but, if need be, these can also be extrapolated from their values at
the previous timestep (FC71)\footnote{FC71 hint that one can also use
  time extrapolation to determine temperature from the values at
  previous timesteps. However, experience has shown that these
  temperature extrapolations can affect the stability and accuracy of
  the result, especially if the solution method is subject to errors
  (such as statistical noise in an MC calculation). Hence, temperature
  extrapolation is usually avoided in practice, and the problem data
  are frozen at the beginning of
  timestep~\citep{Wollaber:08}. Alternatively, one can estimate the
  temperature at the end of the timestep using an additional relatively
  inexpensive deterministic calculation~\citep{Wollaber:09}.} 
Using this approximation in equation~(\ref{eq:temattercoupling2b}), we 
obtain a \emph{linear} equation for $U_r$:    
\begin{equation}
\label{eq:temattercoupling3} 
\frac{1}{\tilde\beta}\frac{\p U_r}{\p t} + \tilde \kappa_a
  c U_r = 2 \pi \int_{-1}^1 \int_0^\infty \tilde
  \kappa_a I \, d\mu d\ve + S \, .
\end{equation}
Using this equation, FC71 derive an approximate equation for $U_r(t)$
for $t\in [t_n, t_{n+1}]$: 
\begin{equation}
\label{eq:ur_imc_photon}
U_r(t) = f_n U_{r,n}^* + 2\pi \frac{1-f_n}{c\tilde\kappa_p} \int_{-1}^1 
\int_0^\infty \tilde \kappa_a I(t) d\mu d\ve \, ,   
\end{equation}
where $U_{r,n}^* = U_{r,n} + \tilde\beta \Delta t_n \bar S$, $\Delta
t_n = t_{n+1} - t_n$, $\bar S$ is the time-averaged value of $S$
within interval $t\in [t_n, t_{n+1}]$:   
\begin{equation}
\label{eq:s_bar}
\bar S = \frac{1}{\Delta t_n} \int_{t_n}^{t_{n+1}} S(t) dt ,
\end{equation}
and $f_n$ is a new variable defined as 
\begin{equation}
\label{eq:fleck_factor_photon}
f_n = \frac{1}{1+\alpha\Delta t_n \tilde\beta c \tilde\kappa_p} \, ,
\end{equation}
where $\alpha$ is a user-defined constant such that $\alpha\in [0.5,
  1]$ for stability. The variable $f_n$ is called the \emph{Fleck factor}.  

Equation~(\ref{eq:ur_imc_photon}) is an important result because
$U_r(t)$ at a given time $t\in[t_n, t_{n+1}]$ depends explicitly on
the $U_{r,n}$ at $t=t_n$ and the intensity at $t\in[t_n,
  t_{n+1}]$. The term $\bar S$ can be treated by using $\bar S$ 
from the previous timestep. Hence, if we approximate the 
transport equation (\ref{eq:te2}) using $\{\kappa_a,
\kappa_s, \vk_s, \beta\} = \{\tilde\kappa_a, \tilde\kappa_s,
\tilde\vk_s, \tilde\beta\}$ and substitute $U_r$ in the resulting
equation with the RHS of (\ref{eq:ur_imc_photon}), we obtain a
transport equation that can be solved independently of the material
energy equation~(\ref{eq:temattercoupling}) within timestep $t_n\le
t\le t_{n+1}$. The resulting transport equation has the following form:    
\begin{eqnarray}
  \label{eq:te3}
\frac{1}{c} \frac{\p I(\mu,\ve)}{\p t} + \mu \frac{\p
  I(\mu',\ve')}{\p r} + \frac{1-\mu^2}{r} \frac{\p
  I(\mu,\ve)}{\p \mu} = \nonumber\\\nonumber\\
  \tilde\kappa_{ea} \tilde b c U_{r,n}^* - \tilde\kappa_{ea}
  I(\mu,\ve) - \tilde\kappa_{es} I(\mu,\ve) - \tilde\kappa_s
  I(\mu,\ve) \nonumber\\\nonumber\\ + 2\pi \frac{\tilde\kappa_a \tilde
  b}{\tilde\kappa_p} \int_{-1}^1 \int_0^\infty \tilde \kappa_{es} I
  (\mu',\ve') \, d\mu' d\ve' \nonumber\\\nonumber\\  + 2 \pi
  \int_{-1}^{+1} \int_0^\infty \tilde\vk_s(\ve',\mu' \to  \ve,\ \mu)
  I(\mu',\ve') d\mu' d\ve' ,     
\end{eqnarray}
where we have introduced two new variables:
\begin{eqnarray}
\label{eq:kappa_ea}
\tilde\kappa_{ea} &=& f_n \tilde\kappa_a \, , \\\nonumber\\
\label{eq:kappa_es}
\tilde\kappa_{es} &=& (1-f_n) \tilde\kappa_a \, , 
\end{eqnarray}
the sum of which equals the total absorption opacity.

We now explore the physical meaning of the new terms on the RHS of
equation~(\ref{eq:te3}). Terms $\tilde\kappa_{ea} \tilde b c U_{r,n}^*
$ and $\tilde\kappa_{ea}I$ look like source and sink terms due to
emission and absorption of particles with absorption opacity
$\kappa_{ea}$ [compare these terms to the 1st and 2nd terms on the RHS
of equation~(\ref{eq:te2})]. Moreover, the terms $\tilde\kappa_{es}I$
and $2\pi \frac{\tilde\kappa_a \tilde b}{\tilde\kappa_p} 
\int_{-1}^1 \int_0^\infty \tilde \kappa_{es} I \, d\mu' d\ve'$ look
like sink and source terms for scattering. Hence, equation
(\ref{eq:te3}) appears to describe the transport of radiation through
matter with absorption  
  opacity, $\tilde\kappa_{ea}$, and an additional scattering opacity,
  $\tilde\kappa_{es}$ (in addition to $\tilde\kappa_s$ and
  $\tilde\vk_s$). For that reason, in this formalism, a portion of true 
absorption and re-emission within a timestep is modeled as an
effective scattering process. Parameters $\tilde\kappa_{ea}$ and
$\tilde\kappa_{es}$ are called by FC71 ``effective'' absorption and
scattering opacities.   

Our next task is to derive the equation for calculating the
temperature at the end of the timestep at $t=t_n$ [assuming that
equation~(\ref{eq:te3}) has already been solved using some
procedure]. If we apply some of the approximations in deriving 
equation~(\ref{eq:ur_imc_photon}) to the material energy
equation~(\ref{eq:temattercoupling}), we obtain the following 
equation:
\begin{eqnarray}
\label{eq:u_m}
U_{m,n+1} = U_{m,n} + 2\pi \rho \Delta t_n \int_{-1}^1 \int_0^\infty 
\tilde\kappa_{ea} \bar I \, d\mu d\ve \nonumber\\\nonumber\\
- c f_n\tilde\kappa_p\Delta t_n \rho U_{r,n}^* + \rho \bar S \Delta
t_n \, ,   
\end{eqnarray}
where $\bar I$ is the value of $I$ averaged over time interval $t_n\le
t\le t_{n+1}$. The integral on the right-hand-side of equation
(\ref{eq:u_m}) represents the total amount of energy absorbed within
time interval $t_n \le t \le t_{n+1} $, the 3rd  term represent the
total energy of emitted particles, and the 4th term accounts for the
energy exchanged due to \emph{physical} (not effective) scattering
within the same interval. All of these quantities can directly be
calculated by just summing the energies of the emitted and absorbed
Monte Carlo particles and the amount of energy exchanged in 
each physical scattering event during a timestep. 

A subtle issue arises here: In the IMC method, effective scatterings are 
introduced in order to model a $1-f_n$ fraction of the total absorptions
and subsequent re-emissions of particles. Since the energies of
absorbed particles do not necessarily coincide with the energies of
the re-emitted radiation particles, the effective scatterings should
generally be inelastic (as evident also from the form of the transport
equation~\ref{eq:te3}), meaning that radiation particles can exchange
energy with material due to effective scatterings. However, in
equation~(\ref{eq:u_m}) for the time update of the internal energy,
there is no term that takes into account the energy exchange due to
\emph{effective} scatterings. Hence, one obvious question to ask is
whether it is possible to have a consistent Monte Carlo interpretation of
equations~(\ref{eq:te3})-(\ref{eq:u_m}) if equation~(\ref{eq:u_m})
does not contain terms that account for the energy exchange due to
effective scatterings? The answer is ``yes'' if the weights of MCPs are
treated in a special way during effective scatterings. An approach
used in IMC photon transport is to assume that the total energy of
an MCP does not change during an effective scattering, while the
energy of individual photons within that MCP is allowed to change
during an effective scattering. Obviously, in this case one has to
change the number of photons within that MCP in order to conserve the
total energy of that MCP during the effective scattering. Using this
treatment, one can execute a consistent Monte Carlo interpretation of
equations~(\ref{eq:te3})-(\ref{eq:u_m}). To the best of our knowledge,
this feature of the IMC method has not been pointed out in the
literature previously. 

Once $U_{m,n+1}$ is obtained using equation~(\ref{eq:u_m}), the
temperature $T_{n+1}$ can be calculated by solving iteratively the
following equation:  
\begin{equation}
\label{eq:Tupdate}
U_{m,n+1} = \int_0^{T_{n+1}} C_v (T') dT' \, ,
\end{equation}
where $C_V$ is the specific heat capacity. Finally, we point out that
none of the approximations made in deriving equation~(\ref{eq:u_m})
violates energy conservation \cite[see, e.g., FC71 or Section~3.2
  of][]{Wollaber:08}.

\subsection{Summary of the Monte Carlo procedure.}

The Monte Carlo procedure for solving equation (\ref{eq:te3}) can be
summarized briefly as follows. Let us assume that $I$ and $T$ are
known at time $t=t_n$, and we wish to determine them at $t = t_{n+1} =
t_n + \Delta t_n$. The temperature $T$ (as well as other relevant
quantities such as the opacity $\kappa_a$, etc.) is represented in
each cell of the spatial computational domain using cell-centered and 
time-centered values. As mentioned above, in many practical
applications the time-centered values $\{\tilde \kappa_a, \tilde
\kappa_p, \tilde\kappa_s, \tilde\vk_s, \tilde b, \tilde\beta\}$ of
$\{\kappa_a, \kappa_p, \kappa_s, \vk_s, b, \beta\}$ in each spatial
zone are given by the data available at $t=t_n$ [although alternatives
  are possible~\citep{Wollaber:09}]. Using these time-centered values,
one calculates the sources for each spatial zone, generates new
particles from the sources, and advances both newly-created and census
MCPs according to the transport equation~(\ref{eq:te3}) by a standard
Monte Carlo procedure (as described in
Section~\ref{sec:MCImplementation}). In the process of advancing MCPs,
one keeps track of the  total energy of emitted, absorbed, and
scattered MCPs. At the end of the  timestep, $U_m$ is advanced in time
according to equation (\ref{eq:u_m}), while the temperature $T$ is
updated using equation (\ref{eq:Tupdate}). 

\section{Extension of the Fleck and Cummings scheme to neutrino
  transport}
\label{sec:IMC_neutrino}

The FC71 scheme is not directly applicable to neutrino transport
because, in the latter case, emission and absorption of radiation
particles not only change $T$, but can also alter the value of
$Y_e$. An additional difficulty arises when one has to evolve
different neutrino types together. In CCSN simulations, one usually
has to solve three different transport equations for three different
species of neutrinos: electron neutrinos ($\nu_e$), electron
anti-neutrinos ($\bar{\nu}_e$), and heavy lepton neutrinos and
antineutrinos, where the latter two are usually lumped together into
one group ($\nu_x$). An additional complication arises due to pair
processes between neutrinos and antineutrinos. Such processes couple
the two neutrino species and add extra non-linearity to the coupling
to matter. Here, we extend the FC71 equations to the more general case
for which there are additional degrees of freedom in $Y_e$ and
multiple neutrino types. For simplicity of illustration, we limit
ourselves to the case of 1D spherically symmetric matter that can
emit, absorb, and scatter radiation. For such a system, the transport
equation for neutrinos of type $i$ is again given by
equation~(\ref{eq:te}), which has to be solved together with the
equations for the change of the internal energy $U_m$ and electron
fraction $Y_e$:  
\begin{eqnarray}
  \label{eq:u0}
  \rho \frac{d U_m} {d t} = 2 \pi \sum_i 
   \int_{-1}^1 \int_0^\infty \kappa_{ai} (I_i - B_i) \, d\mu
   d\ve \nonumber\\\nonumber\\ + \sum_i S_i \, , \\\nonumber\\
  \label{eq:Ye0}
  \rho N_A \frac{d Y_e} {d t} = 2\pi \sum_i s_i \int_{-1}^1
 \int_0^\infty \frac{\kappa_{ai}}{\ve} (I_i - B_i) \,
 d\mu d\ve \, ,  
\end{eqnarray}
where subscript $i$ is used to denote quantities representing
neutrinos of type $i$, and the sum in
equations~(\ref{eq:u0})-(\ref{eq:Ye0}) runs over all neutrino
species. Variable $\ve$ is again the neutrino energy, $N_A$ is the
Avogadro's number, and $s_i$ is a constant equal to $ +1\, , 
-1\, , 0 $ for $\nu_e$, $\bar{\nu}_e$, and $\nu_x$, respectively. 
Function $S_i$ is the
function $S$ defined in formula~(\ref{eq:s}) for neutrino of type $i$.   

In order to handle multiple types of neutrinos, we adopt an
operator-split approach: we evolve different types separately and
independently within a timestep. Moreover, we linearize the pair
processes by assuming that the distribution function of the pair
counterpart not being followed explicitly is given by its local
equilibrium value~\citep[similarly to][]{Janka:89}. This is a good
approximation, since the pair processes are significant only in
high-temperature inner regions, where neutrinos are close to thermal
equilibrium~\citep{Sumiyoshi:12}. Therefore, and hereafter, we focus
on solving the transport equation for a single neutrino species
only. In this case, we will not need to sum over neutrino types in
equations (\ref{eq:u0})-(\ref{eq:Ye0}): 
\begin{eqnarray}
  \label{eq:u}
 &&  \rho \frac{d U_m} {d t} = 2\pi\int_{-1}^1 \int_0^\infty \kappa_{a}
 (I-B) \, d\mu d\ve + S \, , \\\nonumber\\  
  \label{eq:Ye}
 && \rho N_A \frac{d Y_e} {d t} = 2\pi s_i \int_{-1}^1 \int_0^\infty 
  \frac{\kappa_{a}}{\ve} (I-B) \, d\mu d\ve \, .
\end{eqnarray}
In these equations, we have omitted the subscript
$i$, except in $s_i$.  

We now derive some thermodynamic relations which will be used
later. We first represent the time derivative of the specific internal
energy $U_m$ using its partial derivatives with respect to  $T$ and
$Y_e$:  
\begin{equation}
\frac{d U_m}{d t} = \left(\frac{\p U_m}{\p T}\right)_{\rho,Ye}
\frac{dT}{dt} + \left(\frac{\p U_m} {\p Y_e}\right)_{\rho,T}
\frac{dY_e}{dt} \, . 
\end{equation}
Using this equation, we can obtain   
\begin{equation}
\label{eq:dTdt}
\frac{dT}{dt} = \frac{1}{C_V} \left[\frac{d U_m}{d t} - \left(\frac{\p
    U_m}{\p Y_e}\right)_{\rho,T} \frac{dY_e}{dt} \right] \, ,
\end{equation}
where $C_V=\left(\p U_m / \p T \right)_{\rho,Y_e}$ is the
specific heat capacity. Similarly, we split the time-derivative of
${U_r}$: 
\begin{equation}
\label{eq:dBidt}
\frac{d {U_r}}{d t} = \left(\frac{\p {U_r}}{\p T}\right)_{\rho,Ye}
\frac{dT}{dt} + \left(\frac{\p {U_r}}{\p Y_e}\right)_{\rho,T}
\frac{dY_e}{dt}  \, .
\end{equation}
Using (\ref{eq:dTdt}) and (\ref{eq:dBidt}), we obtain the following
expression 
\begin{equation}
  \label{eq:dBidt2}
  \frac{d {U_r}}{d t} = \beta \left(\rho \frac{d U_m}{dt}
  \right) + \zeta \left( \rho N_A \frac{dY_e}{dt} \right)\, ,  
\end{equation}
where\footnote{Note the analogy between this quantity and its
  ``photonic'' counterpart $\beta$ given by
  equation~(\ref{eq:beta}).} 
\begin{equation}
  \label{eq:alphabeta}
  \beta = \frac{1}{\rho C_V} \left(\frac{\p {U_r}}{\p
    T}\right)_{\rho,Ye} \, ,  
\end{equation}  
and
\begin{equation}
  \zeta = \frac{1}{\rho N_A} \left[\left(\frac{\p {U_r}}{\p
      Y_e} \right)_{\rho,T} -\frac{1}{C_V} \left(\frac{\p
      U_m}{\p Y_e} \right)_{\rho,T} \left(\frac{\p
      {U_r}}{\p T} \right)_{\rho,Ye} \right] \, .   
\end{equation}
Note that in numerical simulations, the quantities $C_V$ and $\left(\p U_m
/ \p Y_e \right)_{\rho,T} $ can be calculated using the
EOS\footnote{In neutrino transport simulations in core-collapse
  supernovae, the EOS in nuclear statistical equilibrium (NSE) is
  given as a function of three independent quantities $(\rho, \, T, \,
  Y_e)$, usually in tabulated form.}. The function $U_r$ and its partial 
derivatives with respect to $T$ and $Y_e$ can be calculated (semi)
analytically using the expression for the Fermi-Dirac function
(Appendix A).  

We now define two new variables: 
\begin{equation}
  \label{eq:xi_a}
  \chi_a = \frac{\kappa_{a}}{\ve}\, ,
\end{equation}
and
\begin{equation}
  \label{eq:xi_p}
  \chi_p = \frac{\int_0^\infty\chi_a B d\ve}{\int_0^\infty B d\ve} \, .
\end{equation}
Following FC71, we again approximate the functions $\{\kappa_a, \kappa_p,
\kappa_s, \vk_s, b, \chi_a, \chi_p\}$ with $\{\tilde\kappa_a,
\tilde\kappa_p, \tilde\kappa_s, \tilde\vk_s, \tilde b,
\tilde\chi_a, \tilde\chi_p\}$, the time-centered values of the
former within the time interval $t_n\le t\le t_{n+1}$, and rewrite
equations (\ref{eq:u}-\ref{eq:Ye}) using this approximation: 
\begin{eqnarray}
  \label{eq:u2}
 &&  \rho \frac{d U_m} {d t} = 2\pi \int_{-1}^1 \int_0^\infty \tilde
  \kappa_{a} I d\mu d\ve - c \tilde \kappa_{p} {U_r} + S\, ,
  \\\nonumber\\ 
  \label{eq:Ye2}
 && \rho N_A \frac{d Y_e} {d t} = 2\pi s_i \int_{-1}^1 \int_0^\infty 
  \tilde{\chi}_{a} I \, d\mu d \ve - c s_i
  \tilde{\chi}_{p} {U_r} \, .
\end{eqnarray}
Using (\ref{eq:u2}) and (\ref{eq:Ye2}) and the time-centered values
$\{\tilde \beta$, $\tilde \zeta \}$ of $\{\beta$, $\zeta$\}, we
rewrite equation (\ref{eq:dBidt2}) as     
\begin{eqnarray}
  \label{eq:dBidt3}
  \frac{d {U_r}}{d t} = \tilde \beta \left[ 2\pi \int_{-1}^1
  \int_0^\infty \tilde \kappa_{a} I d \, \mu d \ve - c \tilde
  \kappa_{p} {U_r} + S \right] \nonumber\\\nonumber\\ + \tilde \zeta
  \left[ 2\pi s_i \int_{-1}^1 \int_0^\infty \tilde{\chi}_{a} I d \,
  \mu d \ve - c s_i \tilde{\chi}_{p} {U_r} \right] \, .
\end{eqnarray}
After rearranging terms in this equation, we obtain:
\begin{eqnarray}
  \label{eq:dBidt4}
  \frac{d {U_r}}{d t} = 2\pi\int_{-1}^1 \int_0^\infty (\tilde \beta
  \tilde \kappa_{a} + \tilde \zeta s_i \tilde{\chi}_{a}) I \, d\mu
  d\ve \nonumber\\\nonumber\\
- (\tilde \beta \bar \kappa_{p} + \tilde \zeta s_i 
  \tilde{\chi}_{p}) c {U_r} + \tilde\beta S \, . 
\end{eqnarray}
For brevity, we now introduce the following notation:
\begin{eqnarray}
  \tilde \gamma &=& \tilde \beta \tilde \kappa_{a} + \tilde \zeta s_i
  \tilde{\chi}_{a} \, , \\\nonumber\\ 
  \tilde \gamma_{p} &=& \tilde \beta \tilde \kappa_{p} + \tilde \zeta
  s_i \tilde{\chi}_{p} \, ,  
\end{eqnarray}
and rewrite equation (\ref{eq:dBidt4}) using this notation:
\begin{equation}
  \label{eq:dBidt5}
  \frac{d {U_r}}{d t} = 2\pi \int_{-1}^1 \int_0^\infty \tilde
  \gamma I d \, \mu d \ve - c \tilde \gamma_{p}
  {U_r} + \tilde\beta S \, .  
\end{equation}
Next, we apply the time-averaging operator~(\ref{eq:s_bar}) to 
equation~(\ref{eq:dBidt5}) to get   
\begin{equation}
  \label{eq:dBidt9}
  \frac{U_{r,n+1} - U_{r,n}}{\Delta t_n} = 2\pi\int_{-1}^1
  \int_0^\infty \tilde \gamma \bar{I} \, d\mu d\ve - c \tilde
  \gamma_{p} \bar{U}_r + \tilde\beta \bar S \, .
\end{equation}
Our next task is to eliminate $U_{r,n+1}$ from this equation. In order
to do this, we make one more approximation, 
\begin{equation}
  \label{eq:bbar}
  \bar{U}_r = \alpha U_{r,n+1} + (1 - \alpha) U_{r,n}^* \, ,
\end{equation}
which can also be recast as 
\begin{equation}
  \label{eq:bbar1}
  U_{r,n+1} = U_{r,n}^* + (\bar{U}_r - U_{r,n}^*) / \alpha \, ,  
\end{equation}
where $U_{r,n}^* = U_{r,n}+\tilde\beta\Delta t_n\bar S$ and $\alpha$
is the ``neutrino'' analogue of the user-defined parameter $\alpha \in
[0.5, \ 1]$ of the Fleck \& Cummings scheme for photons discussed in
Section~\ref{sec:IMC_photon}. This parameter controls the degree of
``implicitness'' of the method, with $\alpha=1$ being the most
implicit (since $\bar{U}_r = U_{r,n+1}$ in this case). 
Substituting $U_{r,n+1}$ given by this 
formula into equation~(\ref{eq:dBidt9}), and solving the resulting
equation for $\bar{U}_r$, we obtain    
\begin{equation}
  \label{eq:dBidt10}
  \bar{U}_r = f_{n} U_{r,n}^* + 2\pi\frac{1-f_{n}}{c \tilde
    \gamma_{p}} \int_{-1}^1 \int_0^\infty \tilde \gamma \bar{I} \,
    d\mu d\ve \, ,   
\end{equation}
where 
\begin{equation}
\label{eq:ff_nu}
f_{n} = \frac{1}{1 + \alpha c \Delta t_n \tilde \gamma_{p}}
\end{equation}
is the neutrino analogue of the ``photonic'' Fleck factor given by
formula~(\ref{eq:fleck_factor_photon}). We now make
the final approximation of FC71: we replace the time-averaged
$\bar{U}_r$ and $\bar{I}$ in equation~(\ref{eq:dBidt10}) with their
``instantaneous'' counterparts, $\bar{U}_r = U_r(t)$ and $\bar{I} = I  
(t)$, to obtain:  
\begin{equation}
  \label{eq:dBidt11}
  U_r (t) = f_{n} U_{r,n}^* + 2\pi\frac{1-f_{n}}{c \tilde
    \gamma_{p}} \int_{-1}^1 \int_0^\infty \tilde \gamma I(t) \, d\mu
    d\ve \, ,  
\end{equation}
and rewrite equation (\ref{eq:te}) using the approximation
$\{\kappa_a, \kappa_p, \kappa_s, \vk_s, b, \chi_a, \chi_p\} \simeq
\{\tilde\kappa_a, \tilde\kappa_p, \tilde\kappa_s, \tilde\vk_s, \tilde
b, \tilde\chi_a, \tilde\chi_p\}$ discussed above. We then have:  
\begin{eqnarray}
\label{eq:rt2}
\frac{1}{c} \frac{\p I(\mu,\ve)} {\p t} + \mu \frac{\p
  I(\mu,\ve)}{\p r} + \frac{1-\mu^2}{r} \frac{\p
  I(\mu,\ve)}{\p \mu} \nonumber\\\nonumber\\ = c \tilde
  \kappa_{a} \tilde b {U_r} - (\tilde \kappa_a+\tilde \kappa_s)
  I(\mu,\ve) \nonumber\\\nonumber\\ + 2 \pi \int_{-1}^{+1}
  \int_0^\infty \tilde \vk_s(\ve',\mu' \to \ve,\
  \mu) I(\mu',\ve') d\mu' d\ve' \, .  
\end{eqnarray}
Substituting equation~(\ref{eq:dBidt11}) into equation~(\ref{eq:rt2}),
we obtain the transport equation in a new form:
\begin{eqnarray}
  \label{eq:rt4}
  \frac{1}{c} \frac{\partial I} {\partial t} + \mu \frac{\p I}{\p r} +
  \frac{1-\mu^2}{r} \frac{\p I}{\p \mu} = f_n \tilde \kappa_{a} c
  \tilde b U_{r,n}^*
  \nonumber\\\nonumber\\ - (\tilde \kappa_a + \tilde \kappa_s)I + 
  2\pi\frac{(1-f_n)\tilde \kappa_{a} \tilde b}{\tilde \gamma_{p}}
  \int_{-1}^1 \int_0^\infty \tilde \gamma I \, d\mu d \ve 
  \nonumber\\\nonumber\\
  + 2 \pi \int_{-1}^{+1} \int_0^\infty \tilde
  \vk_s(\ve',\mu' \to \ve,\ \mu)
  I(\mu',\ve') d\mu' d\ve' \, .
\end{eqnarray}
For reasons that will become apparent later, we rewrite this
equation in a slightly different, but equivalent, form:
\begin{eqnarray}
  \label{eq:rt5}
  \frac{1}{c} \frac{\partial I} {\partial t} + \mu \frac{\p I}{\p r} +
  \frac{1-\mu^2}{r} \frac{\p I}{\p \mu} = \tilde \kappa_{ea} c \tilde
  b U_{r,n}^*  \nonumber\\\nonumber\\
  - \tilde \kappa_{ea} I + \tilde \kappa_{es,e} I + \tilde
  \kappa_{es,l} I + \tilde \kappa_s I
  \nonumber\\\nonumber\\+ 
  2\pi \frac{\tilde \kappa_a \tilde b}{\tilde \kappa_p} \int_{-1}^1
  \int_0^\infty \tilde \kappa_{es,e} I \, d\mu d \ve
  + 2\pi \frac{\tilde \kappa_a \tilde b}{\tilde \chi_p} \int_{-1}^1
  \int_0^\infty \tilde \chi_{es,l} I \, d\mu d \ve
  \nonumber\\\nonumber\\ 
  + 2 \pi \int_{-1}^{+1} \int_0^\infty \tilde
  \vk_s(\ve',\mu' \to \ve,\ \mu)
  I(\mu',\ve') d\mu' d\ve' \, ,  
  \nonumber\\  
\end{eqnarray}
where we have introduced a set of new variables:
\bea
\label{eq:kapppa_ese}
\kappa_{es,e} &=& (1-f_n) \frac{\tilde \beta \tilde \kappa_p}{\tilde
  \gamma_p} \kappa_a \, , \\
\label{eq:kapppa_esl}
\kappa_{es,l} &=& (1-f_n) \frac{\tilde \zeta s_i \tilde \chi_p}{\tilde
  \gamma_p} \kappa_a \, , \\
\label{eq:chi_ese}
\chi_{es,e} &=& (1-f_n) \frac{\tilde \beta \tilde \kappa_p}{\tilde
  \gamma_p} \chi_a \, , \\
\label{eq:chi_esl}
\chi_{es,l} &=& (1-f_n) \frac{\tilde \zeta s_i \tilde \chi_p}{\tilde
  \gamma_p} \chi_a \, , 
\eea
and $\kappa_{ea}$ is defined as in formula~(\ref{eq:kappa_ea}).
Equation~(\ref{eq:rt5}), together with boundary and initial ($t=t_n$)
conditions for $I$, determine $I$ during the time interval $t_n \le t
\le t_{n+1}$.

\subsection{Update of $T$ and $Y_e$} 

Having derived the transport equation in a new form~(\ref{eq:rt5}),
our next task is to derive equations for the update of $T$ and $Y_e$
at the end of timestep $t = t_{n+1}$, assuming that some (Monte Carlo)
procedure has been used to solve equation~(\ref{eq:rt5}). We start by
performing the time-averaging integral~(\ref{eq:s_bar}) over
equation~(\ref{eq:u2}):     
\begin{equation}
  \label{eq:u3}
  \rho \frac{U_{m,n+1} - U_{m,n}} {\Delta t_n} = 2\pi\int_{-1}^1
  \int_0^\infty \tilde \kappa_{a} \bar I_i \, d\mu d\ve - c \tilde
  \kappa_{p} \bar{U}_r + \bar S \, .
\end{equation}
To conserve energy, we must approximate this equation precisely the
same way we did in deriving equation (\ref{eq:rt5}). We, therefore,
substitute equation~(\ref{eq:dBidt11}) into equation~(\ref{eq:u3}) to
obtain:  
\begin{eqnarray}
  \label{eq:u4}
  \rho \frac{U_{m,n+1} - U_{m,n}} {\Delta t_n} = 2\pi\int_{-1}^1
  \int_0^\infty \tilde \kappa_{a} \bar I \, d \mu d \ve -
  \nonumber\\\nonumber\\ c \tilde \kappa_{p} \left(f_{n} U_{r,n} +
  2\pi\frac{1-f_{n}}{c \tilde \gamma_{p}} \int_{-1}^1 \int_0^\infty
  \tilde \gamma \bar I \, d \mu d \ve \right) + \bar S \, .
\end{eqnarray}
After rearranging some terms on the RHS of this equation and using
variables defined in
formulae~(\ref{eq:kapppa_ese})-(\ref{eq:chi_esl}), we obtain the
following expression 
\begin{eqnarray}
  \label{eq:u4a}
  \rho \frac{U_{m,n+1} - U_{m,n}} {\Delta t_n} = 2\pi\int_{-1}^1
  \int_0^\infty \tilde \kappa_{ea} \bar I \, d \mu d \ve -
  \nonumber\\\nonumber\\ c f_{n} \tilde \kappa_{p} U_{r,n} +
  2\pi \int_{-1}^1 \int_0^\infty \tilde \kappa_{es,l} \bar I
  \, d \mu  d \ve \nonumber\\\nonumber\\ -
  2\pi \frac{\tilde \kappa_p}{\tilde \chi_p} \int_{-1}^1 \int_0^\infty
  \tilde \chi_{es,l} \bar I \, d \mu d \ve + \bar S \, .
\end{eqnarray}
We solve this equation for $U_{m,{n+1}}$ and obtain:
\begin{eqnarray}
  \label{eq:u5}
  U_{m,n+1} =  U_{m,n} + \frac{\Delta t_n}{\rho} \bigg\{ 2\pi\int_{-1}^1
  \int_0^\infty \tilde \kappa_{ea} \bar I \, d \mu d \ve -
  \nonumber\\\nonumber\\ c f_{n} \tilde \kappa_{p} U_{r,n} +
  2\pi \int_{-1}^1 \int_0^\infty \tilde \kappa_{es,l} \bar I
  \, d \mu  d \ve \nonumber\\\nonumber\\ -
  2\pi \frac{\tilde \kappa_p}{\tilde \chi_p} \int_{-1}^1 \int_0^\infty
  \tilde \chi_{es,l} \bar I \, d \mu  d \ve + \bar S \bigg\} \, .
\end{eqnarray}
Using similar arguments, we obtain a similar expression for $Y_e$:
\begin{eqnarray}
  \label{eq:Ye3}
  Y_{e, n+1} = Y_{e, n} + \frac{\Delta t_n}{\rho N_A} \bigg\{ 2\pi s_i
  \int_{-1}^1 \int_0^\infty \tilde \chi_{ea} \bar I \, d \mu d \ve
  - \nonumber\\\nonumber\\ c s_i f_{n} \tilde \chi_{p} U_{r,n} + 
  2\pi s_i \int_{-1}^1 \int_0^\infty \tilde \chi_{es,e} \bar I \, d
  \mu d \ve \nonumber\\\nonumber\\ - 2\pi s_i \frac{\tilde
  \chi_p}{\tilde \kappa_p} \int_{-1}^1 \int_0^\infty \tilde
  \kappa_{es,e} \bar I \, d \mu d \ve \bigg\}\, ,  
\end{eqnarray}
where $\chi_{ea}=f_n\chi_a$. These last two equations determine how the
values of $U_m$ and $Y_e$ change after each timestep. 

\subsection{Energy and Lepton Number Conservation}
\label{sec:IMC_neutrino_e_l_cons}

The transport equation~(\ref{eq:rt5}) and
equations~(\ref{eq:u5})-(\ref{eq:Ye3}) for the time evolution of the
internal energy $U_m$ and electron fraction $Y_e$ conserve the total
energy and lepton number in the system. This can be demonstrated in
the following way. 
If we apply the operator 
\be
\frac{1}{2} \int_{-1}^1 \int_0^\infty d\mu d\ve
\ee
to the transport equation~(\ref{eq:rt5}) and add the resulting
equation to equation~(\ref{eq:u5}), we obtain the following relation  
\be
\label{eq:en_cons_dis}
\frac{1}{\Delta t_n} \left[\frac{1}{c}\left(I_{0,n+1} - I_{0,n}\right)
  + \rho \left(U_{m,n+1} - U_{m,n}\right)\right] = - \frac{\p {\bar I}_1}{\p
  x} \, ,
\ee
where
\be
I_0 = \frac{1}{2} \int_{-1}^1 \int_0^\infty I d\mu d\ve \, ,
\ee
and
\be
I_1 = \frac{1}{2} \int_{-1}^1 \int_0^\infty \mu I d\mu d\ve \, .
\ee
Clearly, equation~(\ref{eq:en_cons_dis}) is a discretization in time
of the following law:
\be
\frac{\p}{\p t} \left(\frac{I_0}{c} + \rho U_m\right) = - \frac{\p
  I_1}{\p x} \, .
\ee
The two terms inside the brackets are the total energy in radiation
and matter, while the term on the RHS is the radiation energy flux,
meaning that this relation represents the energy conservation law.  

Lepton number conservation is also demonstrated in a similar way. If
we apply the operator   
\be
\frac{1}{2} \int_{-1}^1 \int_0^\infty \frac{1}{\ve} d\mu d\ve
\ee
to the transport equation~(\ref{eq:rt5}) and add the resulting
equation to equation~(\ref{eq:Ye3}), we obtain a variant of
equation~(\ref{eq:en_cons_dis}) for lepton number (instead of
energy), which is a finite-difference representation of the
conservation law for the lepton number. Hence, based on this we
conclude that none of the approximations made in deriving the system of
equations~(\ref{eq:rt5}) and (\ref{eq:u5})-(\ref{eq:Ye3}) 
violate energy and lepton number conservation and that these two
conservation laws are satisfied rigorously. 

\subsection{Monte Carlo Interpretation}

We now give a Monte Carlo interpretation for the transport
equation~(\ref{eq:rt5}) and equations~(\ref{eq:u5})-(\ref{eq:Ye3}) for
time evolution of the internal energy $U_m$ and electron fraction $Y_e$, 
respectively. 

We start with the transport equation~(\ref{eq:rt5}). As in the case of
photon transport discussed in Section~\ref{sec:IMC_photon}, we
interpret the terms $\tilde \kappa_{ea} c \tilde b U_{r,n}^*$ 
and $\tilde \kappa_{ea} I$ on the RHS of equation~(\ref{eq:rt5}) as
source and sink terms due to emission and effective absorption of
MCPs. Moreover, terms $\tilde\kappa_{es,e} I $ and
$2\pi\frac{\tilde\kappa_a \tilde
  b}{\tilde\kappa_p}\int_{-1}^1\int_0^\infty \tilde\kappa_{es,e} I
d\mu d\ve$ look like terms for a sink and source for
scattering. Following FC71, we interpret this scattering as
effective scattering. Analogously, we assume that the total
energies of MCPs are conserved in such effective scatterings, while
the number of leptons in MCPs are allowed to change in order to
conserve the total energy of the MCP. In other words, in such
scatterings, the MCP does not exchange energy with matter, but can
exchange lepton number. 

In addition to these terms, equation~(\ref{eq:rt5}) contains terms
$\tilde\kappa_{es,l} I$ and $2\pi\frac{\tilde\kappa_a \tilde
  b}{\tilde\chi_p}\int_{-1}^1\int_0^\infty \tilde\chi_{es,l} I d\mu
d\ve$. These terms again look similar to the sink and source terms for
effective scatterings, but with a subtle difference: In these
scatterings, the weight of MCPs should be treated differently. Instead
of keeping the total energy of an MCP fixed, here we fix the
total number of leptons in the  MCPs. This is done in order to be
consistent with equation~(\ref{eq:u5}), as will become apparent in the
following. Therefore, in these scatterings, the MCPs exchange energy 
with the matter, but not lepton number.  

In other words, in order to make a Monte Carlo interpretation of
equations~(\ref{eq:rt5}) and (\ref{eq:u5})-(\ref{eq:Ye3}), one has to
introduce \emph{two} types of effective scattering. This feature makes
this scheme slightly different from its counterpart for photons, where
one introduces just \emph{one} type of effective scattering. We refer
to the scattering in which the total energy of MCPs is conserved as
\emph{energy-weight conserving} effective scattering, while the other 
type of scattering that conserves lepton number is called 
\emph{number-weight conserving} effective scattering.  

Let us now consider equations~(\ref{eq:u5}) and (\ref{eq:Ye3}) for the
update of the internal energy $U_m$ and electron fraction $Y_e$,
respectively. Clearly, the 1st and 2nd terms inside the brackets on
the RHS of equation~(\ref{eq:u5}) are responsible for the change of
the internal energy due to absorption and emission of
neutrinos within time interval $t_n \le t \le t_{n+1}$. Similarly,
the 1st and 2nd terms inside the brackets on the RHS 
of equation~(\ref{eq:Ye3}) account for the change of $Y_e$ due to
absorption and emission within the same time interval. Furthermore, 
the 3rd and 4th terms inside the brackets on the RHS of
equation~(\ref{eq:u5}) are the source and sink terms due to
\emph{number-weight} conserving effective scatterings. Analogously,
the 3rd and 4th terms inside the brackets on the RHS of
equation~(\ref{eq:Ye3}) are the source and sink terms due to
\emph{energy-weight} conserving effective scattering within $t_n \le t
\le t_{n+1}$. Finally, the last term on the RHS of
equation~(\ref{eq:u5}) is responsible for energy exchange due to
\emph{physical} scattering, again within $t_n \le t \le t_{n+1}$. All
of these quantities can directly be calculated by summing the energies
(lepton numbers) of emitted and absorbed Monte Carlo particles, and
summing the energy (lepton number) exchanged in (only effective)
scatterings during the timestep. Having calculated $U_{m,{n+1}}$ and
$Y_{e,n+1}$, $T_{n+1}$ can be obtained via the EOS table using the new
values of $U_{m,{n+1}}$ and $Y_{e,n+1}$.

\subsection{Summary of the Monte Carlo procedure}

The Monte Carlo procedure for solving equation (\ref{eq:rt5}) can be
summarized as follows. We assume that $I$, $T$ and $Y_e$ are known at
time $t=t_n$, and we wish to determine them at $t = t_{n+1}$. The
temperature $T$ and electron fraction $Y_e$ (as well as other relevant
quantities, such as the opacity, $\kappa_a$, etc.) are represented on
the spatial computational domain using their cell-centered values in
each of the spatial zones. We assume that the time-centered values
$\{\tilde \kappa_a, \tilde\kappa_p, \tilde b, \tilde\gamma, \,
\dots\}$ of $\{\kappa_a, \kappa_p, b, \gamma, \, \dots\}$ for each
spatial zone are given in terms of the data at the beginning of
timestep at $t=t_n$. Using these time-centered values, we calculate
the sources appropriate to each spatial zone, generate new particles
from the sources, and advance both newly-created and census MCPs
according to the transport equation (\ref{eq:rt5}) by a Monte Carlo
procedure similar to the one described in
Section~\ref{sec:MCImplementation}. In the process of advancing MCPs,
one keeps track of the total energy and lepton number of emitted,
absorbed, and scattered MCPs. At the end of the timestep, we calculate
the updated $Y_e$ with equation~(\ref{eq:Ye3}), while $U_m$ is updated
according to equation~(\ref{eq:u5}). We obtain the new value of $T$
using the EOS table with the new values of $Y_e$ and $U_m$. 

\section{Discrete diffusion scheme for multi-group Monte Carlo
  neutrino transport} 
\label{sec:ddmc}

In the IMC method, when the absorption opacity is high, the Fleck
factor $f_n $ becomes small ($f_n\simeq 0$), and thus $\kappa_{es}
\simeq \kappa_a $ and $\kappa_{ea} \simeq 0$
[cf. equations~(\ref{eq:kappa_ea})-(\ref{eq:kappa_es})], signifying
that most of the absorption (and subsequent re-emission) is replaced
with effective scatterings. In this regime, MCPs undergo Brownian
motion with small ($\simeq 1/\kappa_{es}$) mean-free-path most of the
time. 
The computational cost of each simulated MCP path between
collisions is about equally expensive. 
Thus, simulations with a large scattering cross
section (both effective and physical) can be very time consuming due
to the large number of MCPs paths between scattering events that one has
to simulate.
On the other hand, when the mean-free-path for this effective scattering
is small, then the solution of the transport equation is well
approximated by the solution of a diffusion equation. Several
schemes that aim to make the IMC method more 
computationally efficient at high optical depths by using the diffusion
approximation have been suggested in the literature \citep{Fleck:84,
  Gentile:01}. One of the simplest and most efficient such methods is
the discrete-diffusion Monte Carlo (DDMC) scheme
of~\cite{Densmore:07}.    

\cite{Densmore:07} developed the DDMC scheme for gray radiation
transport without physical scattering in 1D planar geometry for
non-moving matter. In this section, we extend this scheme to the
energy-dependent case with physical scattering (the extension to the
velocity-dependent case presented in
Section~\ref{sec:velocity}). We again assume a 1D spherical static
matter distribution and, for simplicity of illustration, focus on
photon transport (instead of neutrino transport) because the ideas
behind extension to the energy-dependent case do not depend on any aspects
that are specific to photons or neutrinos.  
In the following, we first derive the discretized diffusion equations
in the multi-energy case with physical scattering, and then give a
Monte Carlo interpretation of the relevant diffusion equations. 

We start by introducing the zeroth and first radiation moments $J$ and
$H$~\citep{Mihalas:84}:     
\begin{eqnarray}
\label{eq:j}
J & = & \frac{1}{2} \int_{-1}^1 I d\mu \, , \\
\label{eq:h}
H & = & \frac{1}{2} \int_{-1}^1 I \mu d\mu \ ,
\end{eqnarray}
and apply the operator
\be
\label{eq:ang_int}
\frac{1}{4\pi}\int_{4\pi} d \Omega
\ee
to the IMC photon transport equation~(\ref{eq:te3}) to obtain
\begin{eqnarray}
\label{eq:ddmc_mg_sca_1}
\frac{1}{c}\frac{\p J}{\p t} + \frac{1}{r^2}\frac{\p}{\p r} \left(r^2
  H\right) = f_n \tilde\kappa_a \tilde b \, c U_{r,n}^* - 
  (\tilde\kappa_a+\tilde\kappa_s) J \nonumber\\\nonumber\\ + 4\pi
  (1-f_n)\frac{\tilde\kappa_a\tilde b}{\tilde\gamma_p}\int_0^\infty
  \tilde\gamma J(\ve') d\ve'
  \nonumber\\\nonumber\\
  + \pi \int_{-1}^{+1} \int_{-1}^{+1} \int_0^\infty \tilde
  \vk_s(\ve',\mu' \to \ve,\ \mu)
  I(\mu',\ve') d\mu d\mu' d\ve' \, .  
\end{eqnarray}
Observing that 
\be
2\pi \int_{-1}^{+1} \tilde \vk_s(\ve',\mu' \to \ve,\ \mu) d\mu \equiv
\tilde \vk_s^0 (\ve' \to \ve) 
\ee
does not depend on the ``angle" $\mu$, we rewrite
equation~(\ref{eq:ddmc_mg_sca_1}) using this result:
\bea
\label{eq:ddmc_mg_sca_2}
\frac{1}{c}\frac{\p J}{\p t} + \frac{1}{r^2}\frac{\p}{\p r} \left(r^2
  H\right) = f_n \tilde\kappa_a \tilde b \, c U_{r,n}^* - 
  (\tilde\kappa_a+\tilde\kappa_s) J \nonumber\\\nonumber\\ + 4\pi
  (1-f_n)\frac{\tilde\kappa_a\tilde b}{\tilde\gamma_p}\int_0^\infty
  \tilde\gamma J(\ve') d\ve'
  \nonumber\\\nonumber\\
  + \int_0^\infty \tilde\vk_s^0(\ve'\to\ve) J(\ve') d\ve'\, .    
\eea
We assume that our entire spatial domain $0<r<R$ is divided into
  connected, non-overlapping spatial cells, for which $r\in[r_{j-1/2}, 
  r_{j+1/2}]$, where $j=\{1,\dots,N_r\}$. Furthermore, we designate a
subregion  $0<r<R^\mathrm{DD}$ of this domain, which is covered by  
$j=\{1,\dots,m\}$ cells, for DDMC. The cells with $j<m$ will be called
\emph{interior DDMC} cells, while cell $j=m$ will be called the
\emph{interface DDMC} cell. The discretized diffusion equations for
interior cells are slightly different from those for interface cells.
Therefore, we derive them in two separate steps in the following.

\subsection{Interior DDMC Cells}
\label{sec:interior_ddmc}

For interior DDMC cells, we approximate
equation~(\ref{eq:ddmc_mg_sca_1}) in each spatial cell $j$ by using
  cell-centered values $\{\tilde\kappa_{a,j}, \tilde\kappa_{s,j}, 
  f_{n,j}, U_{r,n,j}^*, b_j, \gamma_j, \gamma_{p,j}, \vk_{s,j}^0\}$ of
  the
  quantities $\{\tilde\kappa_a, \tilde\kappa_s, f_n, U_{r,n}^*, b,
  \gamma, \gamma_p, \vk_s^0\}$:   
\bea
\label{eq:ddmc_mg_sca_3}
\frac{1}{c}\frac{\p J}{\p t} + \frac{1}{r^2}\frac{\p}{\p r} \left(r^2
  H\right) = f_n \tilde\kappa_{a,j} \tilde b \, c U_{r,n,j}^* - 
  (\tilde\kappa_{a,j}+\tilde\kappa_{s,j}) J \nonumber\\\nonumber\\ +
  4\pi (1-f_{n,j})\frac{\tilde\kappa_{a,j}\tilde
  b_j}{\tilde\gamma_{p,j}}\int_0^\infty \tilde\gamma_j J(\ve') d\ve'
  \nonumber\\\nonumber\\ + \int_0^\infty
  \tilde\vk_{s,j}^0(\ve'\to\ve) J(\ve') d\ve'\, . 
\eea
Next, we apply the operator 
\be
\frac{1}{\Delta V_j} \int_{r_{j-1/2}}^{r_{j+1/2}} d V =
\frac{1}{\Xi_j \Delta r_j r^2_j} \int_{r_{j-1/2}}^{r_{j+1/2}} r^2
dr \, ,
\ee
where $\Xi_j=1+\Delta r_j^2/(12r_j^2)$, to
equation~(\ref{eq:ddmc_mg_sca_3}) to obtain
\bea
\label{eq:ddmc_mg_sca_4}
\frac{1}{c} \frac{\p J_j}{\p t} + \frac{1}{\Xi_j \Delta r_j r_j^2}
\left(r^2_{j+1/2} H_{j+1/2} - r^2_{j-1/2} H_{j-1/2}\right)
\nonumber\\\nonumber\\
 = f_n \tilde\kappa_{a,j} \tilde b \, c U_{r,n,j}^* - 
  (\tilde\kappa_{a,j}+\tilde\kappa_{s,j}) J_j \nonumber\\\nonumber\\ +
  4\pi (1-f_{n,j})\frac{\tilde\kappa_{a,j}\tilde
  b_j}{\tilde\gamma_{p,j}}\int_0^\infty \tilde\gamma_j J_j(\ve') d\ve'
  \nonumber\\\nonumber\\ + \int_0^\infty
  \tilde\vk_{s,j}^0(\ve'\to\ve) J_j(\ve') d\ve'\, ,
\eea
where 
\be
J_j = \frac{1}{\Xi_j \Delta r_j r^2_j} \int_{r_{j-1/2}}^{r_{j+1/2}}
r^2 J dr \, 
\ee
and
\be
H_{j\pm1/2} = H(r_{j\pm1/2}) \, .
\ee

We further transform equation~(\ref{eq:ddmc_mg_sca_4}) for the
interior DDMC cells. Using Fick's law~\citep{Pomraning:73},  
\begin{equation}
\label{eq:fick_mg}
H(r) = - \frac{1}{3\kappa_\mathrm{T}} \frac{\p J}{\p r} \, ,
\end{equation}
where $\kappa_T$ is the transport opacity\footnote{The
  transport opacity $\kappa_\mathrm{T}$ is defined as
  $\kappa_\mathrm{T} = \kappa_a + \kappa_s - 2\pi\int_{-1}^1 \mu
  \kappa_s(\mu) \, d\mu$.}, we evaluate the cell-edge $H_{j+1/2}$ of
  $H$ within time interval $t_n\le t\le t_{n+1}$ as: 
\begin{equation}
\label{eq:fick_grid_mg}
H_{j+1/2} = - \frac{1}{3\tilde\kappa_a} \frac{\p J}{\p r}
(r=r_{j+1/2}) \, ,  
\end{equation}
while $H_{j-1/2}$ is calculated similarly at point
$r=r_{j-1/2}$. Note that here we have used the time-centered value
$\tilde\kappa_\mathrm{T}$ of the transport opacity
$\kappa_\mathrm{T}$. By employing a finite-difference  approximation
for equation~(\ref{eq:fick_grid_mg}), we can express $H_{j+1/2}$ in
cell $j$ as  
\begin{equation}
\label{eq:flux1_mg}
H_{j+1/2} = - \frac{2}{3\tilde\kappa^-_{\mathrm{T},j+1/2} \Delta r_j} 
\left(J_{j+1/2} - J_j\right) \, ,
\end{equation}
or in cell $j+1$ as
\begin{equation}
\label{eq:flux2_mg}
H_{j+1/2} = - \frac{2}{3\tilde\kappa^+_{\mathrm{T},j+1/2} \Delta
  r_{j+1}} \left(J_{j+1} - J_{j+1/2}\right) \, ,
\end{equation}
where $J_{j+1/2} = J(r=r_{j+1/2})$ and $\kappa^+_{\mathrm{T},j+1/2}$
is the transport opacity at the inner boundary of cell $j+1$, while
$\kappa^+_{\mathrm{T},j+1/2}$ is that at the outer boundary of cell
$j$. Equating the RHSs of equations~(\ref{eq:flux1_mg}) and
(\ref{eq:flux2_mg}), and solving the resulting equation for
$J_{j+1/2}$, we obtain  
\begin{eqnarray}
\label{eq:phijplus1_mg}
J_{j+1/2} = \frac{\tilde\kappa^+_{\mathrm{T},j+1/2} \Delta r_{j+1} J_j
  + \tilde\kappa^-_{\mathrm{T},j+1/2} \Delta r_j
  J_{j+1}}{\tilde\kappa^-_{\mathrm{T},j+1/2} \Delta r_j +
  \tilde\kappa^+_{\mathrm{T},j+1/2} \Delta r_{j+1}} \, .   
\end{eqnarray}
Then, if we use equation~(\ref{eq:phijplus1_mg}) to evaluate either
equation~(\ref{eq:flux1_mg}) or equation~(\ref{eq:flux2_mg}), we find an
approximate expression for $H_{j+1/2}$: 
\begin{equation}
\label{eq:Fjplus1_mg}
H_{j+1/2} = -\frac{2}{3} \frac{J_{j+1} - J_j}{\tilde\kappa^-_{\mathrm{T},j+1/2}
  \Delta r_j + \tilde\kappa^+_{\mathrm{T},j+1/2} \Delta r_{j+1}} \ . 
\end{equation}
Substituting the RHS of~(\ref{eq:Fjplus1_mg}) and a similar expression
for $H_{j-1/2}$ into equation~(\ref{eq:ddmc_mg_sca_4}), we obtain an
equation for $J_j$ in cell $j$: 
\bea
\label{eq:ddmce1d_mg}
\frac{1}{c}\frac{\p}{\p t} J_j = - \left[\kappa_{L,j} + \kappa_{R,j} +
\tilde\kappa_{a,j} + \tilde\kappa_{s,j} \right] J_j + f_{n,j}
\tilde\kappa_{a,j} \tilde b_j c U_{r,n,j}^* \nonumber\\\nonumber\\
 + \frac{\Xi_{j+1} \Delta r_{j+1} r_{j+1}^2}{\Xi_j \Delta r_j
   r_j^2} \kappa_{L,j+1} J_{j+1} + \frac{\Xi_{j-1} \Delta r_{j-1}
  r_{j-1}^2}{\Xi_j \Delta r_j r_j^2} \kappa_{R,j-1} J_{j-1} 
\nonumber\\\nonumber\\ + 4\pi
(1-f_{n,j})\frac{\tilde\kappa_{a,j}\tilde
  b_j}{\tilde\gamma_{p,j}}\int_0^\infty \tilde\gamma_j J_j(\ve') d\ve' 
 + \int_0^\infty \tilde\vk_{s,j}^0(\ve'\to\ve)
J_j(\ve') d\ve'\, . 
\nonumber\\\nonumber\\ 
\eea
In the last equation, we have introduced two new quantities:
\begin{equation}
\label{eq:kappal_mg}
\kappa_{L,j}=\frac{2 r_{j-1/2}^2}{3\Xi_j \Delta r_j r_j^2}  
  \frac{1}{\tilde\kappa^+_{\mathrm{T},j-1/2} \Delta r_j +
  \tilde\kappa^-_{\mathrm{T},j-1/2} \Delta r_{j-1}} \, , 
\end{equation}
and
\begin{equation}
\label{eq:kappar_mg}
\kappa_{R,j}=\frac{2r_{j+1/2}^2}{3\Xi_j \Delta r_j r_j^2}
  \frac{1}{\tilde\kappa^-_{\mathrm{T},j+1/2} \Delta r_j +
  \tilde\kappa^+_{\mathrm{T},j+1/2} \Delta r_{j+1}} \, , 
\end{equation}
which are called left-leakage $(\kappa_{L,j})$ and right-leakage
$(\kappa_{R,j})$ opacities. The reason they are called ``leakage''
opacities will become apparent below, when we provide a Monte Carlo
interpretation of equation~(\ref{eq:ddmce1d_mg}). 

As a next step, we discretize equation~(\ref{eq:ddmce1d_mg}) into
energy groups:
\bea
\label{eq:ddmce1d_mg_eg1}
\frac{1}{c}\frac{\p}{\p t} J_j(\ve_k) = - \left[\kappa_{L,j}(\ve_k) +
  \kappa_{R,j}(\ve_k) + \tilde\kappa_{a,j}(\ve_k) +
  \tilde\kappa_{s,j}(\ve_k)\right] J_j(\ve_k) \nonumber\\\nonumber\\
+ f_{n,j} \tilde\kappa_{a,j}(\ve_k) \tilde b_j(\ve_k) c U_{r,n,j}^* +
  \frac{\Xi_{j+1} \Delta r_{j+1} r_{j+1}^2}{\Xi_j \Delta r_j
  r_j^2} \kappa_{L,j+1}(\ve_k) J_{j+1}(\ve_k) \nonumber\\\nonumber\\ +
  \frac{\Xi_{j-1} \Delta r_{j-1} r_{j-1}^2}{\Xi_j \Delta r_j 
  r_j^2} \kappa_{R,j-1}(\ve_k) J_{j-1}(\ve_k) \nonumber\\\nonumber\\ 
+ 4\pi (1-f_{n,j})\frac{\tilde\kappa_{a,j}(\ve_k)\tilde
  b_j(\ve_k)}{\tilde\gamma_{p,j}} \sum_l \tilde\gamma_j(\ve_l)
  J_j(\ve_l) \Delta \ve_l \nonumber\\\nonumber\\ + \sum_l
  \tilde\vk_{s,j}^0(\ve_l\to\ve_k) J_j(\ve_l) \Delta\ve_l\, ,
  \nonumber\\\nonumber\\  
\eea
where $\varepsilon_l$ is the value of energy in group $l$ and
$\Delta\ve_l$ is the width of that energy group. We now express the
summations on the RHS of this equation as  
\bea
\label{eq:sum1}
4\pi (1-f_{n,j})\frac{\tilde\kappa_{a,j}(\ve_k)\tilde
  b_j(\ve_k)}{\tilde\gamma_{p,j}} \sum_l \tilde\gamma_j(\ve_l)
J_j(\ve_l) \Delta \ve_l \nonumber\\\nonumber\\ =  4\pi
(1-f_{n,j})\frac{\tilde\kappa_{a,j}(\ve_k)\tilde
  b_j(\ve_k)}{\tilde\gamma_{p,j}} \sum_{l\ne k}
\tilde\gamma_j(\ve_l)J_j(\ve_l) \Delta \ve_l \nonumber\\\nonumber\\ + 
(1-f_{n,j})\tilde\kappa_{a,j}(\ve_k) J_j(\ve_k) \frac{\gamma_j(\ve_k)
  \tilde B_j(\ve_k)\Delta \ve_k}{\int_0^\infty \gamma_j(\ve)\tilde
  B_j(\ve)d \ve} \, ,
\eea
and 
\bea
\label{eq:sum2}
\sum_l \tilde\vk_{s,j}^0(\ve_l\to\ve_k) J_j(\ve_l) \Delta\ve_l =
\nonumber\\\nonumber\\ \sum_{l\ne k} \tilde\vk_{s,j}^0(\ve_l\to\ve_k)
J_j(\ve_l) \Delta\ve_l + \tilde\vk_{s,j}^0(\ve_k\to\ve_k) J_j(\ve_k)
\Delta\ve_k \, . \nonumber\\ 
\eea
Using equations~(\ref{eq:sum1})-(\ref{eq:sum2}) in
equation~(\ref{eq:ddmce1d_mg_eg1}), we obtain
\bea
\label{eq:ddmce1d_mg_eg}
\frac{1}{c}\frac{\p}{\p t} J_{j,k} = - \big[\kappa_{L,j,k} +
  \kappa_{R,j,k} + f_{n,j} \tilde\kappa_{a,j,k}
  + (1-f_{n,j}) \tilde\sigma_{a,j,k} \nonumber\\\nonumber\\ +
  \tilde\sigma_{s,j,k}\big] J_{j,k} + f_{n,j} \tilde\kappa_{a,j,k} 
  \tilde b_{j,k} c U_{r,n,j}^* \nonumber\\\nonumber\\ + 
  \frac{\Xi_{j+1} \Delta r_{j+1} r_{j+1}^2}{\Xi_j \Delta r_j
  r_j^2} \kappa_{L,j+1,k} J_{j+1,k}  +
  \frac{\Xi_{j-1} \Delta r_{j-1} r_{j-1}^2}{\Xi_j \Delta r_j 
  r_j^2} \kappa_{R,j-1,k} J_{j-1,k} \nonumber\\\nonumber\\ 
+ 4\pi (1-f_{n,j})\frac{\tilde\kappa_{a,j,k}\tilde
  b_{j,k}}{\tilde\gamma_{p,j}} \sum_{l\ne k} \tilde\gamma_{j,k}
  J_{j,l} \Delta \ve_l \nonumber\\\nonumber\\ + \sum_{l\ne k}
  \tilde\vk_{s,j}^0(\ve_l\to\ve_k) J_{j,l} \Delta\ve_l\, ,
  \nonumber\\\nonumber\\  
\eea
where the subscript $k$ denotes quantities pertaining to energy group
$k$. In equation~(\ref{eq:ddmce1d_mg_eg}), for brevity we also have
introduced two new variables:
\be
\label{eq:sigma_a}
\tilde\sigma_{a,j,k} = \left[1 - \frac{\tilde\gamma_{j,k} \tilde
  B_{j,k} \Delta\ve_k}{\int_0^\infty \gamma_j(\ve) B_j (\ve)
  d\ve}\right] \tilde\kappa_{a,j,k}
\ee
and
\be
\label{eq:sigma_s}
\tilde\sigma_{s,j,k} = \left[1 - \frac{\tilde\vk_{s,j}^0(\ve_k\to\ve_k)
  \Delta\ve_k}{\kappa_{s,j,k}}\right] \tilde\kappa_{s,j,k} \, .
\ee
It is easy to see that $\tilde\sigma_{a,j,k} < \tilde\kappa_{a,j,k}$
and $\tilde\sigma_{s,j,k} < \tilde\kappa_{s,j,k}$. As we will discuss
below, this property has important implications for the computational
efficiency of our multi-group DDMC scheme.  

We observe that equation~(\ref{eq:ddmce1d_mg_eg}) can be viewed as an
equation for the time evolution of $J_{j,k}$ in cell $j$ and energy
group $k$. Namely, according to equation~(\ref{eq:ddmce1d_mg_eg}),
function $J_{j,k}$ decreases at a rate 
\begin{equation}
\left[\kappa_{L,j,k}+\kappa_{R,j,k}+f_{n,j}\tilde\kappa_{a,j,k} +
  (1-f_{n,j})\tilde\sigma_{a,j,k}+\tilde\sigma_{s,j,k}\right]J_{j,k}c\, ,   
\end{equation}
due to the 1st term on the RHS of that equation and increases at a
rate   
\begin{eqnarray}
\bigg[f_{n,j} \tilde\kappa_{a,j,k} 
  \tilde b_{j,k} c U_{r,n,j}^* + \frac{\Xi_{j+1} \Delta r_{j+1}
  r_{j+1}^2}{\Xi_j \Delta r_j r_j^2} \kappa_{L,j+1,k} J_{j+1,k} +
  \nonumber\\\nonumber\\ 
  \frac{\Xi_{j-1} \Delta r_{j-1} r_{j-1}^2}{\Xi_j \Delta r_j 
  r_j^2} \kappa_{R,j-1,k} J_{j-1,k} \nonumber\\\nonumber\\ 
+ 4\pi (1-f_{n,j})\frac{\tilde\kappa_{a,j,k}\tilde
  b_{j,k}}{\tilde\gamma_{p,j}} \sum_{l\ne k} \tilde\gamma_{j,k}
  J_{j,l} \Delta \ve_l \nonumber\\\nonumber\\ + \sum_{l\ne k}
  \tilde\vk_{s,j}^0(\ve_l\to\ve_k) J_{j,l} \Delta\ve_l\bigg] c \, ,
\end{eqnarray}
due to the rest of the terms on the RHS of
equation~(\ref{eq:ddmce1d_mg}). Now, recalling that function
$J_{j,k}$ represents the number of MC particles in cell $j$ in energy
group $k$, we make the following Monte Carlo interpretation of 
this equation: The terms $ f_{n,j}\tilde\kappa_{a,j,k} \tilde b_{j,k}
c^2 U_{r,n,j}^*$, $ \frac{\Xi_{j+1} \Delta r_{j+1}
  r_{j+1}^2}{\Xi_j \Delta r_j r_j^2} c \kappa_{L,j+1,k} J_{j+1} $,
and $\frac{\Xi_{j-1} \Delta r_{j-1} r_{j-1}^2}{\Xi_j \Delta r_j
  r_j^2} c \kappa_{R,j-1,k} J_{j-1} $ describe the rate of increase of
the number of MCPs in cell $j$ and energy group $k$ due to emission
and leakage from the right and left neighboring cells,
respectively. Moreover, terms $f_{n,j} \tilde\kappa_{a,j,k} J_{j,k} c$,
$\kappa_{L,j,k} J_{j,k} c$ and $\kappa_{R,j,k} J_{j,k} c$ represent
the rate of decrease of $J_{j,k}$ due to absorption of MCPs, and
leakage of MCPs to left and right neighboring cells, respectively. In
addition to these terms, we have terms that are responsible for
physical and effective scattering. Terms $\sigma_{s,j,k} J_{j,k}$ and
$(1-f_{n,j})\sigma_{a,j,k} J_{j,k}$ describe the decrease of $J_{j,k}$
due to physical and effective scattering of MCPs from energy group
$k$ to any other energy group $l$ ($l \neq k$),
respectively. Accordingly, the terms 
\be
4\pi (1-f_{n,j})\frac{\tilde\kappa_{a,j,k}\tilde
  b_{j,k}}{\tilde\gamma_{p,j}} \sum_{l\ne k} \tilde\gamma_{j,k}
J_{j,l} c \Delta \ve_l
\ee
and 
\be
\sum_{l\ne k} \tilde\vk_{s,j}^0(\ve_l\to\ve_k) J_{j,l} \Delta\ve_l c 
\ee
represent the increase rate of $J_{j,k}$ due to scattering of an MCP
from energy group $l\ne k$ to energy group $k$.  

In this picture, DDMC particles have no propagation
angle $\mu$ or position $r$ within a cell, but they always know their
current cell and time. A DDMC particle can either remain in its cell
within a timestep without a collision, or undergo a
``collision.'' Here the term collision refers to an absorption, a
leakage to the left or right neighboring cell, or from one energy
group to another. The quantity    
\be
\label{eq:ddmc_tot_op}
\kappa_{L,j,k} + \kappa_{R,j,k} + f_{n,j} \tilde\kappa_{a,j,k} +
(1-f_{n,j}) \tilde \sigma_{a,j,k} + \tilde\sigma_{s,j,k}\,  
\ee
can be regarded as the total collision opacity. Using this
interpretation, we can perform DDMC transport based on the calculation
of distances, similar to the Monte Carlo procedure
described in Section~\ref{sec:MCImplementation}. However, in DDMC, we
do not calculate distances to the boundaries. Instead, we calculate the
distance  to collision, $d_c$, and distance traveled until the end of
the timestep, $d_t$. Since the distance to collision $d_c$ is based on the 
opacity~(\ref{eq:ddmc_tot_op}), this distance can be calculated
probabilistically using a formula similar to equation~(\ref{eq:d_c}): 
\begin{equation}
\label{eq:d_c_ddmc}
d_c = - \frac{\ln \xi}{\kappa_{L,j,k} + \kappa_{R,j,,k} + f_{n,j}
  \tilde\kappa_{a,j,,k} + (1-f_{n,j}) \tilde\sigma_{a,j,k} +
  \tilde\sigma_{s,j,k}} \, ,   
\end{equation}
where $\xi$ is a random number uniformly distributed in $(0, 1]$, 
and the distance traveled to the end of the timestep is calculated
using relation~(\ref{eq:d_t})\footnote{This procedure is slightly
  different in the continuous absorption method. In this case, we
  again calculate two distances. However, the   
  distance to collision is calculated using the opacity $\kappa_{L,j}
  + \kappa_{R,j} + (1-f_{n,j}) \tilde\kappa_{a,j} +
  \tilde\kappa_{s,j}$ and, thus, a collision can be either
  left-leakage, right-leakage, or physical or effective scattering, while
  the condition for absorption is calculated as described in
  Section~\ref{sec:con_absorption}.}.  

If the time to collision is less than the time remaining in the
timestep, the DDMC particle undergoes a collision, and the time to 
collision is decremented from the time remaining in the
timestep. Again, as we see from the second term on the left side of 
equation~(\ref{eq:ddmce1d_mg}), a ``collision'' can be an absorption, a
left-leakage, a right-leakage, or a leakage from one energy group
to another. The collision type is sampled from the probability of the
collision type that is calculated using the relative magnitudes of the
different ``opacities.'' For example, the probability of left-leakage
can be calculated from 
\be
p_L = - \frac{\kappa_{L,j,k}}{\kappa_{L,j,k} + \kappa_{R,j,k} +
  f_{n,j} \tilde\kappa_{a,j,k} + (1-f_{n,j}) \tilde\sigma_{a,j,k} +
  \tilde\sigma_{s,j,k}} \, .
\ee
If the collision is an absorption, the MCP history
is terminated, as in standard Monte Carlo. If the DDMC particle
undergoes a leakage reaction, it is transferred to the appropriate
neighboring cell, and the simulation continues. If the collision type
is the ``leakage'' from one energy group to another, then the new
neutrino energy is sampled using the functional form of the
differential scattering opacity. If the time to collision is greater
than the time remaining in the timestep, the DDMC particle reaches the
end of the timestep and is stored for simulation in the next
timestep. 


We point out that the DDMC approach is based on the diffusion
approximation to equation~(\ref{eq:te3}), so it should yield accurate
solutions when used in optically-thick regions. As mentioned above,
the DDMC transport process consists of discrete steps that reflect
transfer of MC particles between spatial cells (but not between
spatial locations within a cell, as in a pure MC method). Due to this
property, DDMC can be much more computationally efficient than the
standard Monte Carlo implementation of equation~(\ref{eq:te3}).

It is interesting to note that if the physical scattering is elastic,
then in-scattering and out-scattering terms in
equation~(\ref{eq:ddmce1d_mg}) cancel each other. Hence, the presence
of elastic scattering does not lead to the appearance of any new terms
in equation~(\ref{eq:ddmce1d_mg}) compared to the case when there is no
physical scattering. Instead, the scattering modifies only the values
of the leakage  opacities $\kappa_{L,j}^\mathrm{T}$ and
$\kappa_{R,j}^\mathrm{T}$ which, in turn, are a result of the use of
the transport opacity in Fick's law. For this reason, the DDMC
scheme does not need to perform any special explicit numerical
operation in order to model elastic scattering; the effect of elastic
scattering is taken into account via modification of the values of the
leakage opacities. For this reason, the DDMC scheme leads to the
biggest savings in computational cost in regimes dominated by elastic
scattering. 

There is also another reason for higher efficiency of DDMC compared to
MC schemes, which stems from the following. In a regime where the
absorption opacity is high, the effective scattering opacity
$\tilde \sigma_{es,j,k} = (1-f_{n,j}) \tilde\sigma_{a,j,k}$ will dominate the
collision opacity for DDMC given by
expression~(\ref{eq:ddmc_tot_op}). Hence, the effective mean-free-path 
in this regime is $\simeq 1/ \tilde \sigma_{es,j,k}$. However,
recalling that $\tilde\sigma_{a,j,k} < \tilde\kappa_{a,j,k}$, we see
that the effective mean-free-path for DDMC should be larger compared
to that for the MC scheme, which is given by $1/[(1-f_{n,j}) \tilde
  \kappa_{a,j,k}]$. Depending on the energy group $k$ and its width,
$\tilde\sigma_{a,j,k}$ can be significantly smaller than
$\tilde\kappa_{a,j,k}$. Therefore, from this source alone, we gain a
speed-up by a factor of $\lesssim
\tilde\kappa_{a,j,k}/\tilde\sigma_{a,j,k}$. Similarly, we achieve
speed up from the fact that the inelastic physical scattering
opacity in the DDMC regime, $\tilde\sigma_{s,j,k}$, is smaller than that
for the MC scheme, which is $\tilde\kappa_{s,j,k}$.  

Finally, the biggest speed-up in DDMC comes from the following
assumption at a cost of making one more (but excellent, as will become
clear later) approximation. We split the effective scattering opacity
$\tilde\sigma_{es,j,k}$ into two parts, $a_{j,k}
\tilde\sigma_{es,j,k}$ and $(1-a_{j,k})\tilde\sigma_{es,j,k}$, where
$0 \le a_{j,k} \le 1$, and restrict the first of these two to be
elastic effective scattering, while the second one is free to be
inelastic (as effective scattering would otherwise be). As we
discussed in the above, the presence of an
extra elastic 
scattering source does not increase the cost of doing DDMC
transport. Therefore, by assuming that a $a_{j,k} \tilde\sigma_{es,j,k}$
fraction of effective scattering is elastic (instead of being
inelastic), we achieve computational savings proportional to
$a_{j,k}$. As we demonstrate in Section~\ref{sec:pns_cooling},
depending on the scenario, this can lead to speed-up of calculations by
a factor of $10^2-10^3$ or more.  

Since effective scattering is in general inelastic, an obvious
question arises: Is it a good approximation to treat a fraction $a_{j,k}$
of effective scatterings as elastic in the DDMC region? The answer depends
on the value the Fleck factor $f_{n,j}$ and the parameter $a_{j,k}$. The
only way the inelasticity of effective scattering affects the
transport of 
MCPs is by enabling thermalization of MCPs when they move to a new
cell with different $T$, $Y_e$, or $\rho$. By thermalization we mean
a change of the spectrum of MCPs when they move to a new cell to
reflect the emissivity spectrum of that cell. However, when each MCP
undergoes inelastic scattering at least once after it moves to a new
cell, that MCP acquires an energy spectrum that reflects the emissivity
spectrum of its new cell. In this case, the inelastic nature of
effective scattering should not play any role after MCPs move to a
new cell and before they leak out to another zone. Thus, these
scatterings can be treated as elastic between the two 
events. Alternatively, if MCPs that move to a new cell get absorbed
before they propagate to another zone, the inelastic nature of
effective scattering again should not play a role. Therefore,
for this treatment to be exact, MCPs that leak to a new cell should
undergo effective scattering at least once, or get absorbed before they
move to a different cell. This condition is fulfilled if 
\bea
\kappa_{L,j,k} & \ll & f_{n,j} \tilde\sigma_{a,j,k} \quad \mathrm{and}
\nonumber\\\nonumber\\ 
\kappa_{R,j,k} & \ll & f_{n,j} \tilde\sigma_{a,j,k} ,
\eea
or 
\bea
\label{eq:ddmc_eff_sca_con}
\kappa_{L,j,k} & \ll & a_{j,k} (1-f_{n,j}) \tilde\sigma_{a,j,k} \quad
\mathrm{and} \nonumber\\\nonumber\\
\kappa_{R,j,k} & \ll & a_{j,k} (1-f_{n,j}) \tilde\sigma_{a,j,k} .
\eea
In Section~\ref{sec:pns_cooling}, we demonstrate that these conditions
are met in DDMC regions for appropriately chosen values of
$a_{j,k}$.

\subsection{Interface Cells}

The method which we use for interfacing DDMC with standard MC is the
same as in~\cite{Densmore:07}.

We start by deriving an equation for the cell-centered value of $J_m$ in cell
$m$ on the right boundary of the DDMC region. Using
equation~(\ref{eq:Fjplus1_mg}), we derive an expression for
$H_{m-1/2}$:   
\be
\label{eq:Fjplus1_mg_2}
H_{m-1/2} = -\frac{2}{3} \frac{J_m -
  J_{m-1}}{\tilde\kappa^-_{\mathrm{T},m-1/2} \Delta r_{m-1} +
  \tilde\kappa^+_{\mathrm{T},m-1/2} \Delta r_m} \, .
\ee
Substituting~(\ref{eq:Fjplus1_mg_2}) into
equation~(\ref{eq:ddmc_mg_sca_4}) for cell $j=m$, we obtain 
\bea
\label{eq:ddmc_mg_sca_4_if}
\frac{1}{c} \frac{\p J_m}{\p t} = 
- \left[\kappa_{L,m} + \tilde\kappa_{a,m} + \tilde\kappa_{s,m} \right]
J_m - \frac{r^2_{l+m/2}}{\Xi_m \Delta r_m r_m^2} H_{m+1/2}
\nonumber\\\nonumber\\ + f_n \tilde\kappa_{a,m} \tilde b \, c
U_{r,n,m}^* + \frac{\Xi_{m-1} \Delta r_{m-1} r_{m-1}^2}{\Xi_m
  \Delta r_m r_m^2} \kappa_{R,m-1} J_{m-1}  \nonumber\\\nonumber\\
  4\pi (1-f_{n,m})\frac{\tilde\kappa_{a,m}\tilde
  b_m}{\tilde\gamma_{p,m}}\int_0^\infty \tilde\gamma_m J_m(\ve') d\ve'
  \nonumber\\\nonumber\\ + \int_0^\infty
  \tilde\vk_{s,m}^0(\ve'\to\ve) J_m(\ve') d\ve'\, ,
\eea
where we have made use of equations~(\ref{eq:kappal_mg}) and
(\ref{eq:kappar_mg}). To complete this derivation, we must find an
approximate expression for the flux $H$ at the interface of the DDMC
region. 

Following~\cite{Densmore:07}, we use the asymptotic diffusion-limit
boundary condition~\citep{Habetler:75}: 
\be
\label{eq:bc_habetler75}
\int_0^1 W(\mu) I_b(\mu,t) d\mu = J(r_{m+1/2}) +
\frac{\lambda}{\tilde \kappa_{\mathrm{T},m+1/2}^-} \left[\frac{\p
    J}{\p r}\right]_{r=r_{m+1/2}} \, ,
\ee
where $I_b(\mu,t)$ is the radiation intensity due to Monte Carlo
particles incident on the DDMC region, $\lambda\simeq0.7104$ is a
constant, and $W(\mu)$ is a transcendental function well approximated
by 
\be
W(\mu) \simeq \mu + \frac{3}{2} \mu^2 \ .
\ee
Incident intensity $I_b$ in equation~(\ref{eq:bc_habetler75}) is
weighted by $W(\mu)$, which takes into account the angular
distribution of the MC particles coming into the DDMC region. 

To express $H_{m+1/2}$ using equation~(\ref{eq:bc_habetler75}), we
approximate the derivative on the RHS of
equation~(\ref{eq:bc_habetler75}) with a finite difference:
\be
\label{eq:bc_habetler75_dis}
\int_0^1 W(\mu) I_b(\mu,t) d\mu = J_{m+1/2} +
\frac{2\lambda}{\tilde \kappa_{\mathrm{T},m} \Delta r_m}
(J_{m+1/2}-J_m)\, ,
\ee
where $\tilde\kappa_{\mathrm{T},m}$ is the cell-averaged value of the
transport opacity $\tilde\kappa_\mathrm{T}$ in cell $m$, and
$J_{m+1/2}$ is an appropriately defined cell-edge value of
$J$. Solving equation~(\ref{eq:bc_habetler75_dis}) for $J_{m+1/2}$, we 
obtain 
\be
\label{eq:j_lp}
J_{m+1/2} = \frac{\tilde \kappa_{\mathrm{T},m} \Delta
  r_m}{\tilde \kappa_{\mathrm{T},m} \Delta r_m + 2 \lambda} \int_0^1
  W(\mu) I_b(\mu)d\mu + \frac{2\lambda}{\tilde \kappa_{\mathrm{T},m}
  \Delta r_m + 2 \lambda} J_m \, .
\ee
Next, we use equation~(\ref{eq:flux1_mg}) to represent $H_{m+1/2}$:
\begin{equation}
\label{eq:h_lp}
H_{m+1/2} = - \frac{2}{3\tilde\kappa_{\mathrm{T},m} \Delta r_m}  
\left(J_{m+1/2} - J_m\right) \, ,
\end{equation}
where we again have used the cell-averaged value
$\tilde\kappa_{\mathrm{T},m}$ of the transport opacity
$\tilde\kappa_\mathrm{T}$. Substituting the RHS of
equation~(\ref{eq:j_lp}) into formula~(\ref{eq:h_lp}), we obtain 
\be
H_{m+1/2} = - \frac{2}{3\tilde\kappa_{\mathrm{T},m} \Delta r_m + 2
  \lambda} \left(\int_0^1 W(\mu) I_n(\mu) d\mu + J_m \right) \, .
\ee
Substituting the RHS of the last equation into
equation~(\ref{eq:ddmc_mg_sca_4_if}), we obtain  
\bea
\label{eq:ddmc_mg_sca_4_if_1}
\frac{1}{c} \frac{\p J_m}{\p t} = 
- \left[\kappa_{L,m} + \kappa_{R,m} + \tilde\kappa_{a,m} +
  \tilde\kappa_{s,m} \right] J_m \nonumber\\\nonumber\\  
 + f_{n,m} \tilde\kappa_{a,m} \tilde b_m \, c U_{r,n,m}^* +
 \frac{\Xi_{m-1} \Delta r_{m-1} r_{m-1}^2}{\Xi_m \Delta r_m
   r_m^2} \kappa_{R,m-1} J_{m-1} \nonumber\\\nonumber\\
  + \frac{r^2_{m+1/2}}{\Xi_m \Delta r_m r_m^2} \int_0^1 P(\mu)\mu
  I_b(\mu)d\mu 
  \nonumber\\\nonumber\\ + 4\pi
  (1-f_{n,m})\frac{\tilde\kappa_{a,m}\tilde
    b_m}{\tilde\gamma_{p,m}}\int_0^\infty \tilde\gamma_m J_l(\ve')
  d\ve' \nonumber\\\nonumber\\ + \int_0^\infty
  \tilde\vk_{s,m}^0(\ve'\to\ve) J_m(\ve') d\ve'\, ,
\eea
where the right-leakage opacity is defined as
\begin{equation}
\label{eq:kappar_mg_if}
\kappa_{R,m}=\frac{2r_{m+1/2}^2}{\Xi_m \Delta r_m r_m^2} \frac{1}{3
  \tilde\kappa^-_{\mathrm{T},m+1/2} \Delta r_m + 2\lambda} \, ,  
\end{equation}
instead of equation~(\ref{eq:kappar_mg}), and $P(\mu)$ is defined as   
\be
\label{eq:p_ddmc_if}
P(\mu) = \frac{2}{3\tilde\kappa^-_{\mathrm{T},m+1/2} \Delta r_m +
  2\lambda} \left(1+\frac{3}{2}\mu\right) \, . 
\ee
Finally, we discretize equation~(\ref{eq:ddmc_mg_sca_4_if_1}) in
energy groups to obtain

\bea
\label{eq:ddmc_mg_sca_4_if_2}
\frac{1}{c} \frac{\p J_{m,k}}{\p t} = 
- \left[\kappa_{L,m,k} + \kappa_{R,m,k} + \tilde\kappa_{a,m,k} +
  \tilde\kappa_{s,m,k} \right] J_{m,k} \nonumber\\\nonumber\\  
 + f_{n,m} \tilde\kappa_{a,m,k} \tilde b_{m,k} \, c U_{r,n,m}^* +
 \frac{\Xi_{m-1} \Delta r_{m-1} r_{m-1}^2}{\Xi_m \Delta r_m
   r_m^2} \kappa_{R,m-1,k} J_{m-1,k} \nonumber\\\nonumber\\
  + \frac{r^2_{m+1/2}}{\Xi_m \Delta r_m r_m^2} \int_{0}^1
  P_k(\mu)\mu I_{b,k}(\mu)d\mu \nonumber\\\nonumber\\ + 4\pi
  (1-f_{n,m})\frac{\tilde\kappa_{a,m,k}\tilde
    b_{m,k}}{\tilde\gamma_{p,m}} \sum_{l\ne k} \tilde\gamma_{m,l} J_{m,l}
  \Delta\ve_l \nonumber\\\nonumber\\ + \sum_{l\ne k} 
  \tilde\vk_{s,m}^0(\ve_l\to\ve_k) J_{m,l} \Delta \ve_l \, ,
  \nonumber\\\nonumber\\ 
\eea
where subscript $k$ is again used to denote quantities pertaining to
energy group $k$. 

Equation~(\ref{eq:ddmc_mg_sca_4_if_2}) is similar to
equation~(\ref{eq:ddmce1d_mg_eg}). The only differences are the
expression for the right-leakage opacity $\kappa_{R,m,k}$ and the presence 
of the source due to MC particles coming into the DDMC region. The
flow of energy due to this incoming radiation for a direction $\mu$ is
given by $\mu I_b$. Therefore, $P(\mu)$ can be interpreted as the
probability with which an incident MCP with direction $\mu$ converts into a DDMC
particle. 

Following~\cite{Densmore:07}, we implement the conversion of the MCPs
into DDMC particles (and vice-versa) in two separate ways, depending
on whether the DDMC boundary is at the problem boundary or
not. In the latter case, we use the probability given by
equation~(\ref{eq:p_ddmc_if}) to determine if the incoming MCP is 
converted into a DDMC particle. If converted, it starts transporting
using DDMC in cell $j=m$. Otherwise, the particle returns
isotropically to cell $j=m+1$. The DDMC particles that undergo
right-leakage reactions from cell $j=m$ to cell $j=m+1$ are also
placed isotropically at the boundary of the DDMC region (i.e., at the
inner boundary of the cell $j=m+1$). Note that this angular 
distribution is correct only when the incident intensity is nearly
isotropic. Hence, it is important to choose the boundary between the
DDMC and MC regions where the distribution is sufficiently
isotropic.

Second, if the DDMC region is at the outer boundary of the system,
then the incoming MCPs are regarded as a particle source due to
boundary conditions. In this case, we split incoming MC particles
according to equation~(\ref{eq:p_ddmc_if}): a fraction $P(\mu)$ of
these particles is converted into DDMC particles and begins
transporting using DDMC in the DDMC region, while the remaining
fraction $1-P(\mu)$ is regarded as MC particles escaping the system.  

\section{Velocity-dependent Monte Carlo and DDMC}
\label{sec:velocity}

Thus far we have discussed radiative transfer in material that is
not moving. In this section, we extend the schemes discussed in the
previous section to the case when matter is moving with an arbitrary
velocity.   

Again, we assume a spherically symmetric distribution of matter, split
our spatial computational domain into non-overlapping zones, and
linearize the transport equations within a timestep $t_n \le t\le
t_{n+1}$. Moreover, for simplicity of illustration, we consider the
case with no physical scattering. However, as will become clear in the
following, inclusion of scattering is conceptually simple and our code
is capable of handling physical scattering. 
We also focus on photons because the ideas behind the extension to
the velocity-dependent case is the same for both photons and neutrinos.   
We assume that the radial
component $V_{r,j}$ of the velocity vector ${\bf V}$ (as well as other
information such as temperature, density, etc.) in each cell $j$ does
not change within a transport timestep, while the $\theta$ and $\phi$
components of the velocity are assumed to be zero everywhere. Since MC
and DDMC methods are based on somewhat different techniques, we
separately discuss the extension of each of these to the
velocity-dependent case.     

\subsection{Velocity-dependent MC scheme}
\label{sec:MCvelocity}

Our velocity-dependent MC scheme is based on the mixed-frame formalism
of~\cite{Mihalas:82} and \cite{Hubeny:07}. In this
formalism, emissivities and opacities are defined in a frame comoving
with the fluid, which are then Lorentz-transformed to the Eulerian lab
frame, in which transport is performed. 

Before we describe our velocity-dependent MC algorithm, we present
formulae for the Lorentz transformation between the comoving and the lab
frames for several quantities that will be useful later in the
section. The four-momentum of a massless radiation particle is given
by    
\begin{equation}
M^\alpha = \frac{\ve}{c} (1, {\bf n}) \, ,
\end{equation}
where $\ve$ is the photon (or neutrino) energy, and ${\bf n}$
is a unit spatial 3-vector in the particle propagation direction. In
spherical polar coordinates, $M^\alpha$ has the following
form~\citep{Mihalas:84} 
\begin{equation}
M^\alpha = \frac{\ve}{c} 
\left[1, \, \mu, \, (1-\mu^2)^{1/2} \frac{\cos\varphi}{r}, \,
  (1-\mu^2)^{1/2} \frac{\cos\varphi}{r\sin\theta}\right] \, ,
\end{equation} 
where $\varphi$ is the azimuthal angle. If a particle has energy
$\ve$ and travels in direction $\{\mu,\varphi\}$ as measured
in the lab frame, it will have some other energy  $\ve_0$ and
direction $\{\mu_0,\varphi_0\}$ as measured by an observer attached to a
fluid element moving with velocity vector ${\bf V}$ relative to the
lab frame. (Hereafter, we denote all of the quantities measured in the
comoving frame with subscript 0.) Because $M^\alpha$ is a four-vector,
its components in the two frames moving with respect to each other
with velocity ${\bf V}$ are related by general Lorentz 
transformations. Therefore, we obtain   
\begin{equation}
\label{eq:e_lorentz_gen}
\ve_0 = \gamma \ve \left(1 - \frac{{\bf n}\cdot{\bf V}}{c}\right) \, ,  
\end{equation}
and
\begin{equation}
\label{eq:mu_lorentz_gen}
{\bf n}_0 = \frac{\ve}{\ve_0}\left[{\bf n} - \gamma 
  \frac{{\bf V}}{c} \left(1 - \frac{\gamma}{\gamma+1}
  \frac{{\bf n}\cdot{\bf V}}{c} \right) \right] \, ,
\end{equation}
where $\gamma = (1-V^2/c^2)^{-\frac{1}{2}}$ is the Lorentz factor. In
the spherically symmetric case, for the special case of matter motion
along the radial direction, these relations take the following form:    
\begin{equation}
\label{eq:e_lorentz}
\ve_0 = \gamma \ve \left(1 - \frac{V_r \mu}{c}\right) \, ,
\end{equation}
\begin{equation}
\label{eq:mu_lorentz}
\mu_0  = \frac{\mu-V_r/c}{1-\mu V_r/c} \, ,
\end{equation}
\begin{equation}
\varphi_0  = \varphi \, ,
\end{equation}
where $V_r$ is the radial component of the velocity. Next, we need a
formula for the transformation of opacity $\kappa$ (both for
scattering and absorption), derived first by~\cite{Thomas:30}:  
\be
\label{eq:kappa_lorentz}
\kappa(\mu, \ve) = \frac{\ve_0}{\ve} \kappa_0 (\ve_0) \ .
\ee

From the information about the fluid in each cell (such as
temperature), we calculate the emissivities and opacities of the
material in that cell as measured by an observer comoving with the
fluid. We then calculate how many particles are emitted in the
comoving frame in each cell, and sample the radial coordinates $r_0$,
propagation directions $\mu_0$, energies $\ve_0$, and emission
times $t_0$ of these newly emitted MCPs in the comoving frame using
the comoving frame emissivities in same way we did in
Section~\ref{sec:MCImplementation}. Next, we transform quantities  
$\ve_0$, $\mu_0$, $r_0$, $t_0$ for each MCP to the lab frame,
where we then transport the MCPs. The particle energy is
transformed using formula~(\ref{eq:e_lorentz_gen}), while the angle
$\mu$ is transformed using equation~(\ref{eq:mu_lorentz_gen}). In
order to transform the radial coordinate $r_0$, we assume that, at the
beginning of a timestep $t=t_n$, the radial coordinates of the inner
boundaries of the comoving and the lab frame cells coincide with each
other. Then, the radial location $r_0$ of an MCP in the comoving frame
in cell $j$ is related to the lab-frame radial coordinate $r$ via the
Lorentz transformation:      
\be
r = \gamma_j \left[r_0 + V_{r,j} (t_0-t_n)\right] \, ,
\ee
where $V_{r,j}$ is again the radial velocity of the fluid in cell $j$
measured in the lab frame and $\gamma_j$ is the Lorentz factor in zone
$j$. The MCP emission time is transformed into the lab frame using the 
formula    
\be
t = \gamma_j \left(t_0-t_n+ \frac{V_{r,j} r_0}{c^2}\right) \, .
\ee

Now, having transformed all the necessary information about
particles into the lab frame, the next step is to transport the
particles in this frame. The transport algorithm is similar to that
for static matter described in Section~\ref{sec:MCImplementation}, and
is again based on the calculation of the distances to collision,
spatial, or time boundaries. However, in this case, the distances to
collision (absorption or scattering) need to be calculated using the
lab-frame opacities, which are calculated from their comoving frame
values using formula~(\ref{eq:kappa_lorentz}). If the smallest of the
distances is $d_t$, then the MCP goes into the census for the next
timestep, as in the static case
(Section~\ref{sec:MCImplementation}). If $d_b$ is the smallest
distance, then the MCP moves to the new cell, where we transport the
particle using the lab-frame opacity of the new cell. If 
the MCP is absorbed, then its energy and momentum (and lepton number,
if we are dealing with neutrinos with lepton number) are deposited
into its current cell. If an MCP undergoes scattering, then we first
transform  the energy $\ve$ and angle $\mu$ of the particle into the 
comoving frame, calculate the new values of $\ve_0$ and $\mu_0$ as a
result of scattering (as described in
Section~\ref{sec:MCImplementation}), record how much energy and
momentum is exchanged between the MCP and matter as a result of
scattering (in order to deposit both in that cell at the end of the
current timestep), and transform the new energy $\ve_0$ and angle
$\mu_0$ back to the lab frame to continue the transport of the MCP.   

\subsection{The velocity-dependent DDMC scheme}
\label{sec:DDMCvelocity}

Since it is most natural to formulate the diffusion equations in the
Lagrangian frame, we perform velocity-dependent discrete diffusion
Monte Carlo transport in the comoving frame. We start with the
equation for the energy density $E_0$\footnote{Note that $E_0=4\pi
  J_0/c$, where $J_0$ is defined as in equation~(\ref{eq:j}).} of the
specific intensity in the comoving frame accurate to ${\cal O}(V_r/c)$ 
[cf. equation~95.82 of~\cite{Mihalas:84}]: 
\begin{eqnarray}
\label{eq:E_0}
\frac{DE_0}{Dt}+4\pi\rho_0\frac{\p (r^2F_0)}{\p M_r} -
  \frac{V_r}{r}(3P_0-E_0)  - \frac{D\ln\rho}{Dt}
  (E_0+P_0) \nonumber\\\nonumber\\ + \frac{\p}{\p\ve_0}
  \left\{\ve_0 \left[\frac{V_r}{r}(3P_0-E_0) +
  \frac{D\ln\rho}{Dt}P_0 \right] \right\} \nonumber\\\nonumber\\ =
  \kappa_0 \left(4\pi B - c E_0\right)\, ,  
\end{eqnarray}
where $D/Dt$ is the Lagrangian time derivative, $\partial/\partial
M_r=1/(4\pi r^2\rho)\partial/\partial r$, $F_0$ is the radiation flux
($F_0 = 4\pi H_0$, where $H_0$ is defined as in equation~(\ref{eq:h})),
$P_0$ is the diagonal component of the radiation pressure tensor,
\be
P^{ij} = \frac{1}{c} \int_{4\pi} I n^i n^j d\Omega \, ,
\ee
and
\begin{equation}
\label{eq:dlnrhodt}
\frac{D \ln \rho}{Dt} = - \frac{1}{r^2}\frac{\p}{\p r} \left(r^2
V_r\right) \, . 
\end{equation}

Due to the DDMC approach, we assume isotropy of the radiation in the
comoving frame, which implies that $P_0=1/3E_0$. If we substitute
$1/3E_0$ for $P_0$ in equation~(\ref{eq:E_0}) and drop all terms of
${\cal O}(\lambda_p V_r/lc)$ and higher, where $\lambda_p$ is the
mean-free-path and $l$ is the problem domain size, we obtain an
equation for the evolution of $E_0$ that is valid in the
non-equilibrium diffusion limit for moving matter~\citep{Mihalas:84}:
\begin{eqnarray}
\rho\left\{
\frac{D}{Dt}\frac{E_0}{\rho}+\frac{1}{3}
\left[E_0-\frac{\p}{\p\ve_0}(\ve_0 E_0)\right]
\frac{D}{Dt}\frac{1}{\rho} \right\} \nonumber\\\nonumber\\ +
\frac{1}{r^2}\frac{\p}{\p r} \left(r^2 F_0\right) = \kappa_0
\left(4\pi B - c E_0\right)\, .  
\end{eqnarray}
It is easy to show that the last equation can be cast in the following
form: 
\begin{eqnarray}
\frac{DE_0}{Dt}-E_0 \frac{D\ln\rho}{Dt}
+\frac{\ve_0}{3}\frac{\p E_0}{\p\ve_0}
\frac{D\ln\rho}{Dt}+\frac{1}{r^2}\frac{\p}{\p r}
\left(r^2 F_0\right) \nonumber\\\nonumber\\ = \kappa_0 \left(4\pi B -
c E_0\right) \, . 
\end{eqnarray}
Using equation (\ref{eq:dlnrhodt}) and recalling that
$D/Dt=d/dt+\vec{V}_r\cdot\vec{\nabla}$, we rewrite the last equation
as 
\begin{eqnarray}
\label{eq:vdddmc}
\frac{\p E_0}{\p t}+V_r\frac{\p E_0}{\p r} + E_0 \frac{\p V_r}{\p r} 
+\frac{\ve_0}{3}\frac{\p E_0}{\p\ve_0}
\frac{D\ln\rho}{Dt} \nonumber\\\nonumber\\ +\frac{1}{r^2}\frac{\p}{\p
  r} \left(r^2 F_0\right) = \kappa_0 \left(4\pi B - c E_0\right) \, . 
\end{eqnarray}
This equation incorporates three different velocity-dependent effects:
advection, number density compression/decompression, and Doppler
shift, which are described by the 2nd, 3rd, and 4th terms,
respectively, on the LHS of equation~(\ref{eq:vdddmc}).  

We now compare equation~(\ref{eq:vdddmc}) with the other equations for
$E_0$ (under the same approximation) used in the
literature. \cite{Castor:04} derived an equation for $E_0$ accurate
to ${\cal O}(V/c)$:  
\begin{eqnarray}
  \label{eq:E_castor}
\frac{\partial E_0}{\partial t} + \nabla \cdot ({\bf V} E_0) +
  \left[P_0 - \frac{\partial(\ve P_0)}{\partial \ve}\right] :
  {\nabla}{\bf V} + \nabla \cdot {\bf F}_0
  \nonumber\\\nonumber\\ = 4\pi j_0 - \kappa_a c E_0 \ ,  
\end{eqnarray}
where $j_0 = \kappa_0 B$ and the colon ``:'' operator indicates
summing the product of the tensor on the left with the tensor on the
right over two indices, viz., 
\begin{equation}
  {\bf R}:{\bf S} = \sum_{i,j} R_{ij} S_{ij} \ .
\end{equation}
Rewriting equation~(\ref{eq:E_castor}) in 1D spherical polar
coordinates and substituting $P_0 = 1/3 E_0$, which is valid in the
diffusion limit, we obtain an equation that is equivalent to
equation~(\ref{eq:vdddmc}). 

\cite{Swesty:09} derived an equation for $E_0$ accurate to ${\cal
  O}(V/c)$:   
\begin{equation}
  \label{eq:E_swesty_myra}
\frac{\partial E_0}{\partial t} + \nabla \cdot ({\bf V} E_0) +
\nabla \cdot {\bf F}_0 - \ve
\frac{\partial(P_0:\nabla{\bf V})}{\partial \ve} = S \ , 
\end{equation}
where $S$ is the collision term. If we substitute $P_0 = 1/3 E_0$ into
this equation and rewrite the resulting equation in 1D spherical polar
coordinates, we obtain an equation equivalent to
equation~(\ref{eq:vdddmc}).

\subsubsection{Numerical Implementation}

To implement the DDMC scheme, we first rewrite equation~(\ref{eq:vdddmc}) using the quantities
$J=cE/(4\pi)$ and $H=F/(4\pi)$ that we used in Section~\ref{sec:ddmc}:     
\begin{eqnarray}
\label{eq:vdddmc2}
\frac{1}{c}\frac{\p J_0}{\p t}+\frac{V_r}{c}\frac{\p J_0}{\p r} +
\frac{J_0}{c} \frac{\p V_r}{\p r} +\frac{\ve_0}{3c}\frac{\p
  J_0}{\p\ve_0} \frac{D\ln\rho}{Dt} \nonumber\\\nonumber\\
+\frac{1}{r^2}\frac{\p}{\p r} \left(r^2 H_0\right) = \kappa_0 \left(B
- J_0\right) \, .
\end{eqnarray}
It is easy to show that an IMC version of this equation has the
following form:
\begin{eqnarray}
\label{eq:vdddmc2_imc}
\frac{1}{c}\frac{\p J_0}{\p t}+\frac{V_r}{c}\frac{\p J_0}{\p r} +
\frac{J_0}{c} \frac{\p V_r}{\p r} +\frac{\ve_0}{3c}\frac{\p
  J_0}{\p\ve_0} \frac{D\ln\rho}{Dt}+\frac{1}{r^2}\frac{\p}{\p
  r} \left(r^2 H_0\right) \nonumber\\\nonumber\\ = f_n\kappa_0 \tilde
b U_{r,n} - \tilde \kappa_a J_0 + 4 \pi (1-f_n) \frac{\tilde\kappa_a
  \tilde b}{\tilde\kappa_p} \int_0^\infty \tilde\kappa_a
J(\ve') d\ve' \, . \nonumber\\\nonumber\\
\end{eqnarray}
We now split this equation into \emph{three} separate equations:
\begin{eqnarray}
\label{eq:vdddmc2_imc_a}
\frac{1}{c}\frac{\p J_0}{\p t}+\frac{1}{r^2}\frac{\p}{\p
  r} \left(r^2 H_0\right) = f_n\kappa_0 \tilde b U_{r,n} - \tilde
\kappa_a J_0 \nonumber\\\nonumber\\ + 4 \pi (1-f_n)
  \frac{\tilde\kappa_a \tilde b}{\tilde\kappa_p} \int_0^\infty
  \tilde\kappa_a J(\ve') d\ve' \, ,
\end{eqnarray}
\begin{equation}
\label{eq:vdddmc2_imc_b}
\frac{1}{c}\frac{\p J_0}{\p t} +\frac{\ve_0}{3c}\frac{\p
  J_0}{\p\ve_0} \frac{D\ln\rho}{Dt} = 0 \, ,
\end{equation}
and
\begin{equation}
\label{eq:vdddmc2_imc_c}
\frac{1}{c}\frac{\p J_0}{\p t}+\frac{V_r}{c}\frac{\p J_0}{\p r} +
\frac{J_0}{c} \frac{\p V_r}{\p r} = 0 \, ,
\end{equation}
and solve these three equations in operator-split manner in three
separate steps. Equation~(\ref{eq:vdddmc2_imc_a}) is responsible for the
diffusion of radiation through matter and is the same as
equation~(\ref{eq:ddmc_mg_sca_1}) for DDMC transport for non-moving
matter. Hence, in the first operator-split step, we solve
equation~(\ref{eq:vdddmc2_imc_a}) the same way we did
equation~(\ref{eq:ddmc_mg_sca_1}).   

Equation~(\ref{eq:vdddmc2_imc_b}) is responsible for the Doppler shift
of the radiation energy. In order to solve it, we first rewrite this
equation using a new variable $\epsilon_0 = \ln\ve_0$: 
\begin{equation}
\label{eq:ddmcDopplerShift2}
\frac{\p J_0}{\p t}+\varrho \frac{\p J_0}{\p\epsilon_0} = 0 \, ,   
\end{equation}
where 
\be
\varrho=\frac{1}{3}\frac{D\ln\rho}{Dt} \, .
\ee
Equation~(\ref{eq:ddmcDopplerShift2}) is akin to an advection
equation. Its solution within a timestep (within which $\varrho$ is
assumed constant) at time $t$ is given by   
\begin{equation}
\label{eq:ddmcDopplerShift3}
J_0 (t)= J_0^\mathrm{ini}(\epsilon_0-\varrho (t-t_0)) \, ,
\end{equation}
where $J_0^\mathrm{ini}$ is the value of $J_0$ at initial time $t_0$.  
Due to this simple analytical nature of this solution, there is an
easy way of incorporating equation (\ref{eq:ddmcDopplerShift2})
into the full velocity-dependent DDMC framework. Whenever we perform a
DDMC transport operation on an MCP (i.e., move an MCP from one spatial
cell or energy group to another) within a time interval $\Delta t$,
we shift the energy of the neutrinos (or photons) in that MCP by an
amount corresponding to $\Delta \epsilon_e=\varrho\Delta t$.  

Finally, equation~(\ref{eq:vdddmc2_imc_c}) is responsible for the
change of $J_0$ due to advection and compression/decompression of
radiation together with the fluid. We ``solve'' this equation in the
following way. Once the velocity-independent part of the DDMC
transport and Doppler shift operations are performed, the advection and 
compression/decompression effects are modeled by just changing the
position $r$ of a MC particle by $ V_r(r) \Delta t_n$, where $V_r(r)$ is 
the radial velocity of the matter at point $r$. This operation is
performed at each timestep (after each DDMC transport step), and it 
automatically incorporates both advection and
compression/decompression effects. 

\section{Numerical Implementation and Tests}
\label{sec:tests}

In this Section, we describe some details of the implementation and tests
of our 1D spherically symmetric code. The code uses spherical polar
coordinates, and can handle both equidistant and non-equidistant grids.  

The code is currently parallelized employing
hybrid OpenMP/MPI parallelization using the mesh replication method
(cf. Section~\ref{sec:intro}). It uses the open-source {\tt Cactus
  Computational
  Toolkit}~\citep[][http://www.cactuscode.org]{goodale:03}, which
provides MPI parallelization, input/output, and restart capability.     

\subsection{Non-moving background with fixed $T$ and $Y_e$}
\label{sec:tests1}

As a first test, we consider a scenario in which radiation propagates
through static matter with fixed temperature $T$ and electron fraction
$Y_e$. 

\subsubsection{Static scattering atmosphere}
\label{sec:scat_atmos}

\cite{Hummer:71} found stationary-state solutions of the spherical
analogue of the classical Milne problem. The model consists of a
central radiation source surrounded by a static, spherically
symmetric scattering atmosphere of some radius $R_\mathrm{atm}$. The
source emits radiation isotropically with constant luminosity
$L_0$. The opacity of the atmosphere is assumed to be due only to
isotropic scattering with a simple power-law dependence on radius:
\begin{equation}
\kappa_s = r^{-n} \, , \ 0<r<R_\mathrm{atm} \, , \ n > 1 \, .
\end{equation}
According to the solution of \cite{Hummer:71}, the luminosity $L(r)$
at any distance from the source should be constant
and equal to $L_0$. In the tests we perform, we check whether the
condition $L(r)=L_0$ is fulfilled for all
$0<r<R_\mathrm{atm}$.   

We choose $R_\mathrm{atm} =50$ km and $n=1.1$. The central radial
resolution $\Delta r_0$ is $200$ m, and the entire computational
domain is covered by $N_r=200$ cells with logarithmically increasing
size. We performed four different runs with different MCP weights,
creating $100$, $400$, $1600$, and $6400$ new MCPs in
each timestep for a given luminosity, $L_0 = 6.5 \times 10^{48}$ erg/s

\begin{figure}
\centering
\includegraphics[width=8cm]{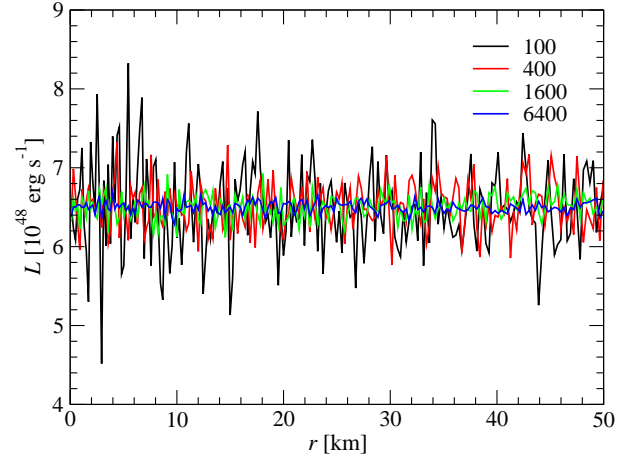}
\caption{The radial profile of the luminosity for a central point
  source emitting into a static scattering atmosphere for four
  different MC runs. In the first run (black line), we emit $100$ MCPs
  at each timestep, while in the second run we produce $400$ MCPs in a
  timestep, etc. In agreement with the analytical solution
  (cf. Section~\ref{sec:scat_atmos}), the luminosity is constant along
  the radial coordinate within statistical errors. According to the
  central limit theorem and as demonstrated in
  Fig.~\ref{fig:delta_l_milne}, these errors decrease as $1/N^{1/2}$,
  where $N$ is the number of MCPs. 
  \label{fig:l_vs_r_milne}} 
\end{figure}

\begin{figure}
\centering
\includegraphics[width=8cm]{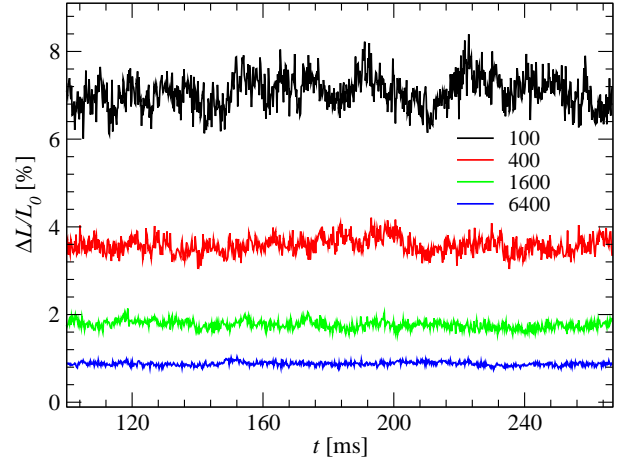}
\caption{Average deviation of the luminosity from its average value in
  each cell as a function of time for radiation from a central point
  source emitting into a static scattering atmosphere. The black line
  corresponds to a run in which we emit $100$ MCPs in each timestep,
  while the red line represents the run with $400$ MCPs, etc. In
  agreement with the central limit theorem, the average deviation decreases as
  $1/N^{1/2}$, where $N$ is the number of MCPs. 
  \label{fig:delta_l_milne}} 
\end{figure}

Figure~\ref{fig:l_vs_r_milne} shows the radial profile of the
luminosity when a steady-state regime is reached. In agreement with
the analytical solution, the luminosity indeed remains $\simeq L_0$ within
statistical errors independent of the distance from the source. Such
errors are expected due to the probabilistic nature of the Monte Carlo
method. Moreover, according to the central limit theorem, these errors
should decrease as $1/\sqrt{N}$, where $N$ is the total number of
simulated MCPs. Figure~\ref{fig:delta_l_milne} shows the relative
deviation $\Delta L(r)/L_0$ of $L(r)$ from its correct value $ L_0$
for these four runs (i.e., here $\Delta L(r) = |L(r)-L_0|$). As we can
see, $\Delta L(r)/L_0$ indeed decreases by $\simeq 2$ as we increase the
number of MCPs by $4$. We have performed additional simulations with
different values of $n$, $\Delta r_0$, $N_r$ and $R_\mathrm{atm}$, and
the code again reproduces the $L(r)=L_0$ solution with statistical
errors that decrease as $1/\sqrt{N}$, as in the case above.  

\subsubsection{Homogeneous sphere}
\label{sec:hom_sphere}

The homogeneous sphere problem is frequently employed to test radiative
transfer codes~\citep{Bruenn:85, Schinder:89, Smit:97, Rampp:02}. This
problem consists of a static homogeneous and isothermal sphere of
radius $R$ that radiates into vacuum. Inside the sphere, the radiation
interacts with the background matter only via isotropic absorption and
thermal emission.   

Despite its simplicity, the problem possesses some important physical
and numerical properties that are often encountered in many practical
applications: There is a sharp discontinuity at the surface of the  
sphere, and this represents a major challenge for finite-difference
methods. However, Monte Carlo methods are well positioned for treating
such discontinuities. Moreover, a situation similar to the stiff
transition from the radiation diffusion regime inside an opaque 
sphere to a free-streaming regime in the ambient vacuum occurs near
a PNS surface in core-collapse supernova simulations. Such a transition
is a source of significant errors in approximate transport schemes
such as, e.g., flux-limited diffusion~\citep{Ott:08}.   

We assume that the sphere of radius $R$ has a constant absorption
opacity $\kappa_a$ and emissivity $B$ in the interior, while in the
ambient vacuum at $r>R$, we have $\kappa_a = B = 0$. For this problem,
the transport equation can be solved
analytically~\citep{Smit:97}:       
\be
\label{eq:an_sol}
I(r,\mu) = B\left(1-e^{-\kappa_a s(r,\mu)}\right) \, ,
\ee
where
\be
\label{eq:an_sol_s}
s(r,\mu)=\left\{
\begin{array}{ll}
r\mu+Rg(r,\mu) & \mathrm{if} \ \ r<R, \quad -1\le\mu\le 1 \, , \\ 
& \\ 
2Rg(r,\mu) & \mathrm{if} \ \ r\ge R, \quad
\sqrt{1-\left(\frac{R}{r}\right)^2} \le\mu\le 1 \, . \\ & \\ 
0 & \mathrm{otherwise} 
\end{array}
\right.
\ee
and
\be
\label{eq:an_sol_g}
g(r,\mu) = \sqrt{1-\left(\frac{r}{R}\right)^2 (1-\mu^2)} \, .
\ee
Note that this solution depends only on three parameters: $\kappa_a$,
$R$, and $B$, where the latter acts as a scale factor for the
solution.    

\begin{figure}
\centering
\includegraphics[width=8cm]{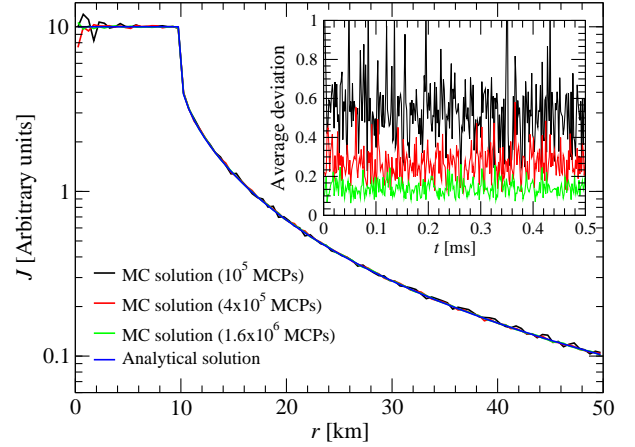}
\caption{The zeroth moment for radiation $J$ as a function of radial
  coordinate $r$ in the homogeneous sphere test for three simulations
  in which $10^5$, $4\times10^5$, and $1.6\times10^6$ new MCPs are
  emitted in each timestep (black, red, and green lines,
  respectively). Also shown (blue line) is the zeroth moment from
  the analytical solution. The inset plot shows the average deviation  
  of the zeroth moment in these three Monte Carlo simulations from the
  analytical solution.
  \label{fig:j_vs_r_hom_sphere}} 
\end{figure}

We perform a set of simulations with $R=10$ km, $\kappa_a = 2.5
\times 10^{-4} \ \mathrm{cm}^{-1}$ and $B=10$ (in CGS units). Our computational
domain has an outer radius of $50$ km and is covered by $100$
equidistant cells. For this setup, we carry out three runs, in which we
choose MCP weights such that $10^5$, $4\times10^5$, or $1.6\times10^6$
new MCPs are emitted in each timestep in a given
simulation (hereafter, simulations A, B, and
C). Figure~\ref{fig:j_vs_r_hom_sphere} shows the zeroth moment $J$ as
a function of the radial coordinate for simulations A, B, and C when
the stationary-state radiation field is reached (black, red, and green lines,
respectively). Also shown (blue line) is the zeroth moment $J$ from
the analytical solution given by
equations~(\ref{eq:an_sol})-(\ref{eq:an_sol_g}). As we can see, the
Monte Carlo solution agrees well with the analytical solution within
statistical errors. The inset plot in Fig.~\ref{fig:j_vs_r_hom_sphere}
shows the time evolution of the average deviation of the zeroth moment in
the Monte Carlo solution from the analytical result when the stationary
state is reached. The average deviation in simulations A, B, and C
fluctuate around $\simeq 0.14$, $\simeq 0.27$, $\simeq 0.53$,
respectively. Hence, in these simulations, when the number of MCPs is
increased by a factor of $4$, the average deviation decreases by a
factor of $\simeq 2$, in agreement with the central limit theorem, and
the solution converges to the analytical result. We have repeated this
simulation set with different values of $\kappa_a$, $B$, $R$ and
different grid resolutions. In all cases, we find excellent agreement
with the analytical result, similar to the case discussed above.     

\subsubsection{Diffusion of a Gaussian Pulse}
\label{sec:diffusion_gauss}

In order to show that our code handles diffusion of radiation properly,
we calculate the diffusion of a Gaussian radiation packet through 
static matter. We assume that radiation interacts with matter only via
isotropic and isoenergetic scattering, and the scattering opacity
$\kappa_s$ is assumed to be constant in space and time.     

The diffusion of a Gaussian packet with initial central position at
$r=r_\mathrm{ini}$ and width $d_0$ in such a medium is described by the
following analytical solution~\citep{Swesty:09,Sumiyoshi:12}
\be
E(r,t) = E_\mathrm{ini}
\left(\frac{t_\mathrm{ini}}{t_\mathrm{ini}+t}\right)^\omega
\exp\left(\frac{-|r-r_\mathrm{ini}|^2} {4D(t_\mathrm{ini}+t)}\right)
\, , 
\ee
where $E(r,t)$ is the radiation energy density at position $r$ and
time $t$ after initial time $t_\mathrm{ini}$. The diffusion
coefficient, $D$, equals $c/3\kappa_s$ and the width of the 
Gaussian pulse, $d_\mathrm{ini}$, as a function of the initial time
equals $(4Dt_\mathrm{ini})^{1/2}$. Parameter
$E_\mathrm{ini}$ is the initial height of the packet and $\omega$ is
related to the number of spatial dimensions, $N_\mathrm{D}$, 
and is equal to $N_\mathrm{D}/2$.  

\begin{figure}
\centering
\includegraphics[width=8cm]{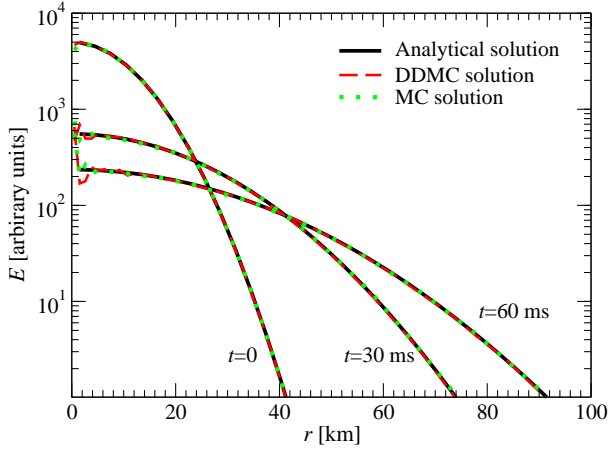}
\caption{The radial profile of the energy density of a Gaussian
  radiation packet diffusing into a uniform medium with a constant
  scattering opacity at three different times. The black line
  corresponds to the analytical solution, the dashed red line is
  obtained using the DDMC scheme, while the dotted green line is
  produced by the MC scheme. Note that the fluctuations in the energy
  density that are particularly visible near the central region at
  $t=30$ ms and $t=60$ ms are simply caused by the random noise
  intrinsic to Monte Carlo methods. This noise is particularly
  pronounced in central zones because those zones have smaller volumes
  and thus contain smaller numbers of MCPs. One can reduce this noise
  by choosing smaller weights for MCPs in central regions. In our
  test, we use constant weight everywhere for the entire radiation
  packet. Moreover, these fluctuation decrease with increasing number
  of MCPs, according to the central limit theorem.
  \label{fig:diffusion_gaussian}} 
\end{figure}

In our runs, we place the packet at the center of our computational
domain with the radial coordinate extending up to $3\times 10^7$ cm
($300$ km). We choose the initial width of the packet to be $10^6$ cm,
while the scattering opacity is set equal to $2 \times 10^{-4}
\ \mathrm{cm}^{-1}$. We run each simulation with the DDMC and MC
schemes and use $2\times 10^5$ MCPs with constant weight in each of
the runs. Figure~\ref{fig:diffusion_gaussian} shows the radial profile
of $E$ at three different times ($0$, $30$, and $60$ ms). The black
line corresponds to the analytical solution, the dashed-red line is
obtained from the DDMC run, and the dotted green line is the MC
solution. Due to diffusion, the radial profile of radiation flattens
with time. During the first $30$ ms of evolution, the central
radiation energy density decreases be a factor of $\sim$20, while in
the next $30$ ms it decreases further by a factor of
$\sim$2. Therefore, the $60$ ms timescale captures the diffusion
timescale of the problem. As we can see in the plot, both the DDMC and
MC methods model diffusion of radiation in scattering medium quite
well. The fluctuations around the analytic solution are due to the
Monte Carlo treatment of transport in a scattering medium, and their
magnitude again decreases as $N^{1/2}$. We have also repeated these
runs with different values of the scattering opacity $\kappa_s$ and
different widths, $d_\mathrm{ini}$, of the Gaussian, and we always
find that both from DDMC and MC schemes agree with the analytical
solution within  statistical errors.

\subsection{Protoneutron star cooling}
\label{sec:pns_cooling}

In this section, we consider the early cooling and deleptonization of
a young non-rotating PNS formed in a CCSN.  This problem is provided 
as a test of neutrino-matter coupling and as a testbed for optimal 
sampling methods and MCP weights.  This particular problem was chosen
because it is a realistic physical context similar to that for which our MC 
and DDMC algorithms were designed. The reader should note, however, that 
there is no analytic or agreed-upon benchmark solution for this problem
and, therefore, that we are testing the behavior and speed of the 
solutions, not the numerical results themselves.

\subsubsection{Numerical Setup}

For our PNS model, we employ the post-core bounce configuration
produced in the collapse of the s20.0 progenitor model
of~\cite{whw:02} with the 2D multi-group multi-angle simulations
of~\cite{Ott:08}\footnote{The 2D data of~\cite{Ott:08} have been
  angle-averaged and then mapped to our grid.}. We start with a
background model at $160$ ms after bounce, and evolve it with our
Monte Carlo code. In doing so, we first evolve the radiation field
with fixed $T$ and $Y_\mathrm{e}$ until it reaches a steady state,
after which we inaugurate coupling to $T$ and $Y_\mathrm{e}$. Our
computational domain extends up to $300$ km, and is covered with 100
fixed logarithmically-spaced radial zones. The central zone has a
width of $500$ m. The timestep is chosen to be the light-crossing time
of the central zone. We assume that the PNS is static and neglect
velocity-dependent effects.   

We include the standard set of neutrino interactions listed
in~\cite{Thompson:03}. We use $48$ logarithmically spaced energy
groups from $2.5$ MeV to $250$ MeV to calculate the ``leakage''
opacities from one energy group to another due to inelastic
scatterings in the DDMC region (however, the energies of MCPs are, of
course, not discretized, since they are selected randomly using the
local emissivity. See discussion in
Section~\ref{sec:interior_ddmc}). Electron neutrinos and antineutrinos
are treated independently, while we combine heavy-lepton neutrinos
($\nu_\mu$, $\bar{\nu}_{\mu}$, $\nu_\tau$, and $\bar{\nu}_{\tau}$)
together into one group.

\emph{Sampling of Monte Carlo particles.} We sample Monte Carlo
particles in each spatial zone by calculating the number of neutrinos
emitted in each zone during each timestep and then by
choosing the weight of MCPs in each zone. We choose the weights of
MCPs based on the following function   
\be
\label{eq:mcp_weight}
\left(\frac{{{\dot T}_\mathrm{emis}}}{T} \frac{{{\dot
      Y}_\mathrm{e,emis}}} {Y_\mathrm{e}} \right)^{0.5}
\left(\frac{1}{\kappa_\mathrm{p}c}\right)^{h} \ ,     
\ee
where $h$ is a constant parameter. Terms ${\dot T}_\mathrm{emis}$ and
${\dot Y}_\mathrm{e,emis}$ are the rates of change of $T$ and
$Y_\mathrm{e}$ that would occur if there were only emission and no
absorption. Term $\kappa_\mathrm{p}$ is the mean absorption opacity;
hence, $1/\kappa_\mathrm{p}c$ represents a timescale for absorption of
radiation. Therefore, the quantity in expression~(\ref{eq:mcp_weight})
is larger in regions that emit strongly with small absorption. Stated
differently, this quantity should have larger values in those regions where
$T$ and $Y_\mathrm{e}$ are likely to undergo significant changes due to
emission and subsequent escape of radiation, implying that sampling
based on quantity~(\ref{eq:mcp_weight}) places particular emphasize on
accurate modeling of PNS cooling and deleptonization. We experimented
with several values of $h$ and found that $h=0.5$ for electron
neutrinos and antineutrinos and $h=0.3$ for $\mu/\tau$ neutrinos lead to
fairly smooth sampling of MCPs, where the largest number of MCPs are
concentrated in zones that are subject to the fastest changes in $T$ and
$Y_e$. However, we did not perform further studies of sampling and we
do not claim that the spatial sampling of MCPs based on
expression~(\ref{eq:mcp_weight}) is the optimal choice for
modeling PNS evolution.   

\emph{The Interface between DDMC and pure Monte Carlo regions.} Since
neutrino absorption and scattering cross sections are strongly
energy-dependent, it is important to take this into account in
choosing the spatial location of the interface between the DDMC and
pure MC regions. To accomplish this, we introduce energy groups and
calculate optical depth for each energy group. We then determine the
spatial location of the interface for an MCP with an energy within a
given group in terms of the optical depth for that group. More
specifically, if the optical depth for the energy group of a given MCP
exceeds some threshold value, $\tau_\mathrm{DDMC}$, then we assume that
this MCP is in the DDMC region. The parameter $\tau_\mathrm{DDMC}$ is
assumed to have the same value for each energy group, meaning that the
interface is located at different radii for different energy
groups. In our simulations, we calculate the optical depth at each
timestep and adjust the interface to its new location corresponding to
the new values of optical depth. In our tests below, we consider
several values of $\tau_\mathrm{DDMC}$ and explore which value
represents an optimal choice for simulations of PNS evolution.

\emph{Treatment of effective scattering.} As mentioned in
Section~\ref{sec:ddmc}, in the DDMC regime, we split the effective
scattering opacity $\tilde\sigma_{es,j,k}$ into two parts, $a_{j,k}
\tilde\sigma_{es,j,k}$ and $(1-a_{j,k})\tilde\sigma_{es,j,k}$, where 
$0 \le a_{j,k} \le 1$, and restrict the first of these two to be 
elastic effective scattering, while the second one is free to be
inelastic (as ``effective-scattering" scattering would otherwise
generally be). As we have discussed in Section~\ref{sec:ddmc}, the
presence of an extra elastic scattering source does not increase the cost
of doing DDMC transport. Therefore, by assuming that an $a_{j,k}
\tilde\sigma_{es,j,k}$ fraction of effective scattering is elastic
(instead of being inelastic), our computational savings are
proportional to $a_{j,k}$.  

We find that it is convenient to determine the parameter $a_{j,k}$ in
terms of the Fleck factor $f_j$ in the following way:
\be
\label{eq:eff_sca_prescription}
a_{j,k} = f_j^\frac{\delta}{1-\delta} ,
\ee
where $\delta$ is a constant ranging from $0$ to $1$. For 
$\delta=0$, all of effective scattering is treated as inelastic, while
for $\delta=1$, all of it is treated as elastic. This prescription has
the advantage that at low optical depth -- where effective scattering
does not dominate calculations -- most effective scattering is treated
as inelastic, while at high optical depth -- where calculations would
otherwise be dominated by inelastic effective scatterings -- a
significantly larger fraction is treated as elastic, which leads to
huge savings in computation. In the following, we explore what values
of $\delta$ are most suitable for simulations of PNS evolution.  

\subsubsection{Results}

In this section, we first present the stationary-state radiation field
results, after which we describe the subsequent fully time-dependent
calculations with coupling to $T$ and $Y_\mathrm{e}$. For these
simulations, unless otherwise noted, we use $\tau_\mathrm{DDMC} = 6$,
$\delta=0.38$ and $\alpha = 1$, and employing $100,000$ MCPs to model
newly-emitted particles at each timestep. 

\begin{figure}
\centering
\includegraphics[width=8cm]{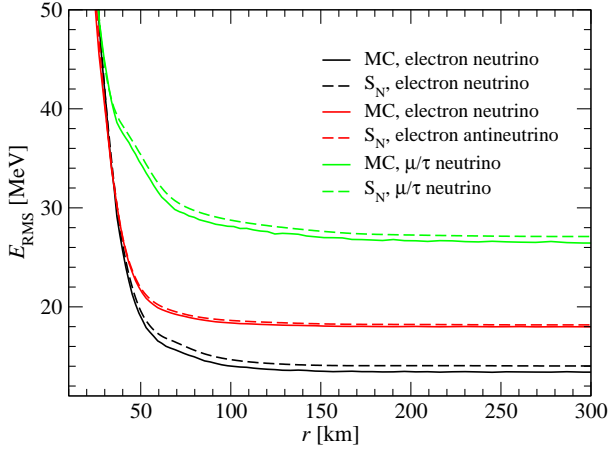}
\caption{The stationary state radial profiles of the RMS neutrino
  energies for the three different types of neutrinos obtained
    with our
  Monte Carlo scheme (solid lines). For comparison, we
  show results obtained by ~\cite{Ott:08} with an $\mathrm{S_N}$ code
  (dashed lines). 
\label{fig:enurms_vs_r}} 
\end{figure}

\begin{figure}
\centering
\includegraphics[width=8cm]{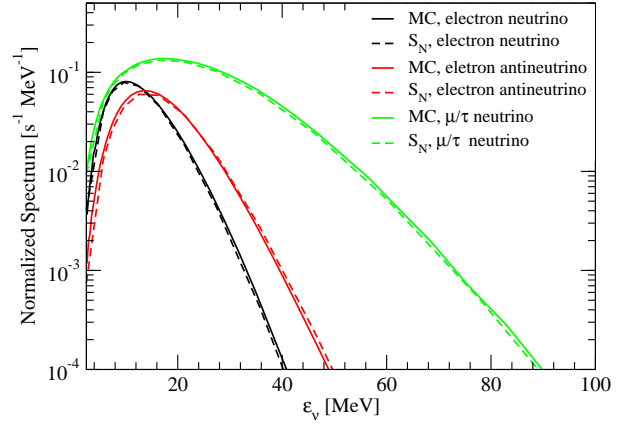}
\caption{The stationary state neutrino spectra at 300 km normalized by
  the total luminosity obtained with our Monte Carlo scheme (solid
  lines). For comparison, we show results obtained by ~\cite{Ott:08}
  with an $\mathrm{S_N}$ code (dashed lines).
\label{fig:spectrum}}
\end{figure}

\emph{Stationary state.} Figure~\ref{fig:enurms_vs_r} shows the radial
profiles of the RMS neutrino energies\footnote{See equation (13)
  of~\cite{Ott:08} for the exact definition of the RMS neutrino energy
  that we employ in our study.} for the three types of neutrinos. The solid
lines are produced by our Monte Carlo code, while the dashed ones are
from the $\mathrm{S_N}$ calculations of~\cite{Ott:08}. The RMS
energies agree well in the inner $ \sim 50$ km region, while for $r
\gtrsim 50$ km, the $\mathrm{S_N}$ code produces RMS energies that are
larger by up to $\simeq 4 \%$. We believe that this difference stems
from the truncation errors in the energy discretization
of~\cite{Ott:08} calculations, which employ only $16$ logarithmic 
energy groups, whereas in our MC cases, we do not use energy
discretization in selecting MCP energies. The corresponding spectral
energy distribution of the neutrino luminosity at $300$ km normalized
by the total luminosity for each neutrino type is shown in
Fig.~\ref{fig:spectrum}, which again demonstrates a good agreement
between our MC calculation and $\mathrm{S_N}$ calculations
of~\cite{Ott:08}\footnote{The reason for showing normalized neutrino
  spectra is the following. We find that while the total luminosities
  of the heavy lepton neutrinos in our MC and $\mathrm{S_N}$
  calculations agree very well, the luminosities of electron neutrinos
  and antineutrinos agree to only $\sim 10-15 \%$. The reason for this
  is that the angular averaging of the 2D background data
  of~\cite{Ott:08} leads to differences in $\rho, \ T,$ and $Y_e$. The
  luminosities of electron neutrinos and antineutrinos are very
  sensitive to the values of $Y_\mathrm{e}$, and differences in the
  latter lead to quantitative differences in the predicted
  luminosities and, thus, to a shift in the values of the absolute
  spectral energy distribution.}.   

\begin{figure}
\centering
\includegraphics[width=8cm]{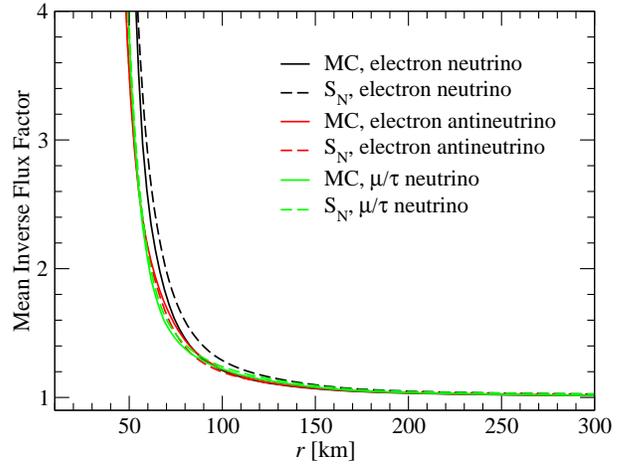}
\caption{The stationary state radial profiles of the mean inverse flux
  factors for the three different types of neutrinos obtained with
    our
  Monte Carlo scheme (solid lines). For comparison, we
  show results obtained by ~\cite{Ott:08} with an $\mathrm{S_N}$ code
  (dashed lines). 
\label{fig:inv_flux}} 
\end{figure}

Figure~\ref{fig:inv_flux} shows the radial profile of the mean inverse
flux factors\footnote{See equation (12) of~\cite{Ott:08} for the exact
  definition of the mean inverse flux factor that we use in our
  analysis.} for the three types of neutrinos. The solid lines again
represent the Monte Carlo results, while the dashed lines are from the
$\mathrm{S_N}$ calculations of~\cite{Ott:08}. Overall, we again find 
good agreement between the two results. The only systematic difference
that we observe is that the $\mathrm{S_N}$ code tends to yield 
inverse flux factors that are larger by $\sim$1 $\%$ than the
corresponding MC data at radii $ \gtrsim 200$ km. This is most likely
caused by the ray-effects intrinsic to $\mathrm{S_N}$
schemes. Nevertheless, we find good overall agreement between our
Monte Carlo and the $\mathrm{S_N}$ calculations of~\cite{Ott:08} for this
stationary-state radiation field.  

\begin{figure}
\centering
\includegraphics[width=8cm]{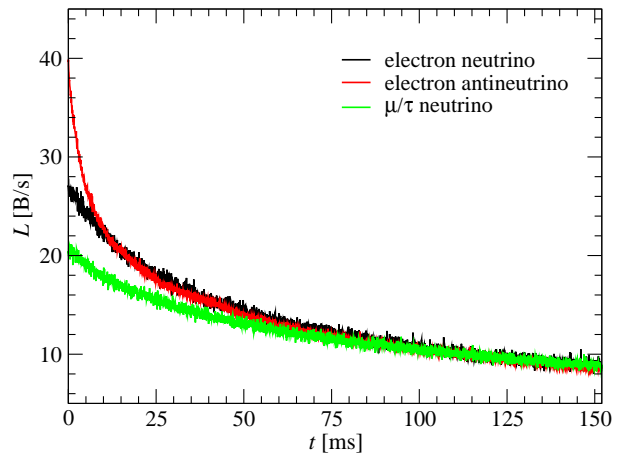}
\caption{Luminosity as a function of time in our PNS evolution
  calculation. 
\label{fig:lum_vs_t}} 
\end{figure}

\emph{Time-dependent calculations.}
Figure~\ref{fig:lum_vs_t} shows the luminosities for the three different
neutrino species as a function of time during the first $150$ ms of
evolution of the PNS model after being mapped to our code.
As expected, the luminosities for all types of neutrinos decrease
gradually during the first $150$ ms of evolution. On a short timescale
of one timestep, the luminosity undergoes significant
fluctuations around its average value. These are expected and are due
to the stochastic nature of MCP emission, absorption, and scattering
processes. These fluctuations decrease significantly if averaged over
longer timescales, such as the PNS light-crossing time, and are practically 
``invisible" on much longer timescales, such as the dynamical timescale of 
the PNS. Moreover, we find that these fluctuations again decrease as
$\sim N^{-1/2}$, where $N$ is the number of MCPs. 

\begin{figure}
\centering
\includegraphics[width=8cm]{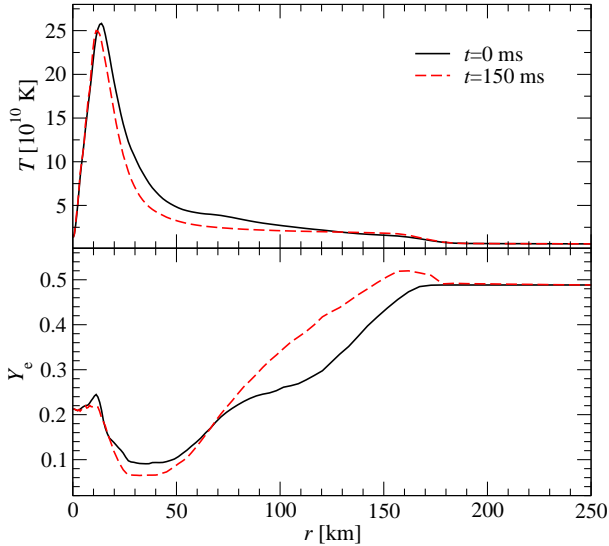}
\caption{The radial profiles of the temperature $T$ and electron
  fraction $Y_e$ at $t=0$ and $t=150$ ms after the start of
  simulation.
\label{fig:tye_vs_t}} 
\end{figure}

Figure~\ref{fig:tye_vs_t} shows the radial profile of the temperature
(upper panel) and electron fraction (bottom panel) at the beginning
($t=0$) and at the end ($t=150$ ms) of our simulation. 
During the early evolution, the temperature decreases
noticeably in the region $r \gtrsim 15$ km due to emission and
diffusion of radiation. There is no significant change in $T$ and
$Y_\mathrm{e}$ in the inner $r \lesssim 15$ km region as the radiation
diffuses out on much longer timescales from that region. The electron
fraction decreases in the region $r \lesssim 60 $ km due to copious 
emission of electron neutrinos, while in the region $r \gtrsim 60 $ km
$Y_\mathrm{e}$ increases as a result of the absorption of
those neutrinos.     

\emph{Quality of energy and lepton number conservation.} As discussed
in Section~\ref{sec:IMC_neutrino_e_l_cons}, our IMC scheme for
neutrinos does not make any approximations that violate energy or
lepton number conservation. Therefore, energy and lepton number should
be conserved up to machine precision in practical calculations using
this scheme. In our calculations, we see that this is indeed the
case, and both energy and lepton number in the system are conserved 
in each timestep to the ${\cal O}(10^{-14})$ precision of the machine.  

\emph{Treatment of effective scattering.} As we mentioned above, we
use prescription~(\ref{eq:eff_sca_prescription}) in order to treat
effective scattering in a computationally efficient way in the DDMC
region. We have performed simulations for different values of
the parameter $\delta$, ranging from $1$ (where all effective
scatterings are elastic) to $0.286$ (where a significant fraction of
effective scatterings are inelastic). We find that there is no
systematic difference in the PNS evolution for values of $\delta$ that
are smaller than $\simeq 0.38$. For example, the total energy
$E_\mathrm{esc}$ (or lepton number $Y_\mathrm{l,esc}$) of neutrinos
that leave the system through the outer boundary in the first
$150$ ms of evolution decreases monotonically by $\sim 2\%$ when
decreasing $\delta$ from $1$ to $0.38$. Further decrease of $\delta$
results in non-systematic variation in both $E_\mathrm{esc}$ and
$Y_\mathrm{l,esc}$ with relative error of only $\sim 0.1\%$, without
any monotonic dependence on $\delta$. The origin of such variations is
likely to be numerical errors and not our prescription for the treatment
of effective scattering\footnote{Indeed, we find that these
  variations tend to increase when we increase the parameter
  $\varsigma$ that controls the remaining fraction of an MCP when it is
  assumed to be absorbed (cf. Section~\ref{sec:con_absorption}) -- the
  larger $\varsigma$ is, the larger are the numerical errors in
  treating MCP absorption.}. This implies that condition
(\ref{eq:ddmc_eff_sca_con}) for the validity of our approximate
treatment of effective scattering is fulfilled for $\delta \lesssim
0.38$ and violated otherwise. On the other hand, because the  cost of
simulations increases with decreasing $\delta$, it is desirable to use
the largest allowed value of $\delta$. Therefore, we conclude that
$\delta=0.38$ is an optimal choice for the treatment of effective
scattering in the DDMC regime, at least in modeling the early phases
of PNS evolution. 

Note that this feature is independent of the number of dimensions. It
should be applicable as long as we have a young PNS surrounded by a
lower density envelope, i.e. in any CCSN scenario.

\begin{figure}
\centering
\includegraphics[width=8cm]{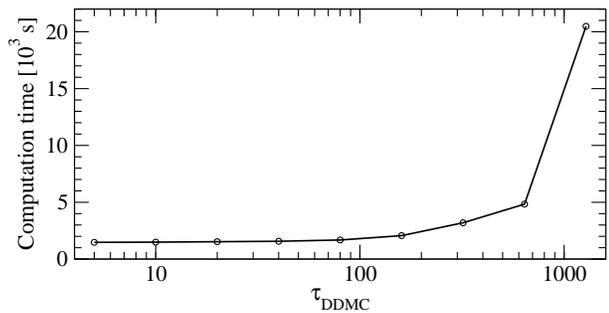}
\caption{Time spent for performing first 500 timesteps as a function
  of parameter $\tau_\mathrm{DDMC}$ for PNS evolution
  simulations. Since at high optical depth the pure Monte Carlo
  calculations are more expensive than the DDMC
  calculations, the computational
  time increases with $\tau_\mathrm{DDMC}$. Beyond
  $\tau_\mathrm{DDMC}>1280$, our simulations become computationally
  too expensive, and we do not consider even
  larger values of $\tau_\mathrm{DDMC}$. 
\label{fig:comp_time_vs_tau_ddmc}} 
\end{figure}

\emph{The interface between DDMC and pure Monte Carlo regions.}
Figure~\ref{fig:comp_time_vs_tau_ddmc} shows the computational time to
perform the first $500$ timesteps as a function of the parameter
$\tau_\mathrm{DDMC}$. We have performed simulations using different
values of $\tau_\mathrm{DDMC}$, ranging from $5$ up to $1280$. Since
Monte Carlo calculations are more expensive than DDMC calculations, the
computational time increases with $\tau_\mathrm{DDMC}$. For example, for
$\tau_\mathrm{DDMC}=5$ the first $500$ timesteps are performed within
$1469$ s, while for $\tau_\mathrm{DDMC}=1280$ this simulation took
$20,480$ s. We have not performed simulations with higher values of
$\tau_\mathrm{DDMC}$, since such simulations become quite expensive,
implying that it is impractical to perform neutrino transport in the
PNS using only the IMC scheme without combining it with the DDMC (or
an alternate) scheme. Interestingly, for this setup, the
computational cost of simulations does not differ much for values of
$\tau_\mathrm{DDMC}$ smaller than $\simeq 100$. This is because the
DDMC scheme yields the largest speed-up in regimes dominated by
scattering. At optical depth of $ \lesssim 100$ (where the Fleck
factor $f$ is $\sim 1$), there
is significant contribution from absorption, which leads to modest
increase of the computational time as we increase $\tau_\mathrm{DDMC}$
from $5$ to $100$. At higher optical depths (where $f\sim
  0$), effective scatterings start to dominate, which leads to steep
increases of computational time with $\tau_\mathrm{DDMC}$.

From a theoretical point of view, the DDMC scheme is accurate if the
underlying approximations -- which are based on Fick's law -- are
fulfilled to a sufficient degree in each zone. Fick's law holds in
a given zone if the mean-free-path of MCPs are significantly smaller
than the grid size. If we assume that the grid size is of the order of
$1 $ km, then for a typical young PNS model, this requirement
translates to $\tau_\mathrm{DDMC} \gtrsim 6$ (for a given energy
group). Therefore, for $\tau_\mathrm{DDMC} \gtrsim 6$ we expect our
DDMC scheme to yield sufficiently accurate results. Indeed, to verify
this premise we have performed a series of long-evolution simulations
for values of $\tau_\mathrm{DDMC}$ in a wide range. We find that the
results (such as luminosity, etc.) indeed agree to within $\sim 1 \%$ for
any value of $\tau_\mathrm{DDMC}$ larger than $\sim 6$. 

\emph{The importance of the implicit scheme.} The simulations
presented in this section are performed using the implicit Monte Carlo
scheme with a ``most implicit'' choice of $\alpha=1$ within the
framework of the IMC (cf. equation~\ref{eq:bbar}). We repeated some of
the runs with the ``less implicit'' value of $\alpha=0.5$, which
results in a Fleck factor that is larger by a factor of $\sim 2$ than
in the $\alpha=1$ case in the diffusive regime
(cf. equation~\ref{eq:ff_nu}). This results in a slight decrease in
solution accuracy, but we do not observe any instabilities. One
obvious question to ask is: would it be possible to perform the same
simulations with an explicit treatment of the emissivity? The answer
is clearly no. With the current timestep set by the light crossing
time for the central zone of width $500$ m, the fully explicit scheme
crashes within a few timesteps. For it to be stable, one would have to
use timesteps of the order of the smallest mean-free-path, which can
be as small as $\lesssim 1$ m in the center of the PNS. Hence, in
order to use a time-explicit scheme, we would need to decrease our
timestep by a factor of at least $\sim 500$, which would make useful
simulations completely impractical. For the above choice of parameters
(i.e., $\tau_\mathrm{DDMC} = 6$, $\delta=0.38$ and $\alpha = 1$, and
employ $100,000$ MCPs to model newly emitted particles at each
timestep), the computational cost of performing the first $150$ ms of
PNS evolution is about $36$ hours on $96$ cores on the Hopper Cray
XTE6 cluster at NERSC.  

\emph{Parallel scaling.} In order to study parallel scaling of
our code, we perform simulations on a number of cores ranging
from $24$ up to $1152$ on the Hopper Cray XTE6 cluster at NERSC. We
implement hybrid OpenMP/MPI parallelization and use 6 OpenMP threads
per each MPI process in our tests.    

\begin{figure}
\centering
\includegraphics[width=8cm]{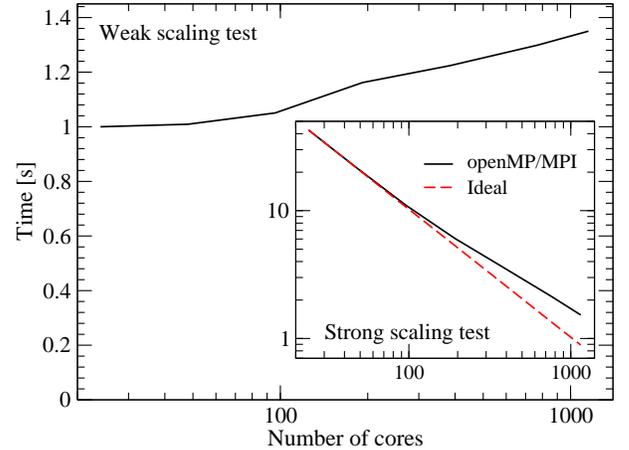}
\caption{The simulation time as a function of the number of
    cores in our strong (main plot) and weak (inset plot) scaling
    tests. In both plots, the $x$-axes show the number of cores, while
    the $y$-axes are the simulations times [in seconds]. The solid
    black
    lines show the simulation time using our code, while the dashed
    red line in the inset plot corresponds to ideal strong scaling.
\label{fig:scaling}} 
\end{figure}

Figure~\ref{fig:scaling} shows the time spent per MPI process
$t_\mathrm{mpi}$ as a function of the number of cores
$N_\mathrm{core}$ in our weak scaling test\footnote{In the strong
scaling test, the problem size remains the same as we increase the  
  number of CPUs so that the amount of work per CPU decreases
  proportionately with the number of cores. Alternatively, in the weak
  scaling test, the problem size on each core remains the
  same, while the problem size increases proportionately with the
  number of cores used.}.   
In this test, we increase the number of MCPs proportionally to the
number of MPI processes, while other problem parameters remain fixed.
Here, the amount of computational work is chosen in such a way that
the time on $24$ cores is equal to $1$ s. This time increases slowly
with $N_\mathrm{core}$ and reaches $1.35$ s for
$N_\mathrm{core}=1152$, a value that is only $35\%$ larger than the
one for $24$ cores. 

The inset plot of Fig.~\ref{fig:scaling} shows the total simulation
time as a function of the number of cores in our strong scaling
test. The simulation time decreases almost linearly (in logarithmic
scale) with $N_\mathrm{core}$ and reaches $1.53$ s at
$N_\mathrm{core}=1152$. This number is only $70 \%$ larger than the
time corresponding to ideal scaling. Based on these results, we
conclude that our code scales nicely up to $1152$ cores, which is more
than sufficient for 1D problems.  

\subsection{Moving background with fixed $T$ and $Y_e$}
\label{sec:tests3}

In this section we describe a test problem that involves the
propagation of radiation in moving matter to demonstrate the
  ability of our code to perform transport in a moving medium. In
this test, we neglect
the change in temperature ($T$) and electron fraction ($Y_e$) of
matter due to emission, absorption, or scattering of radiation. 

\subsubsection{Homologously expanding shell}
\label{sec:hom_exp_shell}

\begin{figure}
\centering
\includegraphics[width=8cm]{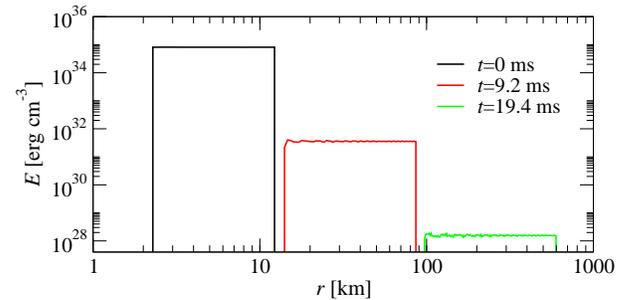}
\caption{The radial profile of the radiation energy density measured
  by a comoving observer at different times in the homologously
  expanding shell test for simulation A1D using $1,280,000$ MC
  particles. See the discussion in the main text for further details
  of the simulation.   
\label{fig:e_profile_ddmc}} 
\end{figure}

\begin{figure}
\centering
\includegraphics[width=8cm]{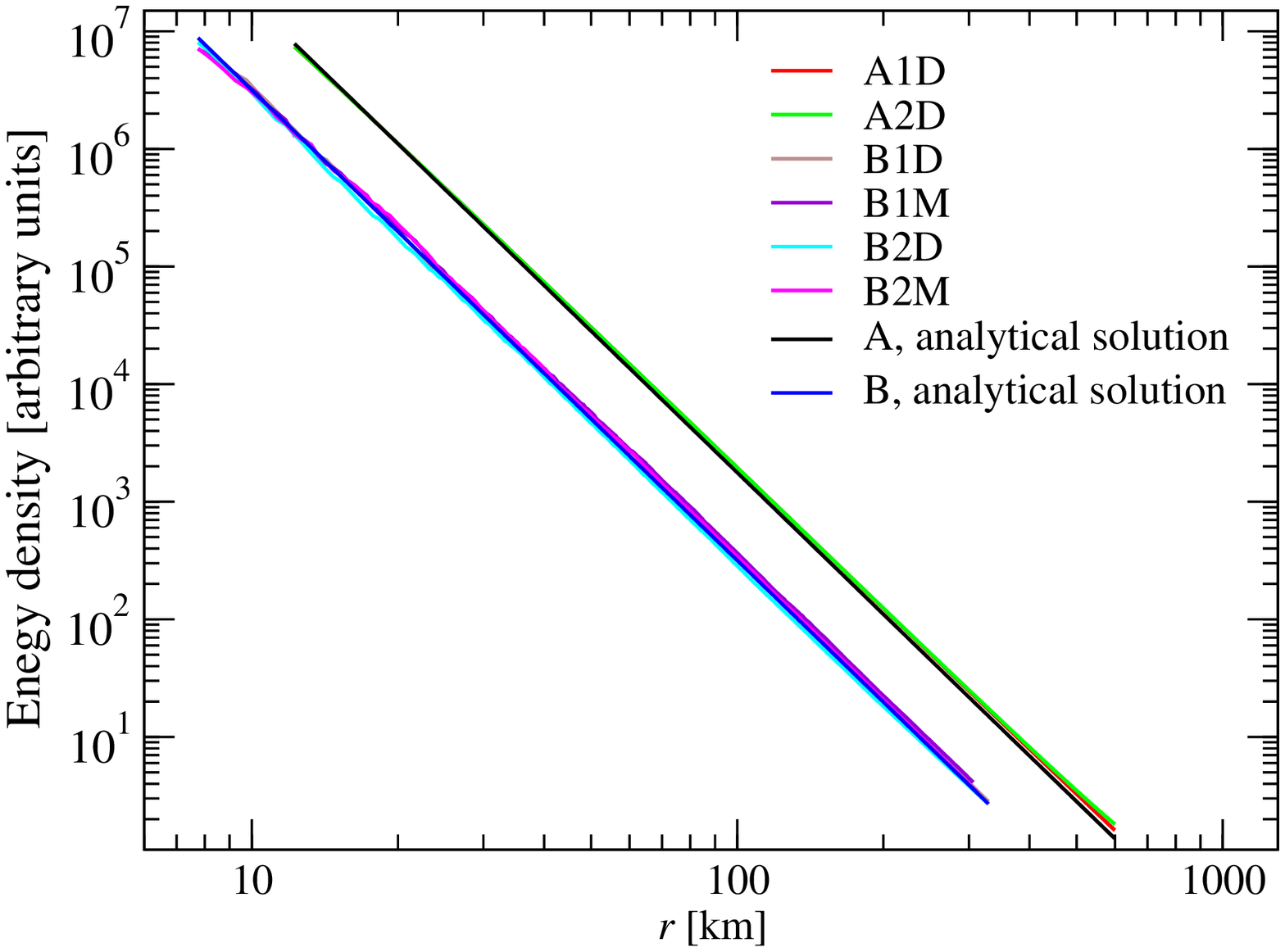}
\caption{The radiation energy density in the comoving frame as a
  function of the radial coordinate inside homologously expanding
  shell
  \label{fig:e_vs_r_mc_ddmc}} 
\end{figure}

A spherically-symmetric shell of matter, in which 
radiation is trapped, is expanding self-similarly with velocity $V_r
\propto r$. For simplicity, we assume that matter interacts with
radiation only through elastic scattering, without any emission or
absorption. According to the analytical solution~\citep{Mihalas:84}, the average
energy of neutrinos $\langle \ve_\nu \rangle$ in the expanding shell
should decrease as 
\be
\langle \ve_\nu \rangle \propto \frac{1}{r} 
\ee 
due to the Doppler shift, while the radiation energy density $E_0$
should decrease according to  
\be
\label{eq:er4scaling}
E_{0} \propto r^{-4} \, , 
\ee
due to both the Doppler shift and expansion of the shell.  

We perform two sets of test simulations. In the first set, we use a
very large constant value of the scattering opacity $\kappa_s$ of $5\cdot
10^3 \ \mathrm{cm}^{-1}$, while in the second set, we use a million
times smaller 
value of $5\cdot 10^{-3}\ \mathrm{cm}^{-1}$. The reason for choosing
two such sets is the following. Generally speaking, for such tests, it
is desirable to have as large an opacity as possible in order to avoid
diffusion of radiation out of the expanding shell, which can lead to the
violation of relation~(\ref{eq:er4scaling}). However, the high
computational cost of performing MC transport at high scattering
opacity limits the maximum value of $\kappa_s$ that we can
simulate. On the other hand, the DDMC scheme is particularly efficient
in the regime dominated by elastic scattering (see the discussion
in Section~\ref{sec:ddmc}). Therefore, we perform simulations
with $\kappa_s=5 \times 10^3 \ \mathrm{cm}^{-1}$ using only the DDMC
scheme, after which we do a set of runs with $\kappa_s=5\times
10^{-3} \ \mathrm{cm}^{-1}$ using both the DDMC and MC schemes. In the
former case, the radiation is completely trapped in the matter, and there
is no diffusion of MCPs from one cell to another during the simulated
time. Instead, the MCPs are transported due only to velocity-dependent
effects caused by the expansion of the shell. Moreover, in this case, the
random numbers are used only for initial sampling of MCPs, while the
velocity-dependent (operator-split) part of the transport is
completely deterministic (see the discussion in
Section~\ref{sec:DDMCvelocity}). Therefore, in this set of runs, we
exclusively test the ability of our DDMC scheme to treat
velocity-dependent effects. In the second set of test runs, since we
use a smaller value of the scattering opacity, the result is affected
also by diffusion of MCPs from the expanding shell. This is prone to
statistical errors arising from the DDMC treatment of
diffusion. Hence, in this case, we explore the ability of the code to 
perform transport of radiation in moving matter when the transport
process itself is affected by statistical errors.  

In both sets of test runs, our shell initially has a radial extent
ranging from $r_\mathrm{min}=2.25$ km to $r_\mathrm{max}=12.25$ km. We
assume that the radiation initially has a constant energy density
profile. 
The shell velocity is given by $v_r=\varpi c r /
R_\mathrm{out}$, where $\varpi$ is a constant, $c$ is the speed of
light, $r$ is the radial coordinate, and $R_\mathrm{out}$ is the
radius of the outer boundary of our computational domain. Our entire
domain is covered by $1200$ radial cells with constant size. In these
simulations, we choose $\varpi$ to be $0.4$ and $0.8$, while
$R_\mathrm{out}$ is chosen to be $600$ km. 
For the first set of test runs, we perform 4 simulations
with $40,000$, $64,000$ $256,000$, $1,280,000$ MC particles for each
of the two values of parameter $\varpi$. For the second set, we
perform 3 simulations with $10,000$, $40,000$, and $164,000$ MCPs
for each value of $\varpi$ using the MC and DDMC schemes.   
We use the following naming convention for our test simulations: the
names of the first set of runs start with ``A,'' while the second set
starts with ``B.'' The second symbol represents the value of $\varpi$
(``1'' for $\varpi=0.4$ and ``2'' for $\varpi=0.8$), while the last
letter indicates whether we use the DDMC or MC scheme (``D'' for DDMC and
``M'' for MC). Thus, for example, the DDMC run with $\varpi=0.4$ from
the first set is denoted as A1D, while the similar run with $\varpi=0.8$
is called A1D.    

Figure~\ref{fig:e_profile_ddmc} shows the radiation energy density in
the comoving frame in each radial cell as a function of the radial
coordinate of those cells at $0$ ms (black line), $9.2$ ms (red line),
and $19.4$ ms (green line) after the start of shell expansion for
model A1D. During this evolution, the radial extent of the shell
increases by a factor of $\sim \! 50$, which leads to a decrease in
radiation energy density by a factor of $\sim \! 5\times
10^6$.
Due to homologous nature of the expansion, the radial profile
of the energy density should stay constant. As we can see in the plot,
our code indeed reproduces its constant radial profile within statistical
errors. These statistical errors are caused by the initial
pseudo-random sampling of MCP positions and energies, and their magnitude
decreases as we increase the number of MCPs.   

Figure~\ref{fig:e_vs_r_mc_ddmc} shows the average energy density in
several outermost zones as a function of the mean radial coordinate of
those zones for simulations A1D and A2D using $1,280,000$ MCPs (red
and green lines, respectively).
The outermost zones are expanding with the fastest velocity compared to
the other parts of the shell and, thus, are affected by the
velocity-dependent effects to the greatest degree. 
This plot also shows energy as a function of $r$ according to the
analytical solution (black line). As we see, the result of these
simulations agrees well with the analytical result during the simulated
time.      

Figure~\ref{fig:e_vs_r_mc_ddmc} also shows the energy density in
several central zones of the expanding shell as a function of the mean
radial coordinate of those zones for the second set of test
runs. Here we focus on the central zones (instead of the
outermost zones, as we did for the models of the first set) because the
scattering opacity in these runs is relatively small. Thus, some
fraction of the trapped radiation diffuses out of the shell during
expansion. The behavior of radiation in the central zones is affected
by this diffusion to a much smaller degree compared to that in the
outer zones. This plot also shows the energy density versus the radial
coordinate from the analytical solution (red line). As we see, 
both the MC and DDMC schemes agree with the analytical result
within the statistical errors during the simulated time. These
statistical errors again decrease as we increase the number of MCPs. 

\section{Conclusion and Future Work}
\label{sec:conclusion}

We have generalized the implicit Monte Carlo (IMC) scheme
of~\cite{Fleck:71} and the discrete-diffusion Monte Carlo (DDMC) method
of~\cite{Densmore:07} to energy- and velocity-dependent neutrino
transport. While the IMC method provides a stable and accurate time
discretization of the transport equation, the DDMC method increases
the computational efficiency of IMC at high optical depths using the
diffusion approximation (as is appropriate).  
 
In the IMC method for photons, one uses the matter energy equation in
order to derive a time discretization of the transport equation in
which the emissivity term is treated semi-implicitly. Since neutrino
emissivity depends not only on energy, but also on the electron fraction,
in order to derive the IMC equations for neutrinos with lepton number
coupling, one has to use not only the energy equation, but also an
equation for the evolution of the electron fraction
$Y_\mathrm{e}$. 
Similar to the photon case, the IMC neutrino transport
equation has new terms that formally describe scattering of
radiation. In the Monte Carlo interpretation of the IMC equation,
these scatterings model a fraction of the absorption and subsequent
re-emission of radiation within a timestep and are called 
effective scatterings. Such scattering effectively decreases the
energy exchanged between matter and radiation within a timestep and
makes the scheme unconditionally stable. However, unlike in the photon case,
the IMC scheme for neutrinos has two types of effective scatterings:
one in which the total energy is conserved (as in photon IMC), 
and another for which the total lepton number is conserved
during effective scattering (this scattering type does not exist in
photon IMC; see the discussion in Section~\ref{sec:IMC_neutrino}). 

In extending the gray DDMC scheme of~\cite{Densmore:07} to the
energy-dependent case, we not only added discretization
in particle energy, but also proposed a new (approximate) way of
treating effective scattering in the DDMC regime to achieve additional
speed up. We split the total effective scattering opacity into two
parts, and the first part is treated as inelastic (as it would
otherwise be), while the other part is treated as elastic. This leads
to extra savings in computational cost because DDMC does not need to
perform any additional operations to model elastic scatterings
(Section~\ref{sec:ddmc}). We parametrize the fraction of the effective
scattering treated as elastic in terms of a new variable $\delta$ and
show that for $\delta \simeq 0.38$, this approximation is excellent
for modeling neutrino transport in systems that consist of a newly-born
PNS with a hot, tenuous surrounding medium.  

Furthermore, we have extended the above schemes to include velocity-dependence.
In the pure (i.e., non-DDMC) Monte Carlo regime, we use the mixed-frame
formalism~\citep{Mihalas:82,Hubeny:07} in which emissivities and
opacities are defined in the comoving frame, while transport is
performed in the lab frame. This is easy to implement and allows one
to take into account velocity-dependent effects correctly for arbitrarily
large (relativistic) fluid velocities, provided that the spatial
resolution and timestep are appropriate for capturing the motion of
the fluid in each cell. In the DDMC regime, we use a somewhat
different approach. Since the diffusion equation is formulated most
naturally in the comoving frame, we start with the equation for the
comoving zeroth moment of radiation accurate to ${\cal O} (V/c) $, and
split it into three equations that are then solved in operator-split
manner. One part deals with diffusion of radiation, the second part
models advection and compression/decompression of radiation, while the
third one takes into account the Doppler shift of radiation
energy. The first equation is identical to the DDMC equations in the
zero-velocity case and, thus, is solved in exactly the same way, while
the second and third parts can be incorporated using simple operations
(Section~\ref{sec:DDMCvelocity}).   

In order to test and validate these schemes, we have implemented them
in our 1D spherically-symmetric code and applied it to several test
problems. In particular, by considering an early evolution of a young
PNS, we show that the IMC method allows much larger timesteps than what
would be possible with a fully time-explicit scheme 
(Section~\ref{sec:pns_cooling}). Moreover, the DDMC scheme leads to
huge saving in computational cost due to the more efficient treatment of
transport at high optical depth. Therefore, in our scheme, we use
the DDMC
method in regions with sufficiently high optical depth, while in the
remaining part, we employ the pure IMC method. We find that the cost 
of performing a PNS evolution with the pure IMC scheme is usually
orders of magnitude higher than the cost with a combination of the IMC and
DDMC schemes (cf. Fig.~\ref{fig:comp_time_vs_tau_ddmc}). All in all,
we conclude that the combination of the IMC and DDMC approaches
represents a robust and accurate method for neutrino transport in
core-collapse supernovae.  

Finally, we point out that our energy- and velocity-dependent scheme
can also be used for photon transport with minimal modifications. The
only changes needed are disabling $Y_e$ changes and swapping in
opacities and emissivities corresponding to photons. 

Having developed a velocity-dependent transport algorithm, our next
task is to couple it to hydrodynamics. Furthermore, in the
future, we plan to generalize our code to the two- and
three-dimensional cases and to extend it to general relativity.     

\section{Acknowledgments}

We are happy to acknowledge helpful exchanges with Timothy Brandt,
Jeffery Densmore, Peter Diener, Roland Haas, Daniel Kasen, Oleg
Korobkin, and Christian Reisswig.  
This work was supported in part by NSF under grant nos. AST-0855535,
OCI-0905046, OCI 0721915, OCI 0725070, OCI 0905046, OCI 0941653,
PIF-0904015, PHY-0960291, and TG-PHY100033, by the DOE under grant
DE-FG02-08ER41544, and by the Sherman Fairchild Foundation. Results
presented in this article were obtained through computations on
machines of the Louisiana Optical Network Initiative under grant
loni\_numrel07, on the Caltech compute cluster ``Zwicky'' (NSF MRI
award No. PHY-0960291), on the NSF Teragrid under grant TG-PHY100033,
and at the National Energy Research Scientific Computing Center
(NERSC), which is supported by the Office of Science of the US
Department of Energy under contract DE-AC03-76SF00098.     

\begin{appendix}

\section{Calculation of $\partial U_{r} / \partial T $ and $\partial
  U_{r} / \partial Y_e $ using the Fermi-Dirac function} 
\label{sec:dur}

Here, we derive analytical expressions for the equilibrium energy
density of neutrinos $U_{r}$ and its partial derivatives with respect
to $T$ and $Y_e$ using the expression for the Fermi-Dirac
function $B (\ve, T, Y_e)$. The Fermi-Dirac function for neutrinos has
the following form:  
\begin{equation}
  \label{eq:planck_function}
B(\ve, T, Y_e) = \frac{\ve^3}{(h c)^3} \frac{1}{\exp \left[
  \frac{\ve}{kT} - \eta (T, Y_e) \right] + 1} \, ,
\end{equation}
where $\eta$ is the degeneracy parameter, which equals to
$(\mu_e - \mu_n + \mu_p)/ {kT}$, $-(\mu_e - \mu_n + \mu_p)/ {kT}$, and
$0$ for electron neutrinos, electron anti-neutrinos, and heavy lepton
neutrinos, respectively, Here, $\mu_e$, $\mu_n$ and $\mu_p$ are the
chemical potentials of the electron, neutron, and proton. 
Using the
definition of the equilibrium energy density of neutrinos $U_{r}$
given by equation~(\ref{eq:u_r}), we obtain:   
\begin{eqnarray}
  \label{eq:dur}
  \frac{\partial U_{r}}{\partial T} &=& \frac{4\pi}{c} \int_0^\infty 
  \frac{\partial B}{\partial T} \, d \ve\, , \\
  \frac{\partial U_{r}}{\partial Y_e} &=& \frac{4\pi}{c} \int_0^\infty 
  \frac{\partial B}{\partial Y_e} \, d \ve \, , \\\nonumber
\end{eqnarray}
where functions $ \frac{\partial B}{\partial T}$ and $\frac{\partial
  B}{\partial Y_e}$, can be obtained using
  formula~(\ref{eq:planck_function}):   
\begin{eqnarray}
\frac{\partial B}{\partial T} &=& \frac{B \exp \left(
  \frac{\ve}{kT} - \eta \right)}{ \exp \left(
  \frac{\ve}{kT} - \eta \right) + 1} \left[
\frac{\ve}{kT^2} + \left( \frac{\partial \eta}{\partial T}
  \right)_{Y_e,\rho} \right] \, , \\
\frac{\partial B}{\partial Y_e} &=&
\frac{B \exp \left(  \frac{\ve}{kT} - \eta \right)}{
  \exp \left( \frac{\ve}{kT} - \eta \right) + 1}
  \left( \frac{\partial \eta}{\partial Y_e} \right)_{T,\rho}\, . \\\nonumber
\end{eqnarray}
Functions $d\eta/dT$ and $d\eta/dY_e$ are expressed in terms of the
derivatives of $\mu_e$, $\mu_n$, and $\mu_p$ that can be calculated
using EOS tables. Finally, the integrals in equation (\ref{eq:dur})
can be  evaluated numerically, or can be expressed in terms of the
Fermi integrals. The latter can be calculated either via direct
numerical integration or by using a series
expansion~\citep{Takahashi:78}.

\end{appendix}


\begin{thebibliography}{85}
\expandafter\ifx\csname natexlab\endcsname\relax\def\natexlab#1{#1}\fi

\bibitem[{Adams \& Larsen(2002)}]{Adams:02}
Adams, M.~L., \& Larsen, E.~W. 2002, Progress in Nuclear Energy, 40, 3

\bibitem[{{Bethe}(1990)}]{Bethe:90}
{Bethe}, H.~A. 1990, Reviews of Modern Physics, 62, 801

\bibitem[{{Bruenn}(1985)}]{Bruenn:85}
{Bruenn}, S.~W. 1985, \apjs, 58, 771

\bibitem[{{Bruenn} {et~al.}(2006){Bruenn}, {Dirk}, {Mezzacappa}, {Hayes},
  {Blondin}, {Hix}, \& {Messer}}]{Bruenn:06}
{Bruenn}, S.~W., {Dirk}, C.~J., {Mezzacappa}, A., {Hayes}, J.~C., {Blondin},
  J.~M., {Hix}, W.~R., \& {Messer}, O.~E.~B. 2006, Journal of Physics
  Conference Series, 46, 393

\bibitem[{{Bruenn} {et~al.}(2009){Bruenn}, {Mezzacappa}, {Hix}, {Blondin},
  {Marronetti}, {Messer}, {Dirk}, \& {Yoshida}}]{Bruenn:09}
{Bruenn}, S.~W., {Mezzacappa}, A., {Hix}, W.~R., {Blondin}, J.~M.,
  {Marronetti}, P., {Messer}, O.~E.~B., {Dirk}, C.~J., \& {Yoshida}, S. 2009,
  Journal of Physics Conference Series, 180, 012018

\bibitem[{Brunner \& Brantley(2009)}]{Brunner:09}
Brunner, T.~A., \& Brantley, P.~S. 2009, J. Comput. Phys., 228, 3882

\bibitem[{Brunner {et~al.}(2006)Brunner, Urbatsch, Evans, \&
  Gentile}]{Brunner:06}
Brunner, T.~A., Urbatsch, T.~J., Evans, T.~M., \& Gentile, N.~A. 2006, J.
  Comput. Phys., 212, 527

\bibitem[{{Buras} {et~al.}(2006{\natexlab{a}}){Buras}, {Janka}, {Rampp}, \&
  {Kifonidis}}]{Buras:06b}
{Buras}, R., {Janka}, H.-T., {Rampp}, M., \& {Kifonidis}, K.
  2006{\natexlab{a}}, \aap, 457, 281

\bibitem[{{Buras} {et~al.}(2006{\natexlab{b}}){Buras}, {Rampp}, {Janka}, \&
  {Kifonidis}}]{Buras:06a}
{Buras}, R., {Rampp}, M., {Janka}, H.-T., \& {Kifonidis}, K.
  2006{\natexlab{b}}, \aap, 447, 1049

\bibitem[{{Burrows} {et~al.}(2007{\natexlab{a}}){Burrows}, {Dessart}, \&
  {Livne}}]{Burrows:07}
{Burrows}, A., {Dessart}, L., \& {Livne}, E. 2007{\natexlab{a}}, in American
  Institute of Physics Conference Series, Vol. 937, Supernova 1987A: 20 Years
  After: Supernovae and Gamma-Ray Bursters, ed. {S.~Immler, K.~Weiler, \&
  R.~McCray}, 370

\bibitem[{{Burrows} \& {Fryxell}(1992)}]{Burrows:92}
{Burrows}, A., \& {Fryxell}, B.~A. 1992, Science, 258, 430

\bibitem[{{Burrows} \& {Goshy}(1993)}]{Burrows:93}
{Burrows}, A., \& {Goshy}, J. 1993, \apjl, 416, L75

\bibitem[{{Burrows} {et~al.}(1995){Burrows}, {Hayes}, \&
  {Fryxell}}]{Burrows:95}
{Burrows}, A., {Hayes}, J., \& {Fryxell}, B.~A. 1995, \apj, 450, 830

\bibitem[{{Burrows} {et~al.}(2006){Burrows}, {Livne}, {Dessart}, {Ott}, \&
  {Murphy}}]{Burrows:06}
{Burrows}, A., {Livne}, E., {Dessart}, L., {Ott}, C.~D., \& {Murphy}, J. 2006,
  \apj, 640, 878

\bibitem[{{Burrows} {et~al.}(2007{\natexlab{b}}){Burrows}, {Livne}, {Dessart},
  {Ott}, \& {Murphy}}]{Burrows:07a}
---. 2007{\natexlab{b}}, \apj, 655, 416

\bibitem[{{Burrows} {et~al.}(2000){Burrows}, {Young}, {Pinto}, {Eastman}, \&
  {Thompson}}]{Burrows:00}
{Burrows}, A., {Young}, T., {Pinto}, P., {Eastman}, R., \& {Thompson}, T.~A.
  2000, \apj, 539, 865

\bibitem[{{Castor}(2004)}]{Castor:04}
{Castor}, J.~I. 2004, {Radiation Hydrodynamics}, ed. {Castor, J.~I.} (Cambridge
  University Press, Cambridge, UK)

\bibitem[{{Demorest} {et~al.}(2010){Demorest}, {Pennucci}, {Ransom}, {Roberts},
  \& {Hessels}}]{Demorest:10}
{Demorest}, P.~B., {Pennucci}, T., {Ransom}, S.~M., {Roberts}, M.~S.~E., \&
  {Hessels}, J.~W.~T. 2010, \nat, 467, 1081

\bibitem[{Densmore \& Larsen(2004)}]{Densmore:04}
Densmore, J.~D., \& Larsen, E.~W. 2004, J. Comput. Phys., 199, 175

\bibitem[{Densmore {et~al.}(2007)Densmore, Urbatsch, Evans, \&
  Buksas}]{Densmore:07}
Densmore, J.~D., Urbatsch, T.~J., Evans, T.~M., \& Buksas, M.~W. 2007, J.
  Comput. Phys., 222, 485

\bibitem[{Fleck \& Canfield(1984)}]{Fleck:84}
Fleck, Jr., J.~A., \& Canfield, E.~H. 1984, J. Comput. Phys., 54, 508

\bibitem[{Fleck \& Cummings(1971)}]{Fleck:71}
Fleck, Jr., J.~A., \& Cummings, Jr., J.~D. 1971, J. Comput. Phys., 8, 313

\bibitem[{Gentile(2001)}]{Gentile:01}
Gentile, N.~A. 2001, J. Comput. Phys., 172, 543

\bibitem[{Gentile(2009)}]{Gentile:09}
Gentile, N.~A. 2009, in In Proc. International Conference on Mathematics,
  Computational Methods \& Reactor Physics (M \& C 2009), Vol. May 3-7, 2009
  (American Nuclear Society, LaGrange Park, IL), 1170

\bibitem[{Godoy \& Liu(2012)}]{Godoy:12}
Godoy, W.~F., \& Liu, X. 2012, Journal of Computational Physics, 231, 4257

\bibitem[{Goodale {et~al.}(2003)Goodale, Allen, Lanfermann, Mass{\'o}, Radke,
  Seidel, \& Shalf}]{goodale:03}
Goodale, T., Allen, G., Lanfermann, G., Mass{\'o}, J., Radke, T., Seidel, E.,
  \& Shalf, J. 2003, in Vector and Parallel Processing -- VECPAR'2002, 5th
  International Conference, Lecture Notes in Computer Science (Berlin:
  Springer)

\bibitem[{{Habetler} \& {Matkowsky}(1975)}]{Habetler:75}
{Habetler}, G.~J., \& {Matkowsky}, B.~J. 1975, J. Math. Phys., 16, 856

\bibitem[{{Hanke} {et~al.}(2011){Hanke}, {Marek}, {Mueller}, \&
  {Janka}}]{Hanke:11}
{Hanke}, F., {Marek}, A., {Mueller}, B., \& {Janka}, H.-T. 2011, ArXiv e-prints

\bibitem[{{Herant} {et~al.}(1992){Herant}, {Benz}, \& {Colgate}}]{Herant:92}
{Herant}, M., {Benz}, W., \& {Colgate}, S. 1992, \apj, 395, 642

\bibitem[{{Herant} {et~al.}(1994){Herant}, {Benz}, {Hix}, {Fryer}, \&
  {Colgate}}]{Herant:94}
{Herant}, M., {Benz}, W., {Hix}, W.~R., {Fryer}, C.~L., \& {Colgate}, S.~A.
  1994, \apj, 435, 339

\bibitem[{Hillebrandt \& Wolff(1985)}]{Hillebrandt:85}
Hillebrandt, W., \& Wolff, R.~G. 1985, {in Nucleosynthesis: Challenges and New
  Developments, ed. W. D. Arnett and J. W. Truran (Chicago, IL: Univ. Chicago,
  Press)}, 131

\bibitem[{{Hubeny} \& {Burrows}(2007)}]{Hubeny:07}
{Hubeny}, I., \& {Burrows}, A. 2007, \apj, 659, 1458

\bibitem[{{Hummer} \& {Rybicki}(1971)}]{Hummer:71}
{Hummer}, D.~G., \& {Rybicki}, G.~B. 1971, \mnras, 152, 1

\bibitem[{{Janka}(1991)}]{Janka:91}
{Janka}, H.-T. 1991, \aap, 244, 378

\bibitem[{{Janka}(1992)}]{Janka:92}
---. 1992, \aap, 256, 452

\bibitem[{{Janka}(2001)}]{Janka:01}
---. 2001, \aap, 368, 527

\bibitem[{{Janka} \& {Hillebrandt}(1989)}]{Janka:89}
{Janka}, H.-T., \& {Hillebrandt}, W. 1989, \aaps, 78, 375

\bibitem[{{Janka} \& {Mueller}(1996)}]{Janka:96}
{Janka}, H.-T., \& {Mueller}, E. 1996, \aap, 306, 167

\bibitem[{{Kalos} \& {Whitlock}(2008)}]{Kalos:08}
{Kalos}, M.~H., \& {Whitlock}, P.~A. 2008, {Monte Carlo Methods: Second Revised
  and Enlarged Edition}, ed. {Kalos, M.~H.~\& Whitlock, P.~A.} (Wiley-VCH
  Verlag)

\bibitem[{{Kasen} {et~al.}(2011){Kasen}, {Woosley}, \& {Heger}}]{Kasen:11}
{Kasen}, D., {Woosley}, S.~E., \& {Heger}, A. 2011, \apj, 734, 102

\bibitem[{{Keil} {et~al.}(2003){Keil}, {Raffelt}, \& {Janka}}]{Keil:03}
{Keil}, M.~T., {Raffelt}, G.~G., \& {Janka}, H.-T. 2003, \apj, 590, 971

\bibitem[{{Kitaura} {et~al.}(2006){Kitaura}, {Janka}, \&
  {Hillebrandt}}]{Kitaura:06}
{Kitaura}, F.~S., {Janka}, H.-T., \& {Hillebrandt}, W. 2006, \aap, 450, 345

\bibitem[{Larsen \& Mercier(1987)}]{Larsen:87}
Larsen, E.~W., \& Mercier, B. 1987, J. Comput. Phys., 71, 50

\bibitem[{Lattimer \& Swesty(1991)}]{Lattimer:91}
Lattimer, J.~M., \& Swesty, F.~D. 1991, {Nucl. Phys. A}, 535, 331

\bibitem[{{Liebend{\"o}rfer} {et~al.}(2004){Liebend{\"o}rfer}, {Messer},
  {Mezzacappa}, {Bruenn}, {Cardall}, \& {Thielemann}}]{Liebendoerfer:04}
{Liebend{\"o}rfer}, M., {Messer}, O.~E.~B., {Mezzacappa}, A., {Bruenn}, S.~W.,
  {Cardall}, C.~Y., \& {Thielemann}, F.-K. 2004, \apjs, 150, 263

\bibitem[{{Liebend{\"o}rfer} {et~al.}(2001){Liebend{\"o}rfer}, {Mezzacappa},
  {Thielemann}, {Messer}, {Hix}, \& {Bruenn}}]{Liebendoerfer:01}
{Liebend{\"o}rfer}, M., {Mezzacappa}, A., {Thielemann}, F.-K., {Messer}, O.~E.,
  {Hix}, W.~R., \& {Bruenn}, S.~W. 2001, \prd, 63, 103004

\bibitem[{{Liebend{\"o}rfer} {et~al.}(2005){Liebend{\"o}rfer}, {Rampp},
  {Janka}, \& {Mezzacappa}}]{Liebendoerfer:05}
{Liebend{\"o}rfer}, M., {Rampp}, M., {Janka}, H.-T., \& {Mezzacappa}, A. 2005,
  \apj, 620, 840

\bibitem[{{Marek} \& {Janka}(2009)}]{Marek:09}
{Marek}, A., \& {Janka}, H.-T. 2009, \apj, 694, 664

\bibitem[{Martin \& Brown(2001)}]{Martin:01}
Martin, W.~R., \& Brown, F.~B. 2001, Trans. Amer. Nucl. Soc., 85, 329

\bibitem[{McClarren \& Hauck(2010)}]{McClarren:10}
McClarren, R.~G., \& Hauck, C.~D. 2010, Journal of Computational Physics, 229,
  5597

\bibitem[{McClarren {et~al.}(2008)McClarren, Holloway, \&
  Brunner}]{McClarren:08}
McClarren, R.~G., Holloway, J.~P., \& Brunner, T.~A. 2008, J. Comput. Phys.,
  227, 2864

\bibitem[{McClarren \& Urbatsch(2009)}]{McClarren:09}
McClarren, R.~G., \& Urbatsch, T.~J. 2009, J. Comput. Phys., 228, 5669

\bibitem[{{Mezzacappa} \& {Bruenn}(1993)}]{Mezzacappa:93a}
{Mezzacappa}, A., \& {Bruenn}, S.~W. 1993, \apj, 410, 740

\bibitem[{{Mezzacappa} {et~al.}(2007){Mezzacappa}, {Bruenn}, {Blondin}, {Hix},
  \& {Bronson Messer}}]{Mezzacappa:07}
{Mezzacappa}, A., {Bruenn}, S.~W., {Blondin}, J.~M., {Hix}, W.~R., \& {Bronson
  Messer}, O.~E. 2007, in American Institute of Physics Conference Series, Vol.
  924, The Multicolored Landscape of Compact Objects and Their Explosive
  Origins, ed. {T.~di Salvo, G.~L.~Israel, L.~Piersant, L.~Burderi, G.~Matt,
  A.~Tornambe, \& M.~T.~Menna}, 234

\bibitem[{{Mihalas} \& {Klein}(1982)}]{Mihalas:82}
{Mihalas}, D., \& {Klein}, R.~I. 1982, J. Comput. Phys., 46, 97

\bibitem[{{Mihalas} \& {Mihalas}(1984)}]{Mihalas:84}
{Mihalas}, D., \& {Mihalas}, B.~W. 1984, {Foundations of radiation
  hydrodynamics}, ed. {Mihalas, D.~\& Mihalas, B.~W.} (New York, Oxford
  University Press)

\bibitem[{{Morel} {et~al.}(2003){Morel}, {Wareing}, {Lowrie}, \&
  {Parsons}}]{Morel:03}
{Morel}, J.~E., {Wareing}, T.~A., {Lowrie}, R.~B., \& {Parsons}, D.~K. 2003,
  Nucl. Sci. Eng., 144, 1

\bibitem[{{Murphy} \& {Burrows}(2008)}]{Murphy:08}
{Murphy}, J.~W., \& {Burrows}, A. 2008, \apj, 688, 1159

\bibitem[{N'Kaoua(1991)}]{NKaoua:91}
N'Kaoua, T. 1991, SIAM J. Sci. and Stat. Comput., 12, 505

\bibitem[{{Nomoto} \& {Hashimoto}(1988)}]{nomoto:88}
{Nomoto}, K., \& {Hashimoto}, M.-A. 1988, \physrep, 163, 13

\bibitem[{{Nordhaus} {et~al.}(2010){Nordhaus}, {Burrows}, {Almgren}, \&
  {Bell}}]{Nordhaus:10}
{Nordhaus}, J., {Burrows}, A., {Almgren}, A., \& {Bell}, J. 2010, \apj, 720,
  694

\bibitem[{{Ott} {et~al.}(2008){Ott}, {Burrows}, {Dessart}, \& {Livne}}]{Ott:08}
{Ott}, C.~D., {Burrows}, A., {Dessart}, L., \& {Livne}, E. 2008, \apj, 685,
  1069

\bibitem[{{Pejcha} \& {Thompson}(2012)}]{Pejcha:12}
{Pejcha}, O., \& {Thompson}, T.~A. 2012, \apj, 746, 106

\bibitem[{{Pomraning}(1973)}]{Pomraning:73}
{Pomraning}, G.~C. 1973, {The equations of radiation hydrodynamics}, ed.
  {Pomraning, G.~C.}

\bibitem[{{Rampp} \& {Janka}(2000)}]{Rampp:00}
{Rampp}, M., \& {Janka}, H.-T. 2000, \apjl, 539, L33

\bibitem[{{Rampp} \& {Janka}(2002)}]{Rampp:02}
---. 2002, \aap, 396, 361

\bibitem[{{Schinder} \& {Bludman}(1989)}]{Schinder:89}
{Schinder}, P.~J., \& {Bludman}, S.~A. 1989, \apj, 346, 350

\bibitem[{{Smit} {et~al.}(1997){Smit}, {Cernohorsky}, \& {Dullemond}}]{Smit:97}
{Smit}, J.~M., {Cernohorsky}, J., \& {Dullemond}, C.~P. 1997, \aap, 325, 203

\bibitem[{{Sumiyoshi} \& {Yamada}(2012)}]{Sumiyoshi:12}
{Sumiyoshi}, K., \& {Yamada}, S. 2012, \apjs, 199, 17

\bibitem[{{Sumiyoshi} {et~al.}(2005){Sumiyoshi}, {Yamada}, {Suzuki}, {Shen},
  {Chiba}, \& {Toki}}]{Sumiyoshi:05}
{Sumiyoshi}, K., {Yamada}, S., {Suzuki}, H., {Shen}, H., {Chiba}, S., \&
  {Toki}, H. 2005, \apj, 629, 922

\bibitem[{{Suwa} {et~al.}(2010){Suwa}, {Kotake}, {Takiwaki}, {Whitehouse},
  {Liebend{\"o}rfer}, \& {Sato}}]{Suwa:10}
{Suwa}, Y., {Kotake}, K., {Takiwaki}, T., {Whitehouse}, S.~C.,
  {Liebend{\"o}rfer}, M., \& {Sato}, K. 2010, \pasj, 62, L49

\bibitem[{{Swesty}(2006)}]{Swesty:06b}
{Swesty}, F.~D. 2006, in Computational Methods in Transport, ed. {F.~Graziani}
  ({Springer}), 469--486

\bibitem[{{Swesty} \& {Myra}(2009)}]{Swesty:09}
{Swesty}, F.~D., \& {Myra}, E.~S. 2009, \apjs, 181, 1

\bibitem[{{Takahashi} {et~al.}(1978){Takahashi}, {El Eid}, \&
  {Hillebrandt}}]{Takahashi:78}
{Takahashi}, K., {El Eid}, M.~F., \& {Hillebrandt}, W. 1978, \aap, 67, 185

\bibitem[{{Takiwaki} {et~al.}(2012){Takiwaki}, {Kotake}, \&
  {Suwa}}]{Takiwaki:12}
{Takiwaki}, T., {Kotake}, K., \& {Suwa}, Y. 2012, \apj, 749, 98

\bibitem[{{Thomas}(1930)}]{Thomas:30}
{Thomas}, L.~H. 1930, Quart. J. Math., 1, 239

\bibitem[{{Thompson} {et~al.}(2003){Thompson}, {Burrows}, \&
  {Pinto}}]{Thompson:03}
{Thompson}, T.~A., {Burrows}, A., \& {Pinto}, P.~A. 2003, \apj, 592, 434

\bibitem[{{Thompson} {et~al.}(2005){Thompson}, {Quataert}, \&
  {Burrows}}]{Thompson:05}
{Thompson}, T.~A., {Quataert}, E., \& {Burrows}, A. 2005, \apj, 620, 861

\bibitem[{{Tubbs}(1978)}]{Tubbs:78}
{Tubbs}, D.~L. 1978, \apjs, 37, 287

\bibitem[{{Wollaber}(2008)}]{Wollaber:08}
{Wollaber}, A.~B. 2008, PhD thesis, {University of Michigan}

\bibitem[{Wollaber \& Larsen(2009)}]{Wollaber:09}
Wollaber, A.~B., \& Larsen, E.~W. 2009, in In Proc. International Conference on
  Mathematics, Computational Methods \& Reactor Physics (M \& C 2009), Vol. May
  3-7, 2009 (American Nuclear Society, LaGrange Park, IL), 1170--1179

\bibitem[{{Woosley} {et~al.}(2002){Woosley}, {Heger}, \& {Weaver}}]{whw:02}
{Woosley}, S.~E., {Heger}, A., \& {Weaver}, T.~A. 2002, Rev. Mod. Phys., 74,
  1015

\bibitem[{{Yakunin} {et~al.}(2010){Yakunin}, {Marronetti}, {Mezzacappa},
  {Bruenn}, {Lee}, {Chertkow}, {Hix}, {Blondin}, {Lentz}, {Bronson Messer}, \&
  {Yoshida}}]{Yakunin:10}
{Yakunin}, K.~N., {et~al.} 2010, Classical and Quantum Gravity, 27, 194005

\bibitem[{{Yamada} {et~al.}(1999){Yamada}, {Janka}, \& {Suzuki}}]{Yamada:99}
{Yamada}, S., {Janka}, H.-T., \& {Suzuki}, H. 1999, \aap, 344, 533

\bibitem[{{Zink}(2008)}]{Zink:08}
{Zink}, B. 2008, ArXiv e-prints, 0810.5349

\end{thebibliography}
\end{document}